\documentclass[12pt]{article}%
\usepackage[nosort]{cite}
\usepackage{graphicx}
\usepackage{multicol}
\usepackage{amsfonts}
\usepackage{amssymb}
\usepackage{amsmath}
\usepackage{heck}
\usepackage{afterpage}
\usepackage{setspace}
\usepackage{verbatim}
\usepackage{color}
\usepackage{longtable}
\usepackage{float}
\usepackage{subcaption}
\usepackage{epsfig}
\usepackage{epstopdf}
\usepackage{adjustbox}
\usepackage{tikz}
\usepackage[margin=1in]{geometry}
\usepackage{titletoc}
\usepackage{hyperref}%
\usepackage{mathrsfs}
\usepackage{dsfont}
\usepackage{algorithm}
\usepackage{algpseudocode}
\usepackage{upgreek}
\usepackage{mathtools}
\usepackage{stmaryrd}
\usepackage{lscape}
\usepackage{amsthm}
\usepackage{tikz-cd}
\usepackage{quiver}
\pdfoutput=1

\usepackage{caption}
\theoremstyle{definition}

\theoremstyle{definition}

\providecommand{\U}[1]{\protect\rule{.1in}{.1in}}

\newsavebox{\mysavebox}

\hypersetup{colorlinks,citecolor=black,filecolor=black,linkcolor=black,urlcolor=black}
\usetikzlibrary{decorations.markings}
\numberwithin{equation}{section}

\usepackage{tikz}
\usetikzlibrary{arrows}
\usetikzlibrary{arrows.meta}
\usetikzlibrary{arrows,decorations.pathmorphing}
\usetikzlibrary{shapes.geometric,calc,arrows, positioning,shapes.misc,decorations.markings}
\tikzset{
  big arrow/.style={
    decoration={markings,mark=at position 1 with {\arrow[scale=2,#1]{>}}},
    postaction={decorate},
    shorten >=0.4pt},
  big arrow/.default=black}

\usetikzlibrary{hobby}
\usetikzlibrary{decorations.markings}

\pgfdeclarelayer{edgelayer}
\pgfdeclarelayer{nodelayer}
\pgfsetlayers{edgelayer,nodelayer,main}
\tikzstyle{none}=[inner sep=0pt]

\tikzstyle{NodeCross}=[draw, shape=circle, cross out, inner sep=0pt, minimum size=10pt,line width=0.25mm]
\tikzstyle{SmallCircle}=[draw, shape=circle, black, fill=black, inner sep=0pt, minimum size=6pt]
\tikzstyle{BigCircle}=[draw, shape=circle, black, fill=black, inner sep=0pt, minimum size=16pt]
\tikzstyle{SmallCircleRed}=[draw, shape=circle, fill={rgb,255: red,191; green,0; blue,0}, inner sep=0pt, minimum size=6pt]
\tikzstyle{BigCircleRed}=[draw, shape=circle, fill={rgb,255: red,191; green,0; blue,0}, inner sep=0pt, minimum size=10pt]
\tikzstyle{SmallCircleBlue}=[draw, shape=circle, fill=blue, inner sep=0pt, minimum size=6pt]
\tikzstyle{BigCircleBlue}=[draw, shape=circle, fill=blue, inner sep=0pt, minimum size=10pt]
\tikzstyle{SmallCirclePurple}=[draw, shape=circle, fill={rgb,255: red,191; green,0; blue,191}, inner sep=0pt, minimum size=6pt]
\tikzstyle{BigCirclePurple}=[draw, shape=circle, fill={rgb,255: red,191; green,0; blue,191}, inner sep=0pt, minimum size=10pt]
\tikzstyle{SmallCircleGreen}=[draw, shape=circle,  fill={rgb,255: red,80; green,200; blue,120}, inner sep=0pt, minimum size=6pt]
\tikzstyle{BigCircleGreen}=[draw, shape=circle,  fill={rgb,255: red,80; green,200; blue,120}, inner sep=0pt, minimum size=10pt]
\tikzstyle{SmallCircleBrown}=[draw, shape=circle,  fill={rgb,255: red,210; green,105; blue,30}, inner sep=0pt, minimum size=6pt]
\tikzstyle{BigCircleBrown}=[draw, shape=circle,  fill={rgb,255: red,210; green,105; blue,30}, inner sep=0pt, minimum size=10pt]
\tikzstyle{Star}=[draw, shape=star, fill=black, star points=8, inner sep=0pt, minimum size=10pt]

\tikzstyle{DashedLine}=[-, densely dashed, line width=0.25mm]
\tikzstyle{DottedLine}=[-, dotted, line width=0.25mm]
\tikzstyle{ThickLine}=[-, line width=0.25mm]
\tikzstyle{RedLine}=[-, draw={rgb,255: red,191; green,0; blue,0}, fill=none, line width=0.5mm]
\tikzstyle{DashedRedLine}=[-, densely dashed, draw={rgb,255: red,191; green,0; blue,0}, fill=none, line width=0.5mm]
\tikzstyle{DottedRed}=[-, dotted, draw={rgb,255: red,191; green,0; blue,0}, fill=none, dotted, line width=0.5mm]
\tikzstyle{BlueLine}=[-, draw=blue, fill=none, line width=0.5mm]
\tikzstyle{DashedBlueLine}=[-, densely dashed, draw=blue, fill=none, line width=0.5mm]
\tikzstyle{DottedBlue}=[-, dotted, draw=blue, fill=none, dotted, line width=0.5mm]
\tikzstyle{PurpleLine}=[-, draw={rgb,255: red,191; green,0; blue,191}, fill=none, line width=0.5mm]
\tikzstyle{DashedPurpleLine}=[-, densely dashed, draw={rgb,255: red,191; green,0; blue,191}, fill=none, line width=0.5mm]
\tikzstyle{DottedPurple}=[-, dotted, draw={rgb,255: red,191; green,0; blue,191}, fill=none, dotted, line width=0.5mm]
\tikzstyle{GreenLine}=[-, draw={rgb,255: red,80; green,200; blue,120}, fill=none, line width=0.5mm]
\tikzstyle{DashedGreenLine}=[-, densely dashed, draw={rgb,255: red,80; green,200; blue,120}, fill=none, line width=0.5mm]
\tikzstyle{DottedGreen}=[-, dotted, draw={rgb,255: red,80; green,200; blue,120}, fill=none, dotted, line width=0.5mm]
\tikzstyle{BrownLine}=[-, draw={rgb,255: red,210; green,105; blue,30}, fill=none, line width=0.5mm]
\tikzstyle{DashedBrownLine}=[-, densely dashed, draw={rgb,255: red,210; green,105; blue,30}, fill=none, line width=0.5mm]
\tikzstyle{DottedBrown}=[-, dotted, draw={rgb,255: red,210; green,105; blue,30}, fill=none, dotted, line width=0.5mm]
\tikzstyle{ArrowLineRight}=[-,  -{Stealth[scale=1.75]}, line width=0.25mm, scale=5]
\tikzstyle{ArrowLineRed}=[-, draw={rgb,255: red,191; green,0; blue,0},  -{Stealth[scale=1.75]}, line width=0.25mm, scale=5]
\tikzstyle{ArrowLineBlue}=[-, draw=blue,  -{Stealth[scale=1.75]}, line width=0.25mm, scale=5]
\tikzstyle{ArrowLinePurple}=[-, draw={rgb,255: red,191; green,0; blue,191},  -{Stealth[scale=1.75]}, line width=0.25mm, scale=5]
\tikzstyle{ArrowLineGreen}=[-, draw={rgb,255: red,80; green,200; blue,120},  -{Stealth[scale=1.75]}, line width=0.5mm, scale=5]
\tikzstyle{ArrowLineBrown}=[-, draw={rgb,255: red,210; green,105; blue,30},  -{Stealth[scale=1.75]}, line width=0.25mm, scale=5]

\tikzset{snake it/.style={decorate, decoration=snake}}
\tikzset{
dashstar/.style={
 dash pattern=on 5pt off 5pt,
 postaction={
  decorate,
  decoration={
   markings,
   mark=between positions 9pt and 1 step 10pt with {
     \node[color=red] {*};
   }
  }
 }
},
dashstarstar/.style={ 
 dash pattern=on 5pt off 10pt,
 postaction={
   decorate,
   decoration={
     markings,
     mark=between positions 10pt and 1
          step 15pt
           with {
            \node at (-2pt,0pt) {\pgfuseplotmark{star}};
            \node at (2pt,0pt) {\pgfuseplotmark{star}};
           }
   }
 }
}
}
\usepackage{pgfplots}
\pgfplotsset{compat=1.16}

\newcommand{\lb}{\left(}
\newcommand{\rb}{\right)}
\newcommand{\lbb}{\left[}
\newcommand{\rbb}{\right]}

\newcommand{\be}{\begin{equation}}
\newcommand{\ee}{\end{equation}}
\newcommand{\ba}{\begin{aligned}}
\newcommand{\ea}{\end{aligned}}
\newcommand{\Z}{{\mathbb Z}}

\newcommand{\C}{{\mathbb C}}

\begin{document}

\begin{flushright}
    UUITP-04/26
\end{flushright}

\thispagestyle{empty}

\date{March 2026}

\title{On the SymTFTs of \\[4mm] Finite Non-Abelian Symmetries}

\institution{Technion}{\centerline{$^{1}$Department of Physics, Technion, Israel Institute of Technology, Haifa, 32000, Israel}}
\institution{PENN}{\centerline{$^{2}$Department of Physics and Astronomy, University of Pennsylvania, Philadelphia, PA 19104, USA}}
\institution{PENNmath}{\centerline{$^{3}$Department of Mathematics, University of Pennsylvania, Philadelphia, PA 19104, USA}}
\institution{UPPSALAPHYS}{\centerline{${}^{4}$Department of Physics and Astronomy, Uppsala University, Box 516, SE-75120 Uppsala, Sweden}}
\institution{CMSA}{\centerline{$^{5}$Center of Mathematical Sciences and Applications, Harvard University,
Cambridge, MA 02138, USA}}
\institution{Jefferson}{\centerline{$^{6}$Jeﬀerson Physical Laboratory, Harvard University, Cambridge, MA 02138, USA}}
\institution{UPPSALAMATH}{\centerline{$^{7}$Mathematics Institute, Uppsala University, Box 480, SE-75106 Uppsala, Sweden}}
\institution{CenterGP}{\centerline{$^{8}$Center For Geometry and Physics, Uppsala University, Box 480, SE-75106 Uppsala, Sweden}}
\institution{VIRGINIATECH}{\centerline{${}^{9}$Department of Physics MC 0435, 850 West Campus Drive, Virginia Tech, Blacksburg, VA 24061, USA}}
\institution{IPMU}{\centerline{$^{10}$Kavli IPMU, University of Tokyo, Kashiwa, Chiba 277-8583, Japan}}

\authors{\vspace{15pt} Oren Bergman\worksat{\Technion}\footnote{e-mail: \texttt{bergman@physics.technion.ac.il}},
Jonathan J. Heckman\worksat{\PENN,\PENNmath}\footnote{e-mail: \texttt{jheckman@sas.upenn.edu}},
Max H\"ubner\worksat{\UPPSALAPHYS,\CMSA,\Jefferson}\footnote{e-mail: \texttt{max@cmsa.fas.harvard.edu}}, \\[4mm]
Daniele Migliorati\worksat{\UPPSALAMATH,\CenterGP}\footnote{e-mail: \texttt{daniele.migliorati@math.uu.se}},
Xingyang Yu\worksat{\VIRGINIATECH}\footnote{e-mail: \texttt{xingyangy@vt.edu}},
and Hao Y. Zhang\worksat{\IPMU}\footnote{e-mail: \texttt{hao.zhang@ipmu.jp}}
}

\longabstract{The $(D+1)$-dimensional symmetry topological field theory (SymTFT$_{D+1}$)
of a $D$-dimensional absolute quantum field theory (QFT$_D$) provides a
topological characterization of symmetry data. In this framework, the SymTFT comes equipped with a physical boundary specifying a relative QFT, and a topological boundary which specifies the global form of symmetries. In general, there need not be a unique bulk theory which encodes this information but it is often helpful to have a more manifest presentation of symmetries in terms of bulk degrees of freedom. For the case of a finite non-Abelian symmetry group $G$, the bulk SymTFT may be described by a Dijkgraaf-Witten TFT with gauge group $G$. This makes manifest the ``electric'' presentation of the symmetry data but can obscure some of the magnetic data as well as non-Abelian structure present in the absolute QFT$_D$ such as symmetry operators which cannot fully detach from the topological boundary. We address these issues for 3D SymTFTs by constructing discrete BF-like theory Lagrangians for finite groups which admit a presentation as an extension by a finite Abelian group and a finite (possibly non-Abelian) group. This enables us to give a streamlined approach to reconstructing the fusion rules of the accompanying Drinfeld center, but also allows us to construct surface-attaching non-genuine line operators associated directly with non-Abelian group elements rather than just their conjugacy classes. We also sketch how our treatment generalizes to higher-dimensional SymTFTs.}

\maketitle
\tableofcontents
\enlargethispage{\baselineskip}

\setcounter{tocdepth}{2}


\section{Introduction}

One of the insights of recent years is the appearance of deep topological structures in the global symmetries of a $D$-dimensional QFT$_D$ \cite{Gaiotto:2014kfa}. These structures can be isolated and packaged in terms of a higher-dimensional bulk symmetry theory,\footnote{See for example \cite{Reshetikhin:1991tc,Turaev:1992hq, Barrett:1993ab, Witten:1998wy, Fuchs:2002cm, Kirillov2010TuraevViroIA,
Kapustin:2010if, Kitaev:2011dxc, Fuchs:2012dt, Freed:2012bs, Kong:2014qka, Kong:2017hcw, Heckman:2017uxe, Freed:2018cec,
Gaiotto:2020iye, Kong:2020cie, Apruzzi:2021nmk, Freed:2022qnc, Kaidi:2022cpf, Antinucci:2022vyk, Lawrie:2023tdz, Baume:2023kkf, Yu:2023nyn, Brennan:2024fgj,Antinucci:2024zjp, DelZotto:2024tae, Argurio:2024oym, Franco:2024mxa, Heckman:2024zdo, Gagliano:2024off, Cordova:2024iti, Cvetic:2024dzu, Bhardwaj:2024igy, Bonetti:2024cjk, Apruzzi:2024htg, Yu:2024jtk, Jia:2025jmn,  Apruzzi:2025mdl, Heckman:2025lmw, Pace:2025hpb, Luo:2025phx, Apruzzi:2025hvs, Torres:2025jcb} and references therein for a partial list of references to foundational early work, as well as more recent generalizations.} relative to which the remaining non-topological features of the original QFT$_D$ are then specified.

\begin{figure}
\centering
\scalebox{0.75}{
\begin{tikzpicture}
\begin{pgfonlayer}{nodelayer}
		\node [style=none] (1) at (-5, 1) {};
		\node [style=none] (2) at (-3, -1) {};
		\node [style=none] (3) at (-3, 3) {};
		\node [style=none] (4) at (-5, -3) {};
		\node [style=none] (5) at (-4, 0.75) {};
		\node [style=none] (6) at (-4, -0.75) {};
		\node [style=none] (7) at (-4, 1) {};
		\node [style=none] (8) at (-4, -1) {};
		\node [style=none] (9) at (5, 1) {};
		\node [style=none] (10) at (7, -1) {};
		\node [style=none] (11) at (7, 3) {};
		\node [style=none] (12) at (5, -3) {};
		\node [style=none] (17) at (1, 1) {};
		\node [style=none] (18) at (3, -1) {};
		\node [style=none] (19) at (3, 3) {};
		\node [style=none] (20) at (1, -3) {};
		\node [style=none] (21) at (0, 0) {};
		\node [style=none] (22) at (-2, 0) {};
		\node [style=none] (23) at (-1, 0.5) {decompress};
		\node [style=none] (24) at (-4, -4) {QFT$_D$};
		\node [style=none] (25) at (4, -4) {SymTFT$_{D+1}$};
		\node [style=none] (30) at (4, 0.75) {};
		\node [style=none] (31) at (4, -0.75) {};
		\node [style=none] (32) at (4, 1) {};
		\node [style=none] (33) at (4, -1) {};
		\node [style=none] (34) at (2, 0.75) {};
		\node [style=none] (35) at (2, -0.75) {};
		\node [style=none] (36) at (2, 1) {};
		\node [style=none] (37) at (2, -1) {};
		\node [style=none] (38) at (6.625, -2.75) {$|\text{Phys}\rangle$};
		\node [style=none] (39) at (1.375, 2.75) {$\langle \text{Top}|$};
	\end{pgfonlayer}
	\begin{pgfonlayer}{edgelayer}
		\filldraw[fill=blue!15, draw=blue!15]  (1, -3) -- (1, 1) -- (3, 3) -- (3, -1) -- cycle;
		\filldraw[fill=red!15, draw=red!15]  (5, -3) -- (5, 1) -- (7, 3) -- (7, -1) -- cycle;
		\filldraw[fill=gray!50, draw=gray!50]  (2, 0.75) -- (2, -0.75) -- (4, -0.75) -- (4, 0.75) -- cycle;
		\draw [style=ThickLine] (1.center) to (4.center);
		\draw [style=ThickLine] (4.center) to (2.center);
		\draw [style=ThickLine] (2.center) to (3.center);
		\draw [style=ThickLine] (3.center) to (1.center);
		\draw [style=BlueLine] (5.center) to (6.center);
		\draw [style=DottedBlue] (7.center) to (5.center);
		\draw [style=DottedBlue] (6.center) to (8.center);
		\draw [style=ThickLine] (9.center) to (12.center);
		\draw [style=ThickLine] (12.center) to (10.center);
		\draw [style=ThickLine] (10.center) to (11.center);
		\draw [style=ThickLine] (11.center) to (9.center);
		\draw [style=ThickLine] (17.center) to (20.center);
		\draw [style=ThickLine] (19.center) to (17.center);
		\draw [style=ArrowLineRight] (22.center) to (21.center);
		\draw [style=ThickLine] (20.center) to (12.center);
		\draw [style=ThickLine] (17.center) to (9.center);
		\draw [style=ThickLine] (11.center) to (19.center);
		\draw [style=DashedLine] (19.center) to (18.center);
		\draw [style=DashedLine] (18.center) to (20.center);
		\draw [style=DashedLine] (18.center) to (10.center);
		\draw [style=BlueLine] (30.center) to (31.center);
		\draw [style=BrownLine] (34.center) to (35.center);
		\draw [style=DottedBrown] (36.center) to (34.center);
		\draw [style=DottedBrown] (35.center) to (37.center);
		\draw [style=DottedBlue] (31.center) to (33.center);
		\draw [style=DottedBlue] (32.center) to (30.center);
	\end{pgfonlayer}
\end{tikzpicture}}
\caption{In the SymTFT framework an absolute $D$-dimensional QFT$_D$ is ``decompressed'' to a relative QFT$_{D}$, (i.e., a physical boundary condition) and a topological boundary condition, with a bulk SymTFT$_{D+1}$ capturing the symmetry category $\mathcal{C}$ of the absolute QFT$_{D}$. Symmetry operators of the absolute theory specify objects in the Drinfeld center $\mathcal{Z}(\mathcal{C})$, and topologically link with defects which stretch from the physical to topological boundaries. A change of polarization in the topological boundary condition amounts to modifying the roles of the symmetry operators and defects. For non-Abelian symmetries, some of the symmetry operators of the absolute QFT$_D$ fail to fully detach from the topological boundary; these are instead captured by a flux tube which extends back to the boundary. The appearance of such non-genuine operators in the bulk SymTFT$_{D+1}$ which are not captured by just $\mathcal{Z}(\mathcal{C})$ is a common feature of non-Abelian symmetries.}
\label{fig:DecomPress}
\end{figure}
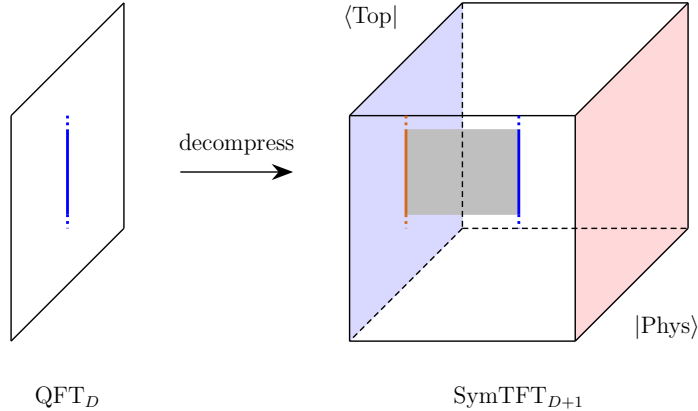

A particularly well-studied class of examples are QFTs with finite global symmetries. The bulk system in this case is a $(D+1)$-dimensional TFT known as the symmetry topological field theory (SymTFT$_{D+1}$). For a QFT$_D$ on a $D$-dimensional manifold $M_D$, the SymTFT$_{D+1}$ is specified on the manifold  $I \times M_D$, with $I = [0,1]$ the unit interval. At one end, we have a relative QFT specifying the so-called physical boundary condition and at the other end we have the so-called topological boundary condition specifying the global form of the QFT$_D$. In the standard TFT formalism, we can view these boundaries as specifying states $\vert \mathrm{phys} \rangle$, $\vert \mathrm{top} \rangle$, and a corresponding partition function $Z = \langle \mathrm{top} \vert \mathrm{phys} \rangle$ (see figure \ref{fig:DecomPress} for a depiction of the SymTFT$_{D+1}$ and a symmetry operator which cannot fully detach from the topological boundary).

This change of perspective is especially helpful in enumerating the symmetry operators, charged defects which link with these operators, as well as possible anomalies obstructing the gauging of global symmetries, as encoded in twist terms of the bulk TFT. A priori, there need not be a unique way of presenting these physical features and different (Morita) equivalent presentations exist. Different presentations are favored depending on how the SymTFT$_{D+1}$ is constructed. For example, when the QFT$_D$ itself arises as a defect theory in some higher-dimensional system, then the latter often determines aspects of the SymTFT$_{D+1}$ and it may be convenient to present the SymTFT$_{D+1}$ in structures originating from this progenitor. Conversely, when manipulating the QFT$_D$, for example by coupling it to gravity, different perspectives and presentations are also of varying utility.

The best studied examples of this sort are when the 0-form symmetry of the absolute QFT$_D$ is a finite Abelian group. In this case, the general class of theories are captured by BF-like theories for each of the cyclic factors appearing in $G = \oplus_i \mathbb{Z} / p_i^{k_i} \mathbb{Z}$, i.e., they come with topological terms of the form $b \cup \delta a$. This is a rather natural class of examples to consider, and also arises in many top down constructions of QFTs.\footnote{See e.g., \cite{Apruzzi:2021nmk, Antinucci:2022vyk, Apruzzi:2023uma, Baume:2023kkf, Yu:2023nyn, Bonetti:2024cjk, Apruzzi:2024htg, Cvetic:2024dzu, Yu:2024jtk, Heckman:2025lmw }. The generic structure of these top down realizations of SymTFTs are often intrinsically  Abelian simply because they descend from generalized Maxwell or Abelian Chern-Simons terms in the corresponding supergravity effective action.}

\begin{figure}
\centering
\scalebox{0.8}{
\begin{tikzpicture}
	\begin{pgfonlayer}{nodelayer}
		\node [style=none] (0) at (-8.5, -2.5) {};
		\node [style=none] (1) at (-7, -1.5) {};
		\node [style=none] (2) at (-8.5, 1.5) {};
		\node [style=none] (3) at (-7, 2.5) {};
		\node [style=none] (4) at (-3, -2.5) {};
		\node [style=none] (5) at (-1.5, -1.5) {};
		\node [style=none] (6) at (-3, 1.5) {};
		\node [style=none] (7) at (-1.5, 2.5) {};
		\node [style=none] (8) at (-5.75, 1) {};
		\node [style=none] (9) at (-5.75, -1) {};
		\node [style=none] (10) at (-5.75, -1.25) {};
		\node [style=none] (11) at (-5.75, 1.25) {};
		\node [style=none] (12) at (-4.25, 1) {};
		\node [style=none] (13) at (-4.25, -1) {};
		\node [style=none] (14) at (-4.25, -1.25) {};
		\node [style=none] (15) at (-4.25, 1.25) {};
		\node [style=none] (16) at (-5.5, -0.5) {};
		\node [style=none] (17) at (-4.5, -0.5) {};
		\node [style=none] (18) at (-5.5, 0.5) {};
		\node [style=none] (19) at (-4.5, 0.5) {};
		\node [style=none] (20) at (1.5, -2.5) {};
		\node [style=none] (21) at (3, -1.5) {};
		\node [style=none] (22) at (1.5, 1.5) {};
		\node [style=none] (23) at (3, 2.5) {};
		\node [style=none] (24) at (7, -2.5) {};
		\node [style=none] (25) at (8.5, -1.5) {};
		\node [style=none] (26) at (7, 1.5) {};
		\node [style=none] (27) at (8.5, 2.5) {};
		\node [style=none] (28) at (5, 1.25) {};
		\node [style=none] (29) at (5, -0.75) {};
		\node [style=none] (30) at (5, -1) {};
		\node [style=none] (31) at (5, 1.5) {};
		\node [style=none] (32) at (2, 0.75) {};
		\node [style=none] (33) at (2, -1.25) {};
		\node [style=none] (34) at (2, -1.5) {};
		\node [style=none] (35) at (2, 1) {};
		\node [style=none] (36) at (2.5, 1.25) {};
		\node [style=none] (37) at (2.5, -0.75) {};
		\node [style=none] (38) at (2.5, -1) {};
		\node [style=none] (39) at (2.5, 1.5) {};
		\node [style=none] (40) at (4.5, 0.75) {};
		\node [style=none] (41) at (4.5, -1.25) {};
		\node [style=none] (42) at (4.5, -1.5) {};
		\node [style=none] (43) at (4.5, 1) {};
		\node [style=none] (44) at (-5, -3.5) {(i)};
		\node [style=none] (45) at (5, -3.5) {(ii)};
        \node [style=none] (45) at (5, -4) {};
	\end{pgfonlayer}
	\begin{pgfonlayer}{edgelayer}
        \filldraw[fill=red!15, draw=red!15]  (-3,1.5) -- (-1.5, 2.5) -- (-1.5, -1.5) -- (-3, -2.5) -- cycle;
        \filldraw[fill=blue!15, draw=blue!15]  (3,2.5) -- (1.5, 1.5) -- (1.5, -2.5) -- (3, -1.5) -- cycle;
        \filldraw[fill=blue!15, draw=blue!15]  (-8.5,1.5) -- (-7, 2.5) -- (-7, -1.5) -- (-8.5, -2.5) -- cycle;
        \filldraw[fill=red!15, draw=red!15]  (8.5,2.5) -- (7, 1.5) -- (7, -2.5) -- (8.5, -1.5) -- cycle;
        \filldraw[fill=gray!50, draw=gray!50]  (2, 0.75) -- (2, -1.25) -- (4.5, -1.25) -- (4.5, 0.75) -- cycle;
        \filldraw[fill=gray!50, draw=gray!50]  (2.5, 1.25) -- (2.5, -0.75) -- (5, -0.75) -- (5, 1.25) -- cycle;
        \filldraw[fill=gray!70, draw=gray!70]  (2.5, 0.75) -- (2.5, -0.75) -- (4.5, -0.75) -- (4.5, 0.75) -- cycle;
		\draw [style=ThickLine] (4.center) to (0.center);
		\draw [style=ThickLine] (0.center) to (2.center);
		\draw [style=ThickLine] (2.center) to (3.center);
		\draw [style=ThickLine] (3.center) to (7.center);
		\draw [style=ThickLine] (7.center) to (6.center);
		\draw [style=ThickLine] (6.center) to (2.center);
		\draw [style=ThickLine] (6.center) to (4.center);
		\draw [style=ThickLine] (4.center) to (5.center);
		\draw [style=ThickLine] (5.center) to (7.center);
		\draw [style=DottedLine] (0.center) to (1.center);
		\draw [style=DottedLine] (1.center) to (3.center);
		\draw [style=DottedLine] (1.center) to (5.center);
		\draw [style=ArrowLineRight, bend left=330, looseness=1.1] (19.center) to (18.center);
		\draw [style=ArrowLineRight, bend right=30, looseness=1.1] (16.center) to (17.center);
		\draw [style=ThickLine] (24.center) to (20.center);
		\draw [style=ThickLine] (20.center) to (22.center);
		\draw [style=ThickLine] (22.center) to (23.center);
		\draw [style=ThickLine] (23.center) to (27.center);
		\draw [style=ThickLine] (27.center) to (26.center);
		\draw [style=ThickLine] (26.center) to (22.center);
		\draw [style=ThickLine] (26.center) to (24.center);
		\draw [style=ThickLine] (24.center) to (25.center);
		\draw [style=ThickLine] (25.center) to (27.center);
		\draw [style=DottedLine] (20.center) to (21.center);
		\draw [style=DottedLine] (21.center) to (23.center);
		\draw [style=DottedLine] (21.center) to (25.center);
		\draw [style=BlueLine] (28.center) to (29.center);
		\draw [style=DottedBlue] (29.center) to (30.center);
		\draw [style=DottedBlue] (31.center) to (28.center);
		\draw [style=BrownLine] (32.center) to (33.center);
		\draw [style=DottedBrown] (33.center) to (34.center);
		\draw [style=DottedBrown] (35.center) to (32.center);
		\draw [style=BrownLine] (36.center) to (37.center);
		\draw [style=DottedBrown] (37.center) to (38.center);
		\draw [style=DottedBrown] (39.center) to (36.center);
		\draw [style=BlueLine] (40.center) to (41.center);
		\draw [style=DottedBlue] (42.center) to (41.center);
		\draw [style=DottedBlue] (43.center) to (40.center);
		\draw [style=PurpleLine] (12.center) to (13.center);
		\draw [style=PurpleLine] (8.center) to (9.center);
		\draw [style=DottedPurple] (11.center) to (8.center);
		\draw [style=DottedPurple] (15.center) to (12.center);
		\draw [style=DottedPurple] (13.center) to (14.center);
		\draw [style=DottedPurple] (9.center) to (10.center);
	\end{pgfonlayer}
\end{tikzpicture}}
\caption{Subfigure (i) depicts two genuine codimension-2 operators whose fusion ring is commutative. Subfigure (ii) depicts two non-genuine codimension-2 operators each fixed to the boundary of a codimension-1 surface which further terminates on the topological boundary condition (blue) of the SymTFT slab. Non-commutative structure is manifest in the fusion ring of such operators. }
\label{fig:permute}
\end{figure}
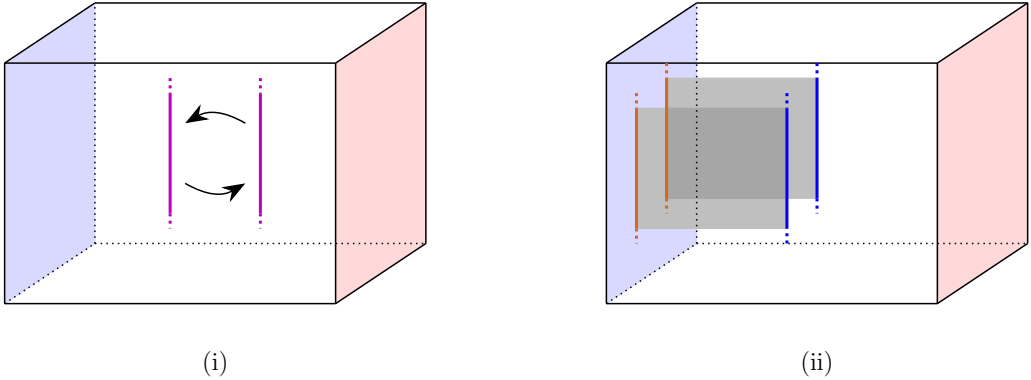

But non-Abelian symmetries also arise in many bottom up and top down realizations of absolute QFTs.\footnote{In top down constructions, these structures frequently arise either from a higher-dimensional ``flavor brane'' or from non-Abelian isometries of the non-compact extra-dimensional geometry itself.} This is possible for $0$-form symmetry operators because they are supported on codimension-one subspaces of the $D$-dimensional spacetime. Note, however, that in the SymTFT$_{D+1}$, any associated genuine topological operator would be supported in codimension-two and would freely commute with other operators  (see subfigure (i) in figure \ref{fig:permute}). An emphasis on genuine topological operators obscures this non-Abelian structure. Indeed, the objects which can fully detach from the topological boundary are labelled by conjugacy classes of $G$ (see \cite{Rudelius:2020orz}), i.e., Gukov-Witten operators \cite{Gukov:2006jk}, not directly the group elements themselves. This clearly obscures some of the non-Abelian structure present in $G$.

\begin{figure}
\centering
\scalebox{0.685}{
\begin{tikzpicture}
	\begin{pgfonlayer}{nodelayer}
		\node [style=none] (46) at (-12, -2.5) {};
		\node [style=none] (47) at (-10.5, -1.5) {};
		\node [style=none] (48) at (-12, 1.5) {};
		\node [style=none] (49) at (-10.5, 2.5) {};
		\node [style=none] (50) at (-6.5, -2.5) {};
		\node [style=none] (51) at (-5, -1.5) {};
		\node [style=none] (52) at (-6.5, 1.5) {};
		\node [style=none] (53) at (-5, 2.5) {};
		\node [style=none] (54) at (-9.25, 0.5) {};
		\node [style=none] (55) at (-9.25, -0.5) {};
		\node [style=none] (56) at (-7.75, 0.5) {};
		\node [style=none] (57) at (-7.75, -0.5) {};
		\node [style=none] (58) at (-3.5, -2.5) {};
		\node [style=none] (59) at (-2, -1.5) {};
		\node [style=none] (60) at (-3.5, 1.5) {};
		\node [style=none] (61) at (-2, 2.5) {};
		\node [style=none] (62) at (2, -2.5) {};
		\node [style=none] (63) at (3.5, -1.5) {};
		\node [style=none] (64) at (2, 1.5) {};
		\node [style=none] (65) at (3.5, 2.5) {};
		\node [style=none] (66) at (-0.75, 0.5) {};
		\node [style=none] (67) at (-0.75, -0.5) {};
		\node [style=none] (68) at (0.75, 0.5) {};
		\node [style=none] (69) at (0.75, -0.5) {};
		\node [style=none] (70) at (0.75, 0.75) {};
		\node [style=none] (71) at (0.75, -0.75) {};
		\node [style=none] (72) at (1.5, 0) {};
		\node [style=none] (74) at (0.225, 0.5) {};
		\node [style=none] (75) at (0.225, -0.5) {};
		\node [style=none] (76) at (5, -2.5) {};
		\node [style=none] (77) at (6.5, -1.5) {};
		\node [style=none] (78) at (5, 1.5) {};
		\node [style=none] (79) at (6.5, 2.5) {};
		\node [style=none] (80) at (10.5, -2.5) {};
		\node [style=none] (81) at (12, -1.5) {};
		\node [style=none] (82) at (10.5, 1.5) {};
		\node [style=none] (83) at (12, 2.5) {};
		\node [style=none] (84) at (7.75, 0.5) {};
		\node [style=none] (85) at (7.75, -0.5) {};
		\node [style=none] (86) at (9.25, 0.5) {};
		\node [style=none] (87) at (9.25, -0.5) {};
		\node [style=none] (88) at (-8.5, -3.5) {(i)};
		\node [style=none] (89) at (0, -3.5) {(ii)};
		\node [style=none] (90) at (8.5, -3.5) {(iii)};
        \node [style=none] (91) at (5, -4) {};
        \node [style=none] (92) at (9.28, -1) {$c$};
        \node [style=none] (92) at (-9.23, -1) {$a$};
        \node [style=none] (92) at (-7.75, -1) {$b$};
        \node [style=none] (92) at (7.77, -1) {$b$};
        \node [style=none] (92) at (0.75, -1.125) {$a$};
        \node [style=none] (92) at (-0.75, -1.125) {$b$};
	\end{pgfonlayer}
	\begin{pgfonlayer}{edgelayer}
        \filldraw[fill=blue!15, draw=blue!15]  (-3.5,1.5) -- (-2, 2.5) -- (-2, -1.5) -- (-3.5, -2.5) -- cycle;
        \filldraw[fill=blue!15, draw=blue!15]  (-12,1.5) -- (-10.5, 2.5) -- (-10.5, -1.5) -- (-12, -2.5) -- cycle;
        \filldraw[fill=red!15, draw=red!15]  (2,1.5) -- (3.5, 2.5) -- (3.5, -1.5) -- (2, -2.5) -- cycle;
        \filldraw[fill=blue!15, draw=blue!15]  (5,1.5) -- (6.5, 2.5) -- (6.5, -1.5) -- (5, -2.5) -- cycle;
        \filldraw[fill=red!15, draw=red!15]  (10.5,1.5) -- (12, 2.5) -- (12, -1.5) -- (10.5, -2.5) -- cycle;
        \filldraw[fill=red!15, draw=red!15]  (-6.5,1.5) -- (-5, 2.5) -- (-5, -1.5) -- (-6.5, -2.5) -- cycle;
        \draw[color=gray, fill=gray!15] (-7.75,0) ellipse (0.21cm and 0.5cm);
        \draw[color=green, fill=green!15] (-9.25,0) ellipse (0.21cm and 0.5cm);
        \draw[color=gray, fill=gray!15] (7.75,0) ellipse (0.21cm and 0.5cm);
        \draw[color=purple, fill=purple!15] (9.25,0) ellipse (0.21cm and 0.5cm);
        \draw[color=gray, fill=gray!15] (0.75,0) ellipse (0.75cm and 0.75cm);
        \draw[color=gray, fill=gray!15] (-0.75,0) ellipse (0.21cm and 0.5cm);
        \filldraw[fill=gray!15, draw=gray!15]  (-0.75,0.5) -- (-0.75, -0.5) -- (0.75, -0.5) -- (0.75, 0.5) -- cycle;
        \draw[color=green, fill=green!7.5] (0.75,0) ellipse (0.21cm and 0.5cm);
		\draw [style=ThickLine] (50.center) to (46.center);
		\draw [style=ThickLine] (46.center) to (48.center);
		\draw [style=ThickLine] (48.center) to (49.center);
		\draw [style=ThickLine] (49.center) to (53.center);
		\draw [style=ThickLine] (53.center) to (52.center);
		\draw [style=ThickLine] (52.center) to (48.center);
		\draw [style=ThickLine] (52.center) to (50.center);
		\draw [style=ThickLine] (50.center) to (51.center);
		\draw [style=ThickLine] (51.center) to (53.center);
		\draw [style=DottedLine] (46.center) to (47.center);
		\draw [style=DottedLine] (47.center) to (49.center);
		\draw [style=DottedLine] (47.center) to (51.center);
		\draw [style=GreenLine, bend left=90, looseness=0.75] (54.center) to (55.center);
		\draw [style=GreenLine, bend right=90, looseness=0.75] (54.center) to (55.center);
		\draw [style=BlueLine, bend left=90, looseness=0.75] (56.center) to (57.center);
		\draw [style=BlueLine, bend right=90, looseness=0.75] (56.center) to (57.center);
		\draw [style=ThickLine] (62.center) to (58.center);
		\draw [style=ThickLine] (58.center) to (60.center);
		\draw [style=ThickLine] (60.center) to (61.center);
		\draw [style=ThickLine] (61.center) to (65.center);
		\draw [style=ThickLine] (65.center) to (64.center);
		\draw [style=ThickLine] (64.center) to (60.center);
		\draw [style=ThickLine] (64.center) to (62.center);
		\draw [style=ThickLine] (62.center) to (63.center);
		\draw [style=ThickLine] (63.center) to (65.center);
		\draw [style=DottedLine] (58.center) to (59.center);
		\draw [style=DottedLine] (59.center) to (61.center);
		\draw [style=DottedLine] (59.center) to (63.center);
		\draw [style=BlueLine, bend right=90, looseness=0.75] (66.center) to (67.center);
		\draw [style=GreenLine, bend left=90, looseness=0.75] (68.center) to (69.center);
		\draw [style=GreenLine, bend right=90, looseness=0.75] (68.center) to (69.center);
		\draw [style=DottedBlue, bend left=90, looseness=0.75] (66.center) to (67.center);
		\draw [bend left=45] (70.center) to (72.center);
		\draw [bend left=45] (72.center) to (71.center);
		\draw (66.center) to (74.center);
		\draw (67.center) to (75.center);
		\draw [bend left, looseness=0.75] (74.center) to (70.center);
		\draw [bend right, looseness=0.75] (75.center) to (71.center);
		\draw [style=ThickLine] (80.center) to (76.center);
		\draw [style=ThickLine] (76.center) to (78.center);
		\draw [style=ThickLine] (78.center) to (79.center);
		\draw [style=ThickLine] (79.center) to (83.center);
		\draw [style=ThickLine] (83.center) to (82.center);
		\draw [style=ThickLine] (82.center) to (78.center);
		\draw [style=ThickLine] (82.center) to (80.center);
		\draw [style=ThickLine] (80.center) to (81.center);
		\draw [style=ThickLine] (81.center) to (83.center);
		\draw [style=DottedLine] (76.center) to (77.center);
		\draw [style=DottedLine] (77.center) to (79.center);
		\draw [style=DottedLine] (77.center) to (81.center);
		\draw [style=BlueLine, bend right=90, looseness=0.75] (84.center) to (85.center);
		\draw [style=PurpleLine, bend left=90, looseness=0.75] (86.center) to (87.center);
		\draw [style=PurpleLine, bend right=90, looseness=0.75] (86.center) to (87.center);
		\draw [style=BlueLine, bend left=90, looseness=0.75] (84.center) to (85.center);
	\end{pgfonlayer}
\end{tikzpicture}}
\caption{Non-commutativity is manifest in the bulk via the spectrum of non-genuine operators and their interactions. Subfigure (i) shows two non-genuine codimension-2 operators supported on the boundary of a single-boundary codimension-1 surface. Subfigure (ii) deforms these such that the surface attaching to $b$ engulfs $a$. Subfigure (iii) shows a contraction of $b$'s surface onto $a$ resulting in $c$ assuming the self-fusion of the codimension-1 surface is invertible. }
\label{fig:ItsActuallyBulkData}
\end{figure}
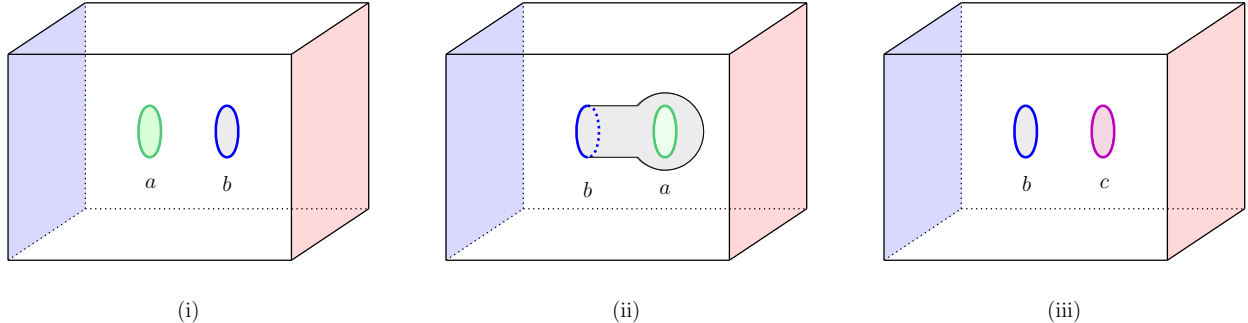

Indeed, on general grounds, one expects that whenever one has a non-Abelian symmetry, there exist symmetry operators of the absolute QFT$_D$ which cannot fully detach from the topological boundary. Rather, they stretch back to the boundary via a ``flux tube'' configuration \cite{Heckman:2024oot}, i.e., they stretch back to the boundary via a codimension-one object in the bulk. Treating the symmetry operator as its own (invertible) quantum system suggests democratically decompressing the QFT$_D$ with a symmetry operator inserted by also decompressing the symmetry operator itself (see figure \ref{fig:DecomPress}). Then, the non-commutative structure of $G$ is manifest when deforming the resulting non-genuine bulk objects past another. Constructed in this fashion, the non-genuine bulk operators attach back to the topological boundary (see subfigure (ii) in figure \ref{fig:permute}). The resulting operators constructed in this way are really codimension-one in the bulk, but with a boundary (terminated by the original 0-form symmetry operator). As such, these flux tubes can be detached from the boundary and are in principle captured by structures in the bulk SymTFT$_{D+1}$. See figure \ref{fig:ItsActuallyBulkData} for a depiction of this procedure.

This motivates some natural questions. For an absolute QFT$_D$ with $0$-form symmetry a finite non-Abelian group $G$:
\begin{itemize}
\item How does the SymTFT track the non-Abelian structure of $G$?
\item Is there a BF-like theory for a non-Abelian $G$?
\end{itemize}

Our aim in this work will be to find a Lagrangian / path integral formulation of a SymTFT$_{D+1}$ with manifest non-Abelian structure in the bulk. With the explicit field content in hand, constructing symmetry operators and flux tube configurations in principle becomes a tractable problem. We primarily focus on the case of 2D QFTs and their corresponding 3D SymTFTs, though we also sketch how the procedure generalizes to a QFT$_D$ and its corresponding SymTFT$_{D+1}$.

Restricting to 3D SymTFTs for finite non-Abelian symmetries, the relevant fusion category for describing the 0-form symmetry group $G$ is $\mathcal{C} = \mathrm{Vec}_G^{\omega}$ consisting of $G$-graded vector spaces with associator $\omega\in H^3(G,U(1))$. Genuine bulk operators are organized into the Drinfeld center $\mathcal{Z}(\mathcal{C})$ whereas the non-genuine codimension-1 bulk operators capture structure of the endomorphism 2-category $\mathrm{Bimod}_2(\mathcal C)$. The latter structure is added on top of the former, indicating that a 3D SymTFT contains more data than its Drinfeld center. In $D > 2$ these distinctions between the Drinfeld center and the extended TFT become even more pronounced.

Now, at an abstract level, the relevant TFT for a finite symmetry $G$ is the Dijkgraaf-Witten theory for the finite group $G$ \cite{Dijkgraaf:1989pz}. That being said, capturing data such as the ``magnetic'' basis of operators as well as non-genuine operators
which cannot fully detach from the boundaries of the SymTFT setup is less manifest in this setup. With these motivations in mind, we seek out a more explicit presentation of the field content and a generalized BF-theory type action. From this point of view, a helpful example to keep in mind is the treatment of $\mathbb{Z}_2^{3}$ gauge theory given in \cite{deWildPropitius:1995cf, He:2016xpi, Kaidi:2023maf, Yu:2023nyn, Franco:2024mxa, Bergman:2024its, Robbins:2025puq}, where the choice of appropriate interaction terms and boundary conditions can be used to build the corresponding SymTFT for e.g., the non-Abelian symmetry $D_4$ (the symmetries of a square) \cite{Franco:2024mxa,Cordova:2024jlk,Bergman:2024its,Cordova:2024mqg,Xue:2025enx}.

We find that a discrete BF-like treatment is indeed available for a broad class of non-Abelian symmetries,
i.e., for finite non-Abelian groups $G$ which fit into a short exact sequence of the form:
\begin{equation}
1 \rightarrow M \rightarrow G \rightarrow A \rightarrow 1,
\end{equation}
where $M$ is an Abelian group, and $A$ is a finite (possibly non-Abelian) group. The treatment we give first constructs a suitable BF-theory for $M$ which is then supplemented by ``on-shell'' degrees of freedom associated with the group $A$. We remark that when the short exact sequence has further structure we can also place the fields valued in $A$ off-shell, and we also identify precisely when there are potential obstructions to doing so. There can in principle be different decompositions of $G$ into $M$'s and $A$'s. This results in different field content, and can provide complementary information on the categorical structure of the resulting SymTFT$_{D+1}$, including the structure of anomalies in the absolute QFT$_D$.

The main technical advance undergirding our construction is that there is an appropriate notion of holonomy both for the fields in $A$, i.e., the ``base'' as well as the ``fiber'' $M$.\footnote{The terminology becomes apparent upon considering the fibration structure of the classifying spaces, i.e., $BM \rightarrow BG \rightarrow BA$.} Whereas the holonomies of $A$ are defined as usual by an appropriate period integral over a 1-cycle in the bulk three-manifold, the holonomies in the fiber direction are in general twisted, as induced by the fibration structure. This sort of twisted structure has been considered in \cite{Baez:2004in} for Lie groupoids and we apply it here in the (comparatively simpler) case of finite groups $G$. These holonomies (twisted as well as ordinary) can then be pieced together to explicitly build the Wilson lines, magnetic dual vortex lines, as well as dyonic lines corresponding to objects in the Drinfeld center, i.e., they automatically come with labels $([g], r_{[g]})$, namely a choice of conjugacy class $[g] \in \mathrm{Conj}(G)$ and a choice of representation for the stabilizer of $[g]$.
A pleasing consequence of having an explicit Lagrangian formulation is that reading off (usually quite unwieldy) combinatorial data such as the fusion rules for the Drinfeld center $\mathcal{Z}(\mathcal{C})$ follows directly from standard manipulations of line operators of BF-like theories; many fusion rules can be read off concisely in the span of a few lines.

But the Lagrangian formulation provides even more. Since we have an explicit basis of fields, we can also directly construct the flux tube configurations necessary to realize the non-genuine line operators of the bulk. In these cases, we are able to directly identify a 2D TFT which has a boundary given by a line operator labelled by a symmetry element $g \in G$. As already mentioned, this goes beyond the data captured by the Drinfeld center $\mathcal{Z}(\mathcal{C})$.

The general formalism we deploy relies on connections for finite gauge groups. On the other hand, there is a well-known way to formulate BF-theories with finite Abelian gauge groups in terms of $U(1)$ connections, and it is natural to ask whether our treatment can be cast in these terms as well. See e.g., reference \cite{Robbins:2025puq} for a recent discussion along these lines. In general, we find that this is not possible in our presentation, but we are able to diagnose precisely when it can and cannot occur. The technical condition we encounter is that precisely when the commutator is contained in the center of $G$, i.e., $[G,G] \subset Z(G)$, a treatment in terms of continuous differential forms is available. When this is not possible, there are obstructions to such an approach.

The rest of this paper is organized as follows. We begin in section \ref{sec:DWSYMTFT} by reviewing Dijkgraaf-Witten TFTs, and discuss their use as SymTFTs. In section \ref{sec:LAGRANGIAN} we turn to discrete BF-like theories which directly manifest non-Abelian structures. 
Section \ref{sec:DiffApproach} establishes when a treatment in terms of continuous differential forms is available, and in section \ref{sec:X-check} we discuss in further detail defects beyond those captured by just the Drinfeld center. In section \ref{sec:HIGHER} we discuss the generalization to higher-dimensional QFTs and their corresponding SymTFTs. Section \ref{sec:CONC} contains our conclusions and future directions. A number of technical details and generalizations are deferred to the Appendices.

\section{Dijkgraaf-Witten SymTFTs}
\label{sec:DWSYMTFT}

The Dijkgraaf-Witten (DW) topological field theory \cite{Dijkgraaf:1989pz} gives us a bulk TFT with gauge group $G$. As such, it provides a convenient starting point for analyzing the corresponding SymTFT for $G$ a zero-form non-Abelian global symmetry of an absolute QFT$_D$. Our aim in this section will be to review some of the structures of DW TFTs, and then motivate our extension of this general formulation to a more explicit BF-like theory presentation, the primary aim of section \ref{sec:LAGRANGIAN}. While one can formulate DW TFTs in any $(D+1)$-dimensional spacetime, we shall primarily focus on 3D SymTFTs, (i.e., those associated with 2D absolute QFTs).

Recall that for a general fusion category $\mathcal{C}$,
the Drinfeld center~$\mathcal{Z}(\mathcal{C})$ is the braided monoidal category
whose objects are pairs $(X, \gamma)$,
with $X \in \mathcal{C}$ and
$\gamma_Y: X \otimes Y \to Y\otimes X$
a half-braiding natural in $Y$
satisfying the hexagon conditions.
Physically, $X$ represents a boundary line operator,
while the half-braiding $\gamma$ expresses the condition
that this operator can move freely in the bulk, i.e., it commutes (up to a phase) with all other boundary lines.
Thus $\mathcal{Z}(\mathcal{C})$ describes all genuine bulk topological line defects.
Conversely, a topological boundary condition of the bulk theory
$\mathcal B \simeq \mathcal \mathcal{Z}(\mathcal C)$ is specified by a
{Lagrangian algebra} object $\mathcal L \subset \mathcal B$
\cite{Kapustin:2010hk,Kitaev:2011dxc}.\footnote{A Lagrangian algebra object is a maximally connected commutative separable algebra object for the modular tensor category $\mathcal{B}\cong \mathcal{Z}(\mathcal{C})$.}  Condensing $\mathcal L$
determines which bulk lines can end on the boundary and yields the
boundary line operator category as the category of
$\mathcal L$-modules:
\begin{equation}
\mathcal C \simeq \mathrm{Mod}_{\mathcal L}(\mathcal B)\,.
\end{equation}
In this sense, for the purpose of encoding and classifying symmetry
realizations on topological boundaries, the SymTFT data consist of the
$(D+1)$-dimensional bulk TFT together with its allowed Lagrangian
algebras $\mathcal L$.

Let us now specialize to a 2D QFT with finite global symmetry group $G$. In this case,
the corresponding fusion category $\mathcal{C}$ describing the symmetry is
\begin{equation}
\mathcal{C} = \mathrm{Vec}_G^{\omega},
\end{equation}
consisting of $G$-graded vector spaces with associator $\omega\in H^3(G,U(1))$, which we physically interpret
as the 't Hooft anomaly of the symmetry.
The corresponding bulk SymTFT is a 3D topological field theory whose genuine line operator category
is equivalent to the Drinfeld center $\mathcal{Z}(\mathcal{C})$\cite{drinfeld1986quantum,Turaev:1992hq, Kitaev:2011dxc, Lan_2018, Lan_2019, Johnson-Freyd:2020usu,Bhardwaj:2023ayw }.\footnote{Mathematically, the SymTFT for a group symmetry $G$ with anomaly $\omega$ is directly characterized by the representation category of the twisted quantum double of $G$, denoted as $\mathrm{Rep}\left(D^{\omega}(G)\right)$, where $D^\omega(G)$ is a quasi-Hopf algebra. It is known there is an equivalence $\mathcal Z(\mathrm{Vec}_G^{\omega}) \simeq \mathrm{Rep}\left(D^{\omega}(G)\right)$, constructed by sending a central pair $(X,\gamma) \in \mathcal Z(\mathrm{Vec}_G^{\omega})$ to its induced $\omega$-Yetter-Drinfeld structure associated to $\mathrm{Rep}\left(D^{\omega}(G)\right)$ and conversely. See, e.g., \cite{Roche:1990hs,drinfeld1986quantum,etingof2015tensor} for more details.} In this setting, the line operators of interest are labeled by
$([g], r_{[g]})$, namely a choice of conjugacy class $[g] \in \mathrm{Conj}(G)$ and a choice of representation for the stabilizer of $[g]$, and with a gauge theory interpretation (as we describe below) we can further interpret these as Wilson, vortex (i.e., 't Hooft lines), and dyonic lines.

To accomplish this, we can use the Dijkgraaf-Witten topological gauge theory with gauge group $G$, which makes much of the non-Abelian structure in $G$ manifest. In this description the Wilson lines and magnetic vortex defects reproduce the simple objects
of the Drinfeld center $\mathcal Z(\mathrm{Vec}_G^\omega)$,
making the Dijkgraaf-Witten theory the canonical bulk realization of the symmetry data for finite $G$.
In the following subsections we briefly review the structure of this theory,
its topological line operators, and the resulting categorical data,
which will later provide a point of comparison for our general Lagrangian construction.

\subsection{Dijkgraaf-Witten Theory for Pedestrians}
\label{DW Wilson lines}

The SymTFT for a finite group symmetry $G$, as discussed above,
admits a categorical description in terms of the Drinfeld center
$\mathcal{Z}(\mathrm{Vec}_{G}^\omega)$.
We now realize this theory concretely through the
Dijkgraaf-Witten topological gauge theory,
whose lattice formulation makes the topological data explicit.

\begin{figure}
    \centering
    \scalebox{0.8}{
    \begin{tikzpicture}
	\begin{pgfonlayer}{nodelayer}
		\node [style=none] (0) at (-0.75, -0.075) {};
		\node [style=none] (1) at (0.75, -0.075) {};
		\node [style=none] (2) at (-1, 0) {};
		\node [style=none] (3) at (1, 0) {};
		\node [style=none] (4) at (0, 1.5) {};
		\node [style=none] (5) at (0, -1.5) {};
		\node [style=none] (6) at (-2, 0) {};
		\node [style=none] (7) at (2.5, 1) {};
		\node [style=none] (8) at (2.5, -1) {};
		\node [style=none] (9) at (5.5, 2) {};
		\node [style=none] (10) at (7.5, 0) {};
		\node [style=none] (11) at (5.5, -2) {};
		\node [style=SmallCircle] (12) at (4.75, 0.5) {};
		\node [style=SmallCircle] (13) at (4.75, -0.5) {};
		\node [style=SmallCircle] (14) at (5.75, 0) {};
		\node [style=none] (15) at (5.75, 1) {};
		\node [style=none] (16) at (5.75, -1) {};
		\node [style=none] (17) at (6.75, 0.5) {};
		\node [style=none] (18) at (3.75, 0.5) {};
		\node [style=none] (19) at (3.75, -0.5) {};
		\node [style=none] (20) at (6.75, -0.5) {};
		\node [style=none] (21) at (4.5, 1) {$i$};
		\node [style=none] (22) at (4.5, -1) {$j$};
		\node [style=none] (23) at (6.4, 0) {$k$};
		\node [style=none] (25) at (6, 1.125) {};
		\node [style=none] (26) at (7, 0.625) {};
		\node [style=none] (27) at (7, -0.625) {};
		\node [style=none] (28) at (6, -1.125) {};
		\node [style=none] (29) at (3.5, -0.5) {};
		\node [style=none] (30) at (3.5, 0.5) {};
		\node [style=none] (31) at (4.25, 0) {$g_{ij}$};
		\node [style=none] (32) at (5.625, -0.5) {$g_{jk}$};
		\node [style=none] (33) at (5.625, 0.6) {$g_{ki}$};
		\node [style=none] (34) at (4.75, -0.125) {};
		\node [style=none] (35) at (5.375, -0.1875) {};
		\node [style=none] (36) at (5.125, 0.3125) {};
        \node [style=none] (37) at (2.5, -2.75) {};
	\end{pgfonlayer}
	\begin{pgfonlayer}{edgelayer}
		\draw [style=ThickLine, in=180, out=90] (6.center) to (4.center);
		\draw [style=ThickLine, in=180, out=0] (4.center) to (7.center);
		\draw [style=ThickLine, in=0, out=-180] (8.center) to (5.center);
		\draw [style=ThickLine, in=-90, out=180] (5.center) to (6.center);
		\draw [style=ThickLine, bend left=15] (0.center) to (1.center);
		\draw [style=ThickLine, bend left=15] (3.center) to (2.center);
		\draw [style=ThickLine, in=-180, out=0] (7.center) to (9.center);
		\draw [style=ThickLine, in=90, out=0] (9.center) to (10.center);
		\draw [style=ThickLine, in=0, out=-90] (10.center) to (11.center);
		\draw [style=ThickLine, in=0, out=180] (11.center) to (8.center);
		\draw [style=ThickLine] (18.center) to (12);
		\draw [style=ThickLine] (12) to (15.center);
		\draw [style=ThickLine] (12) to (14);
		\draw [style=ThickLine] (14) to (13);
		\draw [style=ThickLine] (13) to (19.center);
		\draw [style=ThickLine] (13) to (16.center);
		\draw [style=ThickLine] (14) to (17.center);
		\draw [style=ThickLine] (14) to (20.center);
		\draw [style=ThickLine] (12) to (13);
		\draw [style=DottedLine] (30.center) to (18.center);
		\draw [style=DottedLine] (29.center) to (19.center);
		\draw [style=DottedLine] (16.center) to (28.center);
		\draw [style=DottedLine] (20.center) to (27.center);
		\draw [style=DottedLine] (17.center) to (26.center);
		\draw [style=DottedLine] (15.center) to (25.center);
		\draw [style=ArrowLineRight] (13) to (35.center);
		\draw [style=ArrowLineRight] (14) to (36.center);
		\draw [style=ArrowLineRight] (12) to (34.center);
	\end{pgfonlayer}
\end{tikzpicture}

    }
    \caption{Triangulation of the 3-manifold $X$ with vertices $i,j,k,\dots$ and oriented edges labeled by group elements $g_{ij},g_{jk},g_{ki},\dots$. }
    \label{fig:triangulation}
\end{figure}
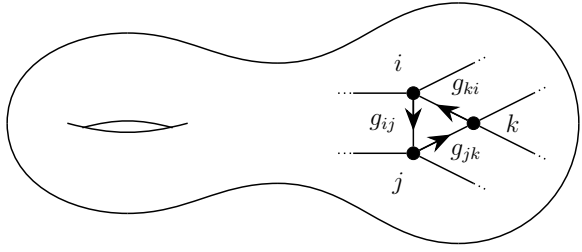

Dijkgraaf-Witten theory can be realized on the lattice as follows \cite{Dijkgraaf:1989pz}: Given a (finite) gauge group $G$, we pick a group cohomology element\footnote{From a physical point of view, this group cohomology element is describing the 't Hooft anomaly of some 2D theory with symmetry group $G$ \cite{Tachikawa:2017gyf,Bhardwaj:2017xup}.} $\omega \in H^3(G,U(1))$. This consists in the datum of a function
\begin{equation}
    \omega: G^3\to U(1) \,,
\end{equation}
such that it satisfies the usual group cocycle condition (and $\omega$ is also defined up to an exact group cocycle). Now, let $X$ be a 3-dimensional manifold and pick a triangulation with vertices labeled by either $i,j,k,\dots$ or $i_0,i_1,i_2,\dots$, see figure \ref{fig:triangulation}. Then, our space of fields is given by a collection of transition functions defined over the edges $g_{ij}\in G$, which are closed,\footnote{In other words, flat $G$-connections (\v{C}ech 1-cocycles).} i.e.,
\begin{equation}
    g_{ij}g_{jk}g_{ik}^{-1}=1 \,,
\end{equation}
where the vertices $i,j,k$ are nearest neighbors. Let us denote this space of fields by $Z^1(X,G)$. Then, we can write the following path integral:
\begin{equation}
\label{eq:PartitionFunction}
    Z(X)=\sum_{g\in Z^1(X,G)} \prod_{i_0<i_1<i_2<i_3}\omega(g_{i_0i_1},g_{i_1i_2},g_{i_2i_3}) \equiv \sum_{g\in Z^1(X,G)} \int_X g^*\omega \,,
\end{equation}
which, up to an overall normalization factor (that we neglect as it is going to be irrelevant for our discussion), gives us precisely the partition function of Dijkgraaf-Witten over a closed 3-manifold $X$. Notice that since $\omega$ is a group cohomology element, we have the following set of gauge transformations
\begin{equation}
    g_{ij} \sim \lambda_i g_{ij} \lambda_j^{-1} \,,
\end{equation}
with gauge parameter $\lambda$. The action $\omega$ is invariant under these transformations, and the physically different field configurations are given by $[g]\in H^1(X,G)$, i.e., $G$-bundles over $X$.

Line operators of this theory are given by a collection of transition functions $g_{ij}$, we can consider the holonomies, which are given by
\begin{equation}
    \oint_\gamma g = g_{i_0i_1}g_{i_1i_2}\dots g_{i_0i_n}^{-1}\,,
\end{equation}
where the vertices $i_0,\dots, i_n$ induce a triangulation for the loop $\gamma$ and which are not gauge invariant. Upon performing a gauge transformations, $g_{ij}\sim \lambda_i g_{ij}\lambda_j^{-1}$, the holonomy of $g$ on $\gamma$ will transform in the adjoint representation of the gauge group
\begin{equation}
    \oint_\gamma g \sim \lambda  \left( \oint_\gamma g \right) \lambda^{-1}\,,
\end{equation}
matching with the fact that the character variety of $G$ is given by $ H^1(X,G)\cong \mathrm{Hom}(\pi_1 X, G)/G$ where $G$ acts via the adjoint action on the space $\mathrm{Hom}(\pi_1 X, G)$.

In order to construct genuine gauge invariant operators, we consider class functions $\chi: G\to \C$ (i.e., characters), and compose them with the holonomies. Since characters are in one to one correspondence with representations of $G$, and the generating ones are given by the irreducible representations, we get the first set of (simple) line operators:
\begin{equation}\label{eq:Wlines}
    ([e],R)\equiv \chi_R \left( \oint g \right) \,,
\end{equation}
where $R$ is an irreducible representation of $G$ and $[e]$ denotes the conjugacy class associated to the unit element $e\in G$, an additional label we justify momentarily. These lines have fusion channels given by $\mathrm{Rep}(G)$:
\begin{equation}
\chi_{R_1} \left( \oint g \right)\chi_{R_2} \left( \oint g \right) = \chi_{R_1\otimes R_2} \left(\oint g \right) = \chi_{\oplus_m n^{(m)}_{12} R_m} \left(\oint g \right) =\sum_m  n^{(m)}_{12}\;\!\chi_{R_m} \left(\oint g \right)\,,
\end{equation}
for some positive integers $n_{12}^{(m)}\in\Z$.

The lines \eqref{eq:Wlines} are topological Wilson lines. Indeed, due to the fact that the transition functions $g_{ij}$ satisfy a cocycle condition, when we deform the path $\gamma$, the value of \eqref{eq:Wlines} will not change. Hence, we can ask the question which are the operators that trigger a non-trivial flux for $\chi_R(\oint_\gamma g)$. These operators must be labeled by conjugacy classes in $G$, since the characters are class functions. Hence, we introduce the magnetic vortex lines of the theory by specifying the holonomy of $\chi_R(\oint_\gamma g)$, when $\gamma$ links them. These vortex lines are labeled by $([g],1)$, and are such that
\begin{equation}
    \langle ([g],1) ([e],R) \rangle \propto \chi_R(g) \,,
\end{equation}
where $\langle\, \dots \rangle$ denotes the correlation function and we assumed that the two operators are supported over two loops that link once. All the possible dyons, as we will see, are then obtained by fusing Wilson lines and magnetic vortex lines. The set of such lines and their fusion algebra combine to the Drinfeld center $\mathcal{Z}(\mathrm{Vec}_G^\omega)$ \cite{drinfeld1986quantum, Kitaev:2011dxc, etingof2015tensor, Lan_2018, Lan_2019, Johnson-Freyd:2020usu, Turaev:1992hq}.

\subsection{The Drinfeld Center $\mathcal{Z}(\mathrm{Vec}_G)$}

To frame the discussion to follow and set notation we now review the construction of the Drinfeld center $\mathcal{Z}(\mathrm{Vec}_G^\omega)$ for trivial action $\omega$ and some of its key properties as a modular tensor category.

To begin, simple objects of $\mathcal{Z}(\mathrm{Vec}_G)$ are labeled by pairs $([g],R)$ where $[g]$ denotes the conjugacy class of $g \in G$, while $R$ is an irreducible representation of the centralizer of $g\in G$, i.e., $N_g = \{h\in G \,|\, hgh^{-1}=g\}$.\footnote{For $\mathcal{C}=\mathrm{Vec}_{G}^{\omega}$ with nontrivial $\omega$, the simple objects of the center are again
labeled by pairs $([g],\pi)$, but with $\pi$ is an irreducible {projective} representation of the centralizer $N_g$
with factor set $\alpha_g \in Z^2(N_g,U(1))$ obtained by
\begin{equation}\label{eq:alpha_g}
\alpha_g(a,b) = \frac{\omega(a,b,g) \omega(g,a,b)}{\omega(a,g,b)}, ~(a,b\in N_g).
\end{equation}
Equivalently,
$\pi$ is an irrep of the twisted group algebra $\mathbb{C}_{\alpha_g}[N_g]$.
When $\omega=1$, $\alpha_g$ is trivial and one recovers ordinary irreps of $N_g$,
which is the case we focus on in this paper.}  The centralizers are all isomorphic for every $g$ in the same conjugacy class. We then denote the abstract group as $N_{[g]}$.

There are two distinguished subsets of simple objects in the Drinfeld center which we refer to as electric Wilson lines and magnetic vortex lines. Regarding the former note that for every group $G$, we have that $N_e = G$. Hence, the set of all Wilson lines are given by $([e], \mathrm{Rep}(G))$. Regarding the latter note that for every centralizer $N_{[g]}$, we always have that the trivial representation is an irreducible representation. Hence, the magnetic vortex lines are given by $([g], 1^{N_{[g]}})$ for $[g] \in \mathrm{Conj}(G)$,
as expected from our previous discussion.

The ribbon structure on $\mathcal{Z}(\mathrm{Vec}_G)$ supplies the modular pair $(\mathbb{S},\mathbb{T})$.\footnote{The terminology here is borrowed from that of a 2D CFT on a torus with complex structure $\tau$ where have the modular transformations $\mathbb{S}: \tau \mapsto -1/\tau$ and $\mathbb{T}: \tau \mapsto \tau +1$.}
In particular, $\mathbb{S}$ computes the correlation of two linked simple lines and $\mathbb{T}$ encodes their twists.
For the untwisted case $\omega=1$, one convenient closed form of $\mathbb{S}$ is:\footnote{Moreover $\mathbb{T}_{([g],R)}=\frac{\chi_R(g)}{\chi_R(e)}$ in the untwisted case,
while for $\omega\neq1$ it picks the additional $\omega$-dependent phase
\cite{Roche:1990hs,etingof2015tensor}.}
\begin{equation}
    \mathbb{S}_{([g],R),([h],T)} = \frac{|G|}{|N_{[g]}||N_{[h]}|}\sum_{l\in G \,:\, glhl^{-1} = lhl^{-1}g} \chi_R(lhl^{-1})\chi_T(l^{-1}gl) \,.
\end{equation}
From a quantum field theory perspective, the datum of the $\mathbb{S}$-matrix is simply the datum of the correlation functions:
\begin{equation}
    \langle([g],R)([h],T)  \rangle = \mathbb{S}_{([g],R),([h],T)}\,.
\end{equation}
For future convenience we record explicit $\mathbb{S}$-matrix elements which follow from straightforward computation:
 \begin{equation}
    \begin{split}
         \mathbb{S}_{([e],R),([h],1^{N_{[h]}})} &= \frac{|G|}{|N_{[h]}|} \chi_R(h) \,,\\
         \mathbb{S}_{([g],1^{N_{[g]}}),([g],1^{N_{[g]}})} &= \frac{|G|}{|N_{[g]}|^2}|\{ l\in G\,|\, lgl^{-1}\in N_g\cap[g] \}| \,,\\
         \mathbb{S}_{([e],R),([e],T)} &= \mathrm{dim}(T)\mathrm{dim}(R) \,.
    \end{split}
 \end{equation}
In particular, from this we see that, since the vacuum expectation value of the Wilson line is simply the dimension of the labeling representation,\footnote{To see this we just need to look at the $\mathbb{S}$-matrix with the identity operator $([e],1^G)$ inserted.} the Wilson lines have trivial charge among each other (as expected from the QFT description). Moreover the correlation function between a magnetic vortex line and a Wilson line is proportional to the evaluation of the character labeling the Wilson line on the equivalence class labeling the vortex line. Finally, we notice that two magnetic vortex lines can have non-trivial linking. In particular, their vacuum expectation value must be given by $\frac{|G|}{|N_{[h]}|}$, so that we can write the following correlation functions:
    \begin{equation}
    \begin{split}
        \langle ([e],R)([h],1^{N_{[h]}}) \rangle &= \langle ([h],1^{N_{[h]}})\rangle\chi_R(h) \,,\\
        \langle([g],1^{N_{[g]}})([g],1^{N_{[g]}}) \rangle &=\langle ([g],1^{N_{[g]}}) \rangle^2 \frac{|\{ l\in G\,|\, lgl^{-1}\in N_g\cap[g] \}|}{|G|} \,, \\
        \langle ([e],R)([e],T) \rangle &= \langle([e],R)\rangle \langle([e],T)\rangle \,.
    \end{split}
    \end{equation}
    Notice that these relations hold in any Dijkgraaf-Witten theory with any gauge group $G$ and trivial topological action $\omega$.

Let us now determine the fusion rules. Since $\mathcal{Z}(\mathrm{Vec}_G)$ is modular, fusion is completely fixed by
the $\mathbb{S}$-matrix through the Verlinde relations.  Physically, this expresses that the
fusion coefficients (up to gauge choices) are determined by
linking data of bulk lines. In the present setting, a remarkable fact is that the linking data are highly constrained by charge conservation arguments: The line operators of Dijkgraaf-Witten theory are labeled by the charges that they have among each other and themselves. The fusion of these operators is then entirely dictated by charge conservation arguments.

Indeed, consider the following scenario, in which we are fusing two lines $X,Y\in \mathcal{Z}(\mathrm{Vec_G})$, and we want to determine if a third line $Z\in\mathcal{Z}(\mathrm{Vec_G})$ enters in their fusion. Concretely, suppose that we have:
\begin{equation}
    F_{XYZ}Z \subseteq X\otimes Y \,,
\end{equation}
where $F_{XYZ}$ denotes the fusion coefficients associated to the triple $X,Y$ and $Z$. Then, fusing both members with $Z^*$ (here the star denotes reversing the orientation of the line and $ZZ^*= 1+\dots$) we get
\begin{equation}
    X\otimes Y\otimes Z^* = F_{XYZ} \mathrm{Id} + \dots \,,
\end{equation}
where the $\dots$ do not contain the identity operator. Now, we measure the charge by linking with a fourth line $W$ both the lefthand and righthand sides:
\begin{equation}
    \mathbb{S}_{W,X\otimes Y \otimes Z^*} = F_{XYZ}\langle  W \rangle + \dots \,.
\end{equation}
Then, we multiply the above expression by $\langle W \rangle$ and we take a sum over all simple objects we get
\begin{equation}
\label{verlinde}
    \sum_{W} \frac{\mathbb{S}_{X,W}\mathbb{S}_{Y,W}\mathbb{S}_{Z,W}^*}{\langle W \rangle} = F_{XYZ} \sum_{W} \langle W \rangle^2 \,,
\end{equation}
where we used the fact that
\begin{equation}
    \sum_{W} \langle W \rangle \mathbb{S}_{X,W} = 0 \,,
\end{equation}
if $X\neq \mathrm{Id}$. Notice that \eqref{verlinde} is precisely the Verlinde formula for the fusion rules of a modular category. In particular, as expected from general physical reasoning, the fusion algebra, given by the fusion rules, is entirely determined by charge conservation. Unpacking this formula \cite{Roche:1990hs} one finds
\begin{equation}
\label{eq:VFormula}
    F_{([g],R),([h],S),([l],T)} = \frac{1}{|G|} \mathop{\mathop{\sum_{g_A\;\! \in\;\!  [g], \;\! g_B\;\! \in\;\!  [h], \;\! g_C\;\! \in\;\!  [l],\;\! h\;\! \in\;\!  G}}_{[g_A,h]=[g_B,h]=[g_C,h]=1}}_{g_Ag_B=g_C}  \chi_R(h)\chi_S(h)\chi_T(h)^* \,.
\end{equation}

Let us now consider the following special case:
\begin{equation}
    F_{([g],1),([e],R),([g],S)} = \sum_{a\in [g]}\sum_{h\in N_a} \chi_R(h)\chi_S(h)^* = \sum_{a\in [g]} |N_a|\langle \mathrm{Res}^G_{N_a}(\chi_R), \chi_S \rangle_{N_a} \,,
\end{equation}
where $ \langle \cdot , \cdot\rangle_{N_a}$ is the inner product on characters for the group $N_a$, and $\mathrm{Res}^G_H$ gives the restriction from $G$ to a subgroup $H$ of a representation. In particular, the sum is zero if and only if $R$ when restricted to $N_a$ does not contain the irreducible representation $S$ in its decomposition. Notice however that we can always find $R\in \mathrm{Rep}(G)$ such that its restriction contains $S$. Indeed, by the Frobenius reciprocity theorem, we have that
\begin{equation}
    \langle \mathrm{Res}^G_{N_a}(\chi_R), \chi_S \rangle_{N_a} = \langle \chi_R, \mathrm{Ind}^G_{N_a}(\chi_S) \rangle_{G} \,,
\end{equation}
where $\mathrm{Ind}^G_{N_a}$ denotes the induced representation from $N_a$ to $G$. Since $\mathrm{Rep}(G)$ is semisimple, we can always find an irreducible representation $R$ such that the above scalar product is non-zero. Hence, we conclude that in order to obtain all the dyons of the theory is enough to take fusion between Wilson and magnetic vortex lines, as expected from a physical perspective.

This completes the characterization of the bulk line operators and their fusion rules in the
Dijkgraaf-Witten TFT. A natural next step, in the SymTFT context, is to understand how such bulk data determine
topological boundary conditions.  In the categorical framework reviewed above,
a topological boundary is specified by a maximal condensable (i.e., Lagrangian) algebra object
inside the modular tensor category $\mathcal{Z}(\mathrm{Vec}_G)$,
whose condensation implements the boundary fusion category and realizes
the symmetry in the lower-dimensional QFT. The complete classification of Lagrangian algebras for an arbitrary $\mathcal{Z}(\mathrm{Vec}_G)$ can be involved, but there are always two canonical ones: The electric boundary condition
\begin{equation}
    \mathcal{L}_{\text{e}}:= \sum_{R}d_{R}([1], R),
\end{equation}
with $d_R$ the dimension of the irrep $R$ of $G$, realizing the $\mathrm{Vec}_G$, i.e., invertible $G$ symmetry; and the magnetic boundary condition
\begin{equation}
    \mathcal{L}_{\text{m}}:=\sum_{[g]}([g], 1),
\end{equation}
realizing $\mathrm{Rep}(G)$ symmetry.

\section{Lagrangian Approach with Discrete Cochains}
\label{sec:LAGRANGIAN}
We now discuss a Lagrangian approach of Dijkgraaf-Witten theory with gauge group $G$. Our goal is a Lagrangian formulation with respect to a maximally simple field content. Whenever $G$ is realized as an extension of groups the field space with respect to which the partition function \eqref{eq:PartitionFunction} is computed can be accordingly decomposed and reparametrized with respect to new fields and relations among these. Further, electric Wilson lines then have a simple description in terms of these fundamental fields and their fusions are straightforwardly computed. Magnetic vortex line operators are presented with respect to a defect prescription, and are less directly covered.


Let us start presenting the general setup.\footnote{See Appendix \ref{group and bundles} for additional discussion.}
Consider a finite group $G$ which fits into the following short exact sequence of groups
\begin{equation}
   1 \rightarrow  M\to G\to A \rightarrow 1 \,,
\end{equation}
and which is classified by two pieces of data
\begin{equation}
\begin{split}
    \beta: A\times A \to M \, ,\\
    \alpha: A \to \mathrm{Aut}(M) \,,
\end{split}
\end{equation}
that have to satisfy the following compatibility conditions
\begin{equation}
\begin{split}
\label{eq: alpha and beta}
\alpha_a(\beta(b,c))\beta(a,bc) = \beta(a,b)\beta(ab,c) \\
    \alpha_a\alpha_b = \mathrm{Ad}^{M}_{\beta(a,b)}\alpha_{ab} \, ,
\end{split}
\end{equation}
for group elements $a,b,c\in A$ and $\text{Ad}^M$ denotes the adjoint action on $M$. In terms of these data, we can reconstruct the multiplication on $G$ as follows:
\begin{equation}
    \begin{bmatrix}
        m_1 \\
        a_1
    \end{bmatrix}\begin{bmatrix}
        m_2 \\
        a_2
    \end{bmatrix} = \begin{bmatrix}
       m_1\alpha_{a_1}(m_2)\beta(a_1,a_2) \\
        a_1a_2
    \end{bmatrix} \,.
\end{equation}
For example, we recover the group law for a semi-direct product of $G = M \rtimes A$ for $\beta = 1$. Now, let us consider a gauge field $g_{ij}\in Z^1(X,G)$. This can be seen as a pair $(m_{ij}, a_{ij})$ of $M$ and $A$-valued fields, which have to satisfy the following cocycle condition
\begin{equation}
\begin{bmatrix}
       m_{ij}\alpha_{a_{ij}}(m_{jk})\beta(a_{ij},a_{jk}) \\
        a_{ij}a_{jk}
    \end{bmatrix}= \begin{bmatrix}
        m_{ik} \\
        a_{ik}
    \end{bmatrix} \, ,
\end{equation}
which are the equations of motion of the system. From the above, we see immediately that the gauge field $g_{ij}\in Z^1(X,G)$ will induce a gauge field $a_{ij}\in Z^1(X,A)$ via the canonical projection map $G\to A$. On the other hand, the equations of motion for $m$ are more complicated, whose set solutions we denote with:
\begin{equation}
    \mathrm{Tw}^1(X,M)_{(a,\alpha,\beta)}=\{ m\in C^1(X,M) \, :   m_{ij}\alpha_{a_{ij}}(m_{jk})\beta(a_{ij},a_{jk})=m_{ik}\, \}\,,
\end{equation}
i.e., the set of twisted 1-cocycles, where $C^1(X,M)$ is the set of $M$-valued 1-cochains on $X$. In terms of these new fields, we have a different presentation of Dijkgraaf-Witten theory as the following path integral:
\begin{equation}
    Z(X) = \sum_{ \substack{a\in Z^1(X,A) \\ m\in\mathrm{Tw}^1(X,M)_{(a,\alpha,\beta)} }} \int_X (m,a)^*\omega\,.
\end{equation}
where $\omega$ is an element in $H^3(G, U(1))$. 

We show that when $M$ is Abelian, we can indeed construct an explicit BF-like Lagrangian. In our treatment, fields valued in $A$ will generically be kept on-shell, though even this restriction can be relaxed when $A$ is also Abelian.

\subsection{Untwisted Lagrangian}
\label{subsec:Untwisted}

Our goal in this section will be to first expand on the Lagrangian presentation of our theory, introducing dual variables and then present line operators of our theory with respect to these fundamental fields. To make progress we will now specialize to the case where the action $\omega$ is trivial, i.e., the case of an ``untwisted Lagrangian''.\footnote{Most of the techniques that we are going to develop will hold also if the action $\omega$ is non-trivial but it does not depend from the field $m$. This means that the group cocycle $\omega$ factors as follows:
\be
\omega: G^{ 3}\xrightarrow{(\pi,\pi,\pi)}A^{ 3} \to U(1)\,,
\ee where $\pi: G\to A$ is the projection defining the short exact sequence.} For example, when decomposing solvable groups with respect to their subnormal series such a setup occurs in the ultimate step of the decomposition. We will return to such more general cases in concrete examples later.

Under the assumption of Abelian $M$, the equations of motion for the field $m$ simplify to:
\begin{equation}
     \delta_\alpha m_{ijk} = m_{ij} - m_{ik} +\alpha_{a_{ij}}m_{jk} = -\beta (a_{ij}, a_{jk})\,,
\end{equation}
or, in a more concise presentation,
\begin{equation}\label{eq:twisted-m-eom}
    \delta_\alpha m + a^*\beta = 0 \,,
\end{equation}
where $a^*$ is a pullback from $A$ to $X$.

Our assumption that $M$ is Abelian now implies that the conjugation in the second line of \eqref{eq: alpha and beta} is trivial, and as a consequence $\alpha$ is a group homomorphism $A \rightarrow \mathrm{Aut}(M)$ and $\beta$ is a group 2-cocycle: $\beta\in H^2(A,M)$. The path integral of Dijkgraaf-Witten theory with trivial action is then
\begin{equation}\label{eq:actionDWtriv}
    Z(X) = \sum_{\substack{a\in Z^1(X,A) \\ m\in\mathrm{Tw}^1(X,M)_{(a,\alpha,\beta)} }} 1 \,,
\end{equation}
where now the space of twisted cocycles is given by the simpler form
\begin{equation}
   \mathrm{Tw}^1(X,M)_{(a,\alpha,\beta)} =  {\{ m\in C^1(X,M) \,:\, \delta_\alpha m + a^*\beta=0 \}}\,.
\end{equation}
In this presentation, the gauge transformations of the fundamental fields take the following form:
\begin{equation}
\begin{split}
    a_{ij} &\to a_{ij}' := \kappa_i a_{ij}\kappa_j^{-1} \,,\\
    m_{ij} &\to \alpha_{\kappa_i}(m_{ij}) + (\epsilon_i - \alpha_{ a_{ij}'}(\epsilon_j)) - \alpha_{ a_{ij}'}(\beta(\kappa_j,\kappa_j^{-1}))\,,\\
       \delta_\alpha m_{ijk} + a^*\beta_{ijk} &\to \alpha_{\kappa_i}(\delta_\alpha m_{ijk} + a^*\beta_{ijk})\,,
\end{split}
\end{equation}
where $\kappa_i $ and $\epsilon_i$ are a set of gauge parameters associated with $A$ and $M$, respectively. Further, by definition, $\alpha_{\kappa_i}\in\mathrm{aut}(M)$ as specified by the gauge parameter $\kappa_i$ via the action $\alpha$.

Next, introduce a Lagrange multiplier to impose the equations of motion for the field $m$ in the Lagrangian description of Dijkgraaf-Witten theory. Call this multiplier $\hat{m}$. Then, we want to write:
\begin{equation}
    Z(X) = \sum_{\substack{ a\in Z^1(X,A) \\ m,\hat{m}\in C^1(X,M)}} \exp\left(2\pi i \int_X \langle \hat{m}, \delta_\alpha m +a^*\beta \rangle_M \right) \,,
\end{equation}
where $\langle \cdot , \cdot\rangle_M$ is a gauge invariant, non-degenerate pairing. Since we are assuming $M$ is Abelian, $M\cong \oplus_i \Z_{p_i^{k_i}}$ for some primes $p_i$ (repetitions are allowed) and some positive integer $k_i$. Hence, the field $m$ is given by a tuple $m = (m^1, m^2 ,\dots, m^n)$, where $m^i \in C^1(X,\Z_{p_i^{k_i}})$. Then the non-degenerate pairing is given by:
\begin{equation}
    \langle \hat{m}, \delta_\alpha m + a^*\beta\rangle_{M} = \hat{m}^T\cup \mathbf{Q}( \delta_\alpha m +a^*\beta) \,,
\end{equation}
where $\mathbf{Q} = \mathrm{diag}(\dots , 1/p_i^{k_i} ,\dots) \equiv \mathrm{diag}(\dots, q_i, \dots)$, and both the cup product and the differential act component-wise. Notice that this pairing is gauge invariant provided that $\hat{m}$ transforms as follows:
\begin{equation}
    \hat{m}_{ij} \to \alpha_{\kappa^{-1}_j}^\mathrm{op}(\hat m_{ij}) + \alpha^\mathrm{op}_{\kappa_ia_{ij}\kappa_j^{-1}}(\hat\epsilon_i) - \hat{\epsilon}_j \,,
\end{equation}
where the $\alpha^\mathrm{op}$ denotes the induced action on $\hat{m}$ by $\alpha$ which is given by
\begin{equation}
    \alpha^\mathrm{op}_\kappa = \mathbf{Q}\alpha_\kappa\mathbf{Q}^{-1}\,.
\end{equation}
The action $\alpha^\mathrm{op}$ on $\hat{m}$ is a right action, meaning that $\alpha_\kappa^\mathrm{op} \alpha_\rho^\mathrm{op}(\hat{m}) = \alpha^\mathrm{op}_{\rho\kappa}(\hat{m}) $. Then, $\delta^\text{op}_{\alpha}$ is the adjoint to $\delta_\alpha$ with respect to the pairing $\langle \cdot, \cdot \rangle_M$, and it has the following properties:
\begin{equation}
\begin{split}
    \delta^\mathrm{op}_{\alpha} \hat m_{ijk} &= \alpha^\mathrm{op}_{a_{jk}}(\hat m_{ij}) - \hat m_{ik} + \hat m_{jk} \,,\\
    \delta^\mathrm{op}_{\alpha} \hat{m}_{ijk} &\to \alpha^\mathrm{op}_{\kappa_k^{-1}}(\delta^\mathrm{op}_{\alpha}\hat{m}_{ijk})\,.
\end{split}
\end{equation}
In particular, we see that the adjoint differential always transforms in a covariant way.

Before proceeding, some further remarks are in order. First, note that a similar dualization trick (which would introduce variables $\hat a$ and take $a$ off-shell) would in general result in the combination $\delta_\alpha m + a^*\beta$ not transforming covariantly and, to preserve gauge invariance, gauge transformations would need to be modified \cite{Kaidi:2022cpf,Kaidi:2022uux}. The resulting structures are poorly understood and we will instead opt to keep $a$ in general on-shell. Second, the group $G$ can fit into several short exact sequences. This results in equivalent presentations of the theory, the simplest example of this kind is $G=D_4$ which admits a semi-direct product presentation $\Z_4\rtimes \Z_2$ where $\beta$ is trivial and a presentation as $\Z_2^3$ extending according to some non-trivial $\beta$ (but trivial $\alpha$).\footnote{The $\Z_2^3$ presentation manifestly gives rise to 3D SymTFT with $\mathbb{Z}_2 \times \mathbb{Z}_2 \times \mathbb{Z}_2$ gauge symmetry and Chern-Simons-like / BF-like quadratic terms and cubic interactions with all fields off-shell \cite{Wang:2014oya,He:2016xpi,Kaidi:2023maf,Yu:2023nyn,Franco:2024mxa}.}

Now that we have introduced a new degree of freedom, we can study the line operators given by $\hat{m}$ which will furnish a Lagrangian description of another sector (of magnetic vortex lines) for the SymTFT of $G$. To proceed then, consider the equation of motion for the dual field $\hat{m}$ given by
\begin{equation}
    \delta^\mathrm{op}_{\alpha} \hat m  = 0\,.
\end{equation}
In particular, this equation is telling us that the pair $(\hat{m}, a)$ always describes the gauge field for the semi-direct product $M\rtimes_{\alpha^\mathrm{op}} A$. Indeed, upon integrating out the field $m$, we end up with the SymTFT for the semi-direct product $M\rtimes_{\alpha^\mathrm{op}} A$ with a 't Hooft anomaly given precisely by $\beta$.\footnote{Clearly this holds only in 3D, which we remind the reader is our setup. In higher dimensions, the analog of this statement is that the pair $(\hat m, a)$ describes the gauge field of a higher group with trivial Postnikov invariant, but non-trivial action. Furthermore the higher group has a 't Hooft anomaly described by $\beta$. We will comment further on this in section \ref{sec:HIGHER}.}

We next turn to construct line operators from the field content $m$, $\hat{m}$ and $a$. For this we first introduce the notion of a twisted integral. To motivate this integral consider the holonomies of the total field $g\in Z^1(X,G)$ given by
\begin{equation}
    \oint_\gamma g =  \prod_{j = 0}^n g_{i_j i_{j+1}}  \quad (i_{n+1} \equiv i_0)\,,
\end{equation}
where $i_0,\dots, i_n$ are the (ordered) vertices on $\gamma$ induced by the triangulation on $X$.\footnote{A more rigorous way of writing the RHS is $g_{i_0i_1}g_{i_1i_2}\dots g_{i_{n-1}i_n}g_{i_0i_n}^{-1}$, so that there is a well-defined orientation on the simplex without any loops. However, in the rest of the paper, we ignore this subtlety and simply use cyclic indexing.}
If we now write down $g$ as a pair of fields, we end up with
\begin{equation}
    \oint_\gamma g = \begin{bmatrix}
        \oint_\gamma^{(\alpha,\beta) }m\\
        \oint_\gamma a
    \end{bmatrix} \,,
\end{equation}
where the top line contains the twisted integral $\oint_\gamma^{(\alpha,\beta)} $ which when summing the field $m$ takes the explicit form
\begin{equation}
\begin{split}
    \oint_\gamma^{(\alpha,\beta)} m &= m_{i_0 i_1} +  \sum_{j = 1}^{n-1} \left[ \alpha_{\prod_{k = 0}^{j-1} a_{i_k i_{k+1}}} m_{i_{j} i_{j+1}} + \beta(\prod_{k = 0}^{j-1} a_{i_k i_{k+1}}, a_{i_{j} i_{j+1}})\right] \ \ (i_{n+1} \equiv i_0)\,.
\end{split}
\end{equation}
Under gauge transformations, the twisted integral transforms as:
\begin{equation}
\begin{split}
        &\oint_\gamma^{(\alpha,\beta)} m  \sim \alpha_{\kappa_{i_0}} \oint_\gamma^{(\alpha,\beta)} m + \epsilon_{i_0} + \beta(\kappa_{i_0}, \oint_\gamma a) - \alpha_{(\oint_\gamma a )'}( \epsilon_{i_0} + \beta(\kappa_{i_0},\kappa_{i_0}^{-1}) ) + \beta(\kappa_{i_0}\oint_\gamma a,\kappa_{i_0}^{-1}) \,,
\end{split}
\end{equation}
where we remind that $(\kappa_{i_0}, \epsilon_{i_0})$ are a pair of gauge parameters for $A$ and $M$, respectively, and $(\oint_\gamma a )' = \kappa_{i_0} (\oint_\gamma a )\kappa_{i_0}^{-1}$ is the transformed integral corresponding to the subgroup $A$.

In particular, if $a$ has trivial holonomy, then the above gauge transformation simplifies
\begin{equation}
    \oint_\gamma^{(\alpha,\beta)} m  \sim \alpha_{\kappa_{i_0}} \oint_\gamma^{(\alpha,\beta)} m \,.
\end{equation}
For the field $\hat m$ we simply need to set $\beta$ to zero, and we denote its twisted integral as follows:
\begin{equation}
    \oint_\gamma ^{(\alpha^\mathrm{op}, 0)} \hat m \equiv \oint_\gamma ^{\alpha^\mathrm{op}} \hat m \,.
\end{equation}

\subsubsection{Genuine Line Operators}

With this established, we can now write a complete list of the line operators of the theory expressed with respect to the fundamental fields $m,\hat m, a$ of the Lagrangian and which we will therefore refer to as Lagrangian line operators. We have the simple Wilson lines given by the characters for the field $a$
\begin{equation}
\label{wilson a}
\chi_R \left(\oint a \right) \,,
\end{equation}
where $R$ is an irreducible representation of $A$. For the line operators associated to $m$ and $\hat{m}$, we instead proceed as follows: Pick an element in $\xi\in M$, and consider its orbit $\mathrm{Orb}^\mathrm{op}(\xi)$ under the action of $\alpha^\mathrm{op}$,
\begin{equation}
    [\xi] := \mathrm{Orb}^\mathrm{op}(\xi) = \{ \zeta \in M : \zeta = \alpha_\kappa^\mathrm{op} (\xi) \,, \, \kappa \in A \} \,.
\end{equation}
 Notice that we can pick a representative $\xi_i$ in each orbit $[\xi]$, so that $M$ decomposes as a disjoint union of orbits: $M = \sqcup_i [\xi_i]$. Then, the line operators associated to $m$ are given by
\begin{equation}
\label{eq:WOps}
    W_{[\xi_i]}(\gamma)=\delta \left(\oint_\gamma a \right) \sum_{\zeta\in[\xi_i]} e^{2\pi i \oint_\gamma^{(\alpha,\beta)}\zeta^T\mathbf{Q}m} \,.
\end{equation}
This set of operators must reproduce the remaining characters associated to $G$, that are not described by the operators associated to $a$. In \eqref{eq:WOps} the notation $\delta(\dots)$ denotes a Dirac delta function. For the magnetic vortex line, we can proceed in an analogous way. First, we consider the orbit of $\hat\xi\in M$ under the action of $\alpha$:
\begin{equation}
    [\hat\xi] := \mathrm{Orb}(\hat\xi) = \{ \hat\zeta \in M : \hat\zeta = \alpha_\kappa (\hat\xi) \,, \, \kappa \in A \} \,.
\end{equation}
If we pick a representative $\hat{\xi}_i$ in each orbit $[\hat\xi_i]$, we have a decomposition $M=\sqcup_i [\hat\xi_i]$. Then we can write down the magnetic vortex line operators as follows:
\begin{equation}
    H_{[\hat\xi_i]}(\gamma) = \delta \left(\oint_\gamma a \right) \sum_{\hat\zeta\in[\hat\xi_i]} e^{2\pi i \oint_\gamma^{\alpha^\mathrm{op}}\hat m^T \mathbf{Q}\hat\zeta}\,.
\end{equation}
The fusion of these operators is simply given by taking ordinary products, and we obtain all Lagrangian dyons by considering such products. Notice also that the Dirac delta function on the holonomies of $a$ is gauge invariant ensuring that the twisted integrals transform in a covariant way.

From this description we are able to recover the $\mathbb{S}$-matrix elements for these operators. Clearly $\chi_R(\oint a)$ will have trivial $\mathbb{S}$-matrix elements with both $H_{[\hat\xi]}$ and $W_{[\xi]}$, as an insertion of $\chi_R(\oint a)$ in the path integral will not modify the equations of motion for $m$ and $\hat{m}$, leading to the formula:
\begin{equation}
    \langle \chi_R(\oint a)  W_{[\xi]}\rangle = \langle\chi_R(\oint a)\rangle\langle W_{[\xi]}\rangle = \mathrm{dim}(R)|[\xi]| \,,
\end{equation}
where $|[\xi]|$ denotes the cardinality of $[\xi]$ and similarly for the magnetic vortex operator:
\begin{equation}
    \langle \chi_R \left(\oint a \right)  H_{[\hat\xi]}\rangle = \langle\chi_R \left(\oint a \right)\rangle\langle H_{[\hat\xi]}\rangle = \mathrm{dim}(R)|[\hat\xi]| \,.
\end{equation}
Hence, we simply need to discuss the linking between the magnetic vortex and the Wilson line operators. This time $\hat{m}$ and $m$ are mutually dual variables, and so we expect $H_{[\hat\xi]}$ and $W_{[\xi]}$ to be mutually charged under each other. Indeed, inserting the operator $H_{[\hat\xi]}(\sigma)$ will modify the equation of motions to
\begin{equation}
    \delta_\alpha m + a^*\beta \in [\hat \xi] \, \mathrm{PD}(\sigma)\,,
\end{equation}
where $\mathrm{PD}(\sigma)$ denotes the Poincare' dual of $\sigma$. Substituting this into $W_{[\xi]}(\gamma)$ we get, assuming $\sigma$ and $\gamma$ link once,
\begin{equation}
    \langle H_{[\hat\xi]}(\sigma) W_{[\xi]}(\gamma)\rangle =|[\hat\xi]|\sum_{ \zeta\in[\xi]} e^{2\pi i \zeta^T\mathbf{Q}\hat \zeta} \,.
\end{equation}
Equivalently, we can also integrate out $m$ instead of $\hat{m}$, then we have the following equations of motion:
\begin{equation}
    \delta^{\mathrm{op}}_\alpha \hat m \in [\xi]\,\mathrm{PD}(\gamma) \,,
\end{equation}
which leads similarly to the following correlation functions:
\begin{equation}
    \langle H_{[\hat\xi]}(\sigma) W_{[\xi]}(\gamma)\rangle =|[\xi]|\sum_{\hat \zeta \in [\hat \xi] } e^{2\pi i \zeta^T\mathbf{Q}\hat \zeta} \,.
\end{equation}


The remaining line operators, which we will refer to as non-Lagrangian, consist of the line operators that have a non-trivial charge under the line operators \eqref{wilson a}. These vortex operators are defined by a defect description: remove the line on which they are supported from the spacetime and prescribe non-trivial holonomies for the operators $\chi_R(\oint a)$ whenever they link the removed loop. This prescription has the following immediate consequences:
\begin{itemize}
    \item These non Lagrangian vortex operators are labeled by conjugacy classes in $A$. We will refer to these as $D_{[s]}$, where $[s]$ denotes a conjugacy class in $A$.
    \item Consequently, they will have zero $\mathbb{S}$-matrix element with both $H_{[\hat \xi]}$ and $W_{[\xi]}$.
    \item The remaining line operators of the theory (both purely magnetic and dyonic) are obtained via fusion with the Lagrangian operators: $\mathrm{Lagrangian} \otimes D_{[s]}$.
\end{itemize}
With this only the linkings among the operators of type $D_{[s]}$ remains to be specified and, once they are, the Drinfeld center governing the full set of genuine line operators is fully fixed following the charge conservation arguments which resulted in the Verlinde formula \eqref{eq:VFormula}.

However, note that this linking data supplements our Lagrangian approach. In practice, when computing (for example) the Drinfeld center fusion tables, we therefore first exhaust fusions between Lagrangian line operators using the formalism above, and then determine the remaining fusions using the Verlinde formula \eqref{eq:VFormula}.

\subsubsection{Non-Genuine Line Operators}
\label{ssec:NGen}

\begin{figure}
\centering
\scalebox{0.8}{
\begin{tikzpicture}
	\begin{pgfonlayer}{nodelayer}
		\node [style=none] (0) at (-5.75, 0) {};
		\node [style=none] (1) at (-2.75, 0) {};
		\node [style=none] (2) at (-3, 0) {};
		\node [style=none] (3) at (-5.5, 0) {};
		\node [style=none] (4) at (-1.5, 0) {};
		\node [style=none] (5) at (1.5, 0) {};
		\node [style=none] (6) at (1.25, 0) {};
		\node [style=none] (7) at (-1.25, 0) {};
		\node [style=none] (8) at (2.75, 0) {};
		\node [style=none] (9) at (5.75, 0) {};
		\node [style=none] (10) at (5.5, 0) {};
		\node [style=none] (11) at (3, 0) {};
		\node [style=none] (12) at (-4.25, 0) {};
		\node [style=none] (13) at (-0.1, 1) {};
		\node [style=none] (14) at (-0.1, 0) {};
		\node [style=none] (15) at (0.1, 0) {};
		\node [style=none] (16) at (4.25, 0) {};
		\node [style=none] (17) at (4.25, 1) {};
		\node [style=none] (18) at (0.1, 1) {};
		\node [style=SmallCircle] (19) at (-4.25, 0) {};
		\node [style=SmallCircle] (20) at (0, 1) {};
		\node [style=SmallCircle] (21) at (4.25, 1) {};
		\node [style=none] (22) at (-10, 0) {};
		\node [style=none] (23) at (-7, 0) {};
		\node [style=none] (24) at (-7.25, 0) {};
		\node [style=none] (25) at (-9.75, 0) {};
		\node [style=none] (26) at (-8.5, 0) {};
		\node [style=none] (27) at (-8.5, -1.25) {(i)};
		\node [style=none] (28) at (-4.25, -1.25) {(ii)};
		\node [style=none] (29) at (0, -1.25) {(iii)};
		\node [style=none] (30) at (4.25, -1.25) {(iv)};
		\node [style=none] (31) at (-2, -1.75) {};
		\node [style=none] (32) at (-9.25, -0.375) {DBC};
		\node [style=none] (33) at (-5, -0.375) {DBC$^{}$};
		\node [style=none] (34) at (-3.35, -0.375) {DBC$^{\:\!(g)}$};
		\node [style=none] (35) at (-0.75, -0.375) {DBC$^{}$};
		\node [style=none] (36) at (0.9, -0.375) {DBC$^{\:\!(g)}$};
		\node [style=none] (37) at (3.25, -0.375) {DBC$^{}$};
		\node [style=none] (38) at (5.375, -0.375) {DBC$^{\:\!(g)}$};
		\node [style=SmallCircleBrown] (39) at (4.25, 0) {};
		\node [style=none] (40) at (-4.25, 0.5) {$O_g$};
		\node [style=none] (41) at (0, 1.5) {$O_g$};
		\node [style=none] (42) at (4.25, 1.5) {$O_g$};
		\node [style=none] (43) at (4.25, -0.475) {\text{$J_g$}};
		\node [style=none] (44) at (4.75, 0.5) {$B_g$};
	\end{pgfonlayer}
	\begin{pgfonlayer}{edgelayer}
		\filldraw[fill=gray!20, draw=gray!20]  (-7.25, 0) -- (-9.75, 0) -- (-9.75, 2) -- (-7.25,2) -- cycle;
		\filldraw[fill=gray!20, draw=gray!20]  (-3, 0) -- (-5.5, 0) -- (-5.5, 2) -- (-3,2) -- cycle;
		\filldraw[fill=gray!20, draw=gray!20]  (3, 0) -- (5.5, 0) -- (5.5, 2) -- (3,2) -- cycle;
		\filldraw[fill=gray!20, draw=gray!20]  (0.1, 0) -- (1.25, 0) -- (1.25, 2) -- (0.1,2) -- cycle;
		\filldraw[fill=gray!20, draw=gray!20]  (-1.25, 0) -- (-0.1, 0) -- (-0.1, 2) -- (-1.25,2) -- cycle;
		\filldraw[fill=gray!20, draw=gray!20]  (-0.1, 1) -- (-0.1, 2) -- (0.1, 2) -- (0.1,1) -- cycle;
		\draw [style=BlueLine] (3.center) to (12.center);
		\draw [style=BlueLine] (7.center) to (14.center);
		\draw [style=BlueLine] (11.center) to (16.center);
		\draw [style=GreenLine] (12.center) to (2.center);
		\draw [style=GreenLine] (15.center) to (6.center);
		\draw [style=GreenLine] (16.center) to (10.center);
		\draw [style=GreenLine] (18.center) to (15.center);
		\draw [style=BlueLine] (13.center) to (14.center);
		\draw [style=DottedBlue] (3.center) to (0.center);
		\draw [style=DottedBlue] (7.center) to (4.center);
		\draw [style=DottedBlue] (11.center) to (8.center);
		\draw [style=DottedGreen] (10.center) to (9.center);
		\draw [style=DottedGreen] (6.center) to (5.center);
		\draw [style=RedLine] (21) to (16.center);
		\draw [style=BlueLine] (25.center) to (26.center);
		\draw [style=DottedBlue] (25.center) to (22.center);
		\draw [style=DottedGreen] (2.center) to (1.center);
		\draw [style=BlueLine] (26.center) to (24.center);
		\draw [style=DottedBlue] (24.center) to (23.center);
	\end{pgfonlayer}
\end{tikzpicture}
}
\caption{In subfigure (i) we sketch the reference configuration with Dirichlet boundary condition DBC. In subfigure (ii) we insert an invertible symmetry operator $O_g$ labeled by $g\in G$ across which the Dirichlet boundary condition jumps from DBC to DBC$^{\:\!(g)}$. In subfigure (iii) we deform this insertion into the SymTFT.  In subfigure (iv) we fuse / fold the boundary condition onto itself resulting in a surface $B_g$ terminating on the boundary on the junction $J_g$. }
\label{fig:ThxKantaro}
\end{figure}
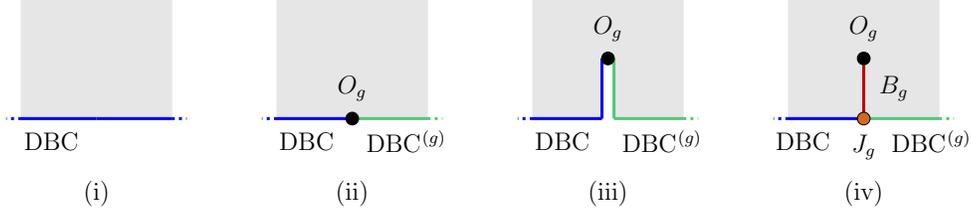

We next turn to discuss non-genuine operators of the DW theory. To begin we consider a SymTFT setup and consider the DW theory on a 3-manifold with topological boundary and physical boundary, such that overall, we are describing a 2D QFT with 0-form symmetry $G$. Symmetry operators of the QFT, labeled by group elements $g\in G$, localize as codimension-1 operators to the topological boundary. Across such symmetry operators the Dirichlet boundary condition fixing the $G$-background of the QFT jumps according to $g\in G$. This boundary insertion can be deformed into the bulk by folding the topological boundary condition onto itself (see figure \ref{fig:ThxKantaro}).\footnote{We thank K. Ohmori for discussion on this operation. See also \cite{Fuchs:2013gha}.} The result is a non-genuine bulk codimension-2 operator $O_g$ attached to a codimension-1 bulk surface $B_g$ terminating on a junction $J_g$, all of which are labeled by group elements.

The surface $B_g$ trivially contains $G$ codimension-1 operators (with respect to the surface $B_g$). To see this, one can insert operators $O_{g'}^{L},O_{g''}^{R}$ on the left and the right, prior to folding, which after folding live on the codimension-1 surface. If $O_{g'}^{L},O_{g''}^{R}$ coincide after folding on $B_g$ and if $g'=g g'' g^{-1}$ then the resulting $O_{g',g''}$ which is obtained by fusing $O_{g'}^{L},O_{g''}^{R}$ is a line \textit{within} the surface $B_g$. When the topological boundary condition has the algebra $L$ condensed then $B_g$ is an $L-L$ bimodule labeled by $g\in G$.

The configuration of $(O_g,B_g,J_g)$ can be further manipulated. Denote by $C(g)$ the commutant of $g\in G$ and by $H_g\subset C(g)$ any subgroup. Then $H_g$-worth of lines can be condensed without changing the non-genuine operator $O_g$. The triple becomes $(O_g,B_g/H_g,J_g)$ in obvious notation. The bimodule obtained by gauging $C(g)$ is the ``simplest" way of deforming $O_g$ into the bulk.

Finally note that while $(O_g,B_g/H_g,J_g)$ is invariant with respect to bulk gauge transformations it does transform covariantly with respect to background gauge transformations, by which we mean gauge transformations applied to the Dirichlet boundary conditions. The configuration $(O_g,B_g/C(g),J_g)$ is gauge-invariant even under background transformations in $C(g)$. We give concrete examples in section \ref{sec:Pure2Cocycle}, \ref{ssec:FormalGauge} and a universal mathematical construction in section \ref{sec:X-check}.

\subsection{Semi-Direct Product of Cyclic Groups}
\label{sec: semidirect}

We now make our discussion concrete and specialize to the class of examples with Abelian group $A\cong \Z_k$, module $M\cong \Z_n$, and $\beta=0$, so $G\cong \Z_n\rtimes_\alpha \Z_k$ for some $\alpha$. The fundamental fields have the following equations of motions:
\begin{equation}
    \delta_\alpha m = \delta_\alpha^{\mathrm{op}} \hat m = 0 \,,
\end{equation}
where, $a\in Z^1(X,\Z_k)$ on the 3-manifold $X$. The Lagrangian is
\begin{equation}
    \mathcal{ L} = \frac{2\pi i}{n} \hat m \cup \delta_\alpha m \,,
\end{equation}
where the action $\alpha:\Z_k\rightarrow \text{Aut}(\Z_n)$ maps $x\in \Z_k$ to multiplication by $q^x$ on $\Z_n$ and under the following conditions
\begin{equation}
\begin{split}
    \gcd(q, n) &= 1 \,,\\
    q^k &= 1\mod n \,,
\end{split}
\end{equation}
this gives an order $k$ action, the last condition assures $\alpha^k=\text{Id}\in \text{Aut}(\Z_n)$. The adjoint action of $\Z_k$ on $\Z_n$ in $G$ is simply $\alpha$, however the opposite differential $\delta^\mathrm{op}_\alpha$ still has a different (but equivalent) combinatorial description than $\delta_\alpha$. Moreover, $\mathbf{Q}$ is simply the number $1/n$. 

According to our prescription the Lagrangian line operators of this theory are given by three different sectors. We give their expression with respect to the Lagrangian field content and equate them to the corresponding Drinfeld center element.
\begin{itemize}
    \item Electric $\Z_k$ Wilson lines: Taking the holonomies of $a$ we obtain
    \begin{equation}
    W_t \equiv e^{ \frac{2\pi i}{k} t \oint_\gamma a} =([e], 1^G_t) \,,
    \end{equation}
    which is well-defined as $a$ is closed, being on-shell. The 1-dimensional representations  $1^G_t$  of $G$ are labeled by $t\in \Z_k$. In particular, $1_0^G = 1^G$ is the trivial representation.
    \item Electric $\Z_n$ Wilson lines: The remaining set of operators in the sector $([e],\mathrm{Rep}(G))$ are given by the $\alpha$-twisted integral for $m$:
    \begin{equation}
        W_{[r]} \equiv \delta \left(\oint a \right) \sum_{l\in[r]} e^{\frac{2\pi i}{n} l\oint^\alpha m}  = ([e], R^G_{[r]}) \,,
    \end{equation}
    where $r\in M$ and $[r] = \{ l\in M : \, l = \alpha^x r \,, x\in \Z_k \}$. The notation $R^G_{[r]}$ denotes a representation of $G$ of dimension given by the cardinality of $[r]$, and is labeled by the conjugacy class $[r]$. The way from the conjugacy class $[r]$ to reconstruct the precise representation is discussed in Appendix \ref{app:RepTheory}. Basically, all the representations of $\Z_n \rtimes \Z_k$ with the conjugation action $\alpha$ is in one-to-one correspondence with representations $\sigma_{i, \rho}$, where $i$ is the index $\chi_i$ that is a representation of the Abelian group $M$, and $\rho$ is a representation of the subgroup $A_i$ of $A$ that is fixed by the character $\chi_i$.
    \item Magnetic $\Z_n$ vortex lines: By exchange symmetry between $m$ and $\hat{m}$, we have also the $\alpha$-twisted integral for $\hat m$:
     \begin{equation}
        H_{[r]} \equiv \delta \left(\oint a \right) \sum_{l\in[r]} e^{\frac{2\pi i}{n} l\oint^\alpha \hat m}  = ([r],1^{\Z_n\rtimes_\alpha \Z_{\mathrm{Stab}(r)}}) \,,
    \end{equation}
    where $\mathrm{Stab}(r)$ is the order of the following subgroup of $\{ y\in \Z_k : \, \alpha^y r = r  \}\subset \Z_k$. The action $\alpha$ is given precisely by the restriction of $\alpha:\Z_k \to \mathrm{Aut}(\Z_n)$ to this subgroup. This is a sector of magnetic vortex line operators.
\end{itemize}

The set of Wilson lines $W_t$ and $W_{[r]}$ will reproduce the fusion of the representation ring $\mathrm{Rep}(\Z_n\rtimes_\alpha \Z_k)$, and, by symmetry, the set of operators $W_t$ and $H_{[r]}$ will also reproduce the fusion of $\mathrm{Rep}(\Z_n\rtimes_\alpha \Z_k)$. This means that from the SymTFT perspective, if we gauge the $\Z_n$ symmetry on the boundary, we obtain again a semi-direct product global symmetry $\Z_n\rtimes_\alpha\Z_k$ with the same action $\alpha$ and the new $\Z_n$ lines given by the electromagnetic dual of the old $\Z_n$. In higher dimensions, we instead get a special case of an $n$-group with trivial Postnikov class but non-trivial action of the 0-form symmetry on the $n$-form symmetry.

The fusion (via multiplication) between the $W_{[r]}$ and the $H_{[r]}$ sectors will produce all possible dyons of the theory which are labeled by $([r], \mathrm{Rep}(\Z_n\rtimes_\alpha \Z_{\mathrm{Stab}(r)}))$, as we will see explicitly when we discuss more concrete examples.

Let us end this section with some comments on the non-Lagrangian sector of the theory which are defined via a defect prescription. The vortex operators that we fail to describe with this Lagrangian formulation are the quantum dual of the operators $([e],R^G)=\chi_R(\oint a)$.
These operators are the ones that trigger a non-trivial flux for $([e],R^G)$, and therefore are labeled by conjugacy classes in $A$. Furthermore, to these operators we can stack all the Lagrangian ones, obtaining all the possible line operators of this theory.

In the semi-direct product case at hand, the basic vortex lines in the non-Lagrangian sector are going to be labeled by $([s^l],1)$, where $s\in G$ is the generator of $\Z_k$ in $G$ and $l=1,\dots, k-1$. Before proceeding, we are going to assume that the action of $\Z_k$ on $\Z_n$ is free (i.e., has no fixed points except for the identity element). We then have an operator $D_{[s^l]}$ which corresponds to the Drinfeld center element $([s^l],1^{\Z_k})$ and all the remaining dyons are obtained by considering the fusions
\be
D_{[s^l]}\otimes W_t\,,
\ee
for $t\in \Z_k$. This matches the Drinfeld center fusion channel
\begin{equation}
    ([s^l],1_t^{\Z_k}) = ([s^l],1^{\Z_k})\otimes ([e],1_t^G)  \,.
\end{equation}
Hence, in this case, the only non-Lagrangian fusion rules that we need to determine from purely algebraic data (i.e., by the $\mathbb{S}$-matrix using the Verlinde formula, which we recall is simply a statement on charge conservation), correspond to $([s^l],1^{\Z_k})\otimes ([s^t],1^{\Z_k})$. This is a huge simplification, as once we know how the $D_{[s^l]}$ operators fuse among themselves then we are done, as the remaining fusions are Lagrangian, except for those corresponding to:
\begin{equation}
    ([s^l],1^{\Z_k})\otimes ([e], R^G_{[r]})  = ([s^l],1^{\Z_k})\otimes  ([r],1^{\Z_n\rtimes_\alpha \Z_{\mathrm{Stab}(r)}}) = ([s^l],1^{\Z_k})\otimes([r], 1^{\Z_n}_i)=\bigoplus_t ([s^l],1_t^{\Z_k}) \,,
\end{equation}
where $i=0,\dots, n-1$ labels the representation of $\Z_n$.

 Indeed, we can consider the Verlinde formula for the fusion symbols which in this case reduces to:
\begin{equation}
    F_{([s^l],1^{\Z_k}),([r],1_i^{\Z_n}),([s^l],1^{\Z_k}_t)} = \frac{1}{nk}\sum_{g\in[r]}\sum_{h\in[s]} 1 = 1 \,,
\end{equation}
for any $t = 0,\dots, k-1$. Notice that actually this result holds also if $M$ is a general Abelian finite group.

Another general fusion rule which we can derive is the following:
\begin{equation}
\label{eq:easyfus}
    ([s^l],1^{\Z_k})\otimes([s^{k-l}],1^{\Z_k}) = ([e], 1^G)+\sum_{[r]} ([e],R^G_{[r]}) + \sum_{[r]}\sum_{l=0}^{n-1} ([r], 1^{\Z_n}_l)\,.
\end{equation}
Indeed, we first notice that we have the following:
 \begin{equation}
     F_{([s^l],1^{\Z_k}),([s^{k-l}],1^{\Z_n}),([e],R^G)} = \frac{n}{nk}\sum_{l\in\Z_k\subset G} \chi^*_R(l) \,,
 \end{equation}
 where $\Z_k$ is the subgroup of $G$ stabilizing $s\in G$. The fusion symbol is going to be $1$ if and only if $R$ corresponds to the representation associated to some operator $W_{[r]}$ or the trivial representation, and is going to be zero otherwise, i.e., if $R$ corresponds to the representation described by some operator $W_t$. By symmetry, we must have also that all the $H_{[r]}$ operators must enter in the fusion with coefficient 1.

 Finally, let us now focus our attention on the dyons. The Verlinde formula in this case reduces to the following:
 \begin{equation}
     F_{([s^l],1^{\Z_k}),([s^{k-l}],1^{\Z_n}),([r],1^{\Z_n}_t)} = \frac{1}{nk}\sum_{g\in [s^l] \,,\, h\in[s^{k-l}] \, :\, gh\in[r]} 1 = 1\,,
 \end{equation}
 where to get this result we noted that the formula does not depend on $t$ and we know by the symmetry argument above that for $t=0$ the result must be 1. By comparing the quantum dimension of the LHS and RHS of equation \eqref{eq:easyfus}, we know that we have no other operator contributing to this fusion channel. This proves the above formula.

\subsubsection{Example: $S_3$}
\label{ssec:S3Example}
We are going to match the Drinfeld Center operators for the group
\be
S_{3}=\langle  r,s \,|\,r^3=e\,,~s^2=e\,, srs^{-1}=r^{-1} \rangle\,.
\ee
This is a semidirect product $\Z_3 \rtimes_\alpha \Z_2$ where the action $\alpha$ is simply the multiplication by $2$. $\alpha^{\mathrm{op}}$ is also given by multiplication by $2$, and $\mathbf{Q}$ is just the number $1/3$. The Lagrangian is
\begin{equation}
    \mathcal{L}_{S_3} = \frac{2\pi i}{3} \hat{m} \cup \delta_\alpha m \,.
\end{equation}
The conjugacy classes with the respective stabilizers are given by table \ref{ZS3}. A simple element in the Drinfeld center is labeled by the pair $(C,\rho)$, where $C$ is a conjugacy class and $\rho$ is an irreducible representation of the stabilizer. A complete list of the fusion rules for the Drinfeld center of $S_3$ is given in Appendix \ref{App: S3 fusion}. Direct comparison shows that our method reproduces the correct fusion rules. We will now discuss and explicitly rewrite some of these fusion channels in the Lagrangian formalism.

{\renewcommand{\arraystretch}{1.35}
\begin{table}[]
\centering
\begin{tabular}{|c|c|c|}
\hline
\textbf{Conjugacy class $[g]$} & \textbf{Centralizer $C_{S_3}(g)$} & \textbf{Irreps of $C_{S_3}(g)$} \\
\hline
$[e]=\{e\}$ & $S_3$ & $1^{S_3},\ 1_{\text{sign}}^{S_3},\ 2^{S_3}$ \\
\hline
$[s]=\{s,\ sr,\ sr^2\}$ & $\langle s \rangle \cong \mathbb{Z}_2$ & $1^{\mathbb{Z}_2}$, $1^{\mathbb{Z}_2}_-$ \\
\hline
$[r]=\{r,\ r^2\}$ & $\langle r \rangle \cong \mathbb{Z}_3$ & $1^{\mathbb{Z}_3}$, $1^{\mathbb{Z}_3}_\omega$ ,$1^{\mathbb{Z}_3}_{\omega^2}$ \\
\hline
\end{tabular}
\caption{Conjugacy classes of $S_3$, their centralizers, and the irreducible representations of each centralizer. Each pair $(C,\rho)$ labels a simple object of the Drinfeld center $Z(S_3)$. Here $\omega=\exp(2\pi i/3)$. Recall, that in the last column the notation indicates the dimension of the representation of the group listed in the exponent. If there are multiple representation of the same dimension the index distinguishes these, e.g., in the first row `sign' indicates the 1D sign representation of $S_3$ and the remaining rows the index gives the charge of the representation.}
\label{ZS3}
\end{table}}

\paragraph{Sector:\,$([e], \mathrm{Rep} (S_3))$.} The line operators in this sector are given by the characters of $S_3$. According to our general discussion in section \ref{sec: semidirect}, we obtain the line operators in the first three columns of table \ref{identity sector S3}.

{\renewcommand{\arraystretch}{1.35}
\begin{table}[]
\centering
\begin{tabular}{|c|c|}
\hline
\textbf{Operators in the QFT} & \textbf{Element in the Drinfeld Center} \\
\hline
 $1$ & $ ([e],1^{S_3})$  \\
\hline
 $W_s=e^{i\pi\oint a}$ & $([e],1_{\text{sign}}^{S_3})$  \\
\hline
 $W_{[r]}=(e^{i\frac{2\pi}{3} \oint^\alpha m} + e^{i\frac{2\pi}{3} \oint^\alpha 2m})\delta \left(\oint a \right)$& $([e],2^{S_3})$ \\
\hline
 $H_{[r]}=(e^{i\frac{2\pi}{3} \oint^\alpha \hat{m}} + e^{i\frac{2\pi}{3} \oint^\alpha 2\hat{m}})\delta \left(\oint a \right) $& $ ([r], 1^{\mathbb{Z}_3}) $ \\
 \hline
 $  (e^{i\frac{2\pi}{3}\oint^\alpha m} e^{i\frac{2\pi}{3}\oint^\alpha \hat{m}} + e^{i\frac{2\pi}{3}\oint^\alpha 2m} e^{i\frac{2\pi}{3}\oint^\alpha 2\hat{m}})\delta \left(\oint a \right)$& $ ([r], 1^{\mathbb{Z}_3}_\omega) $ \\
 \hline
 $ (e^{i\frac{2\pi}{3}\oint^\alpha m}e^{i\frac{2\pi}{3}\oint^\alpha 2\hat{m}} + e^{i\frac{2\pi}{3}\oint^\alpha 2m}e^{i\frac{2\pi}{3}\oint^\alpha \hat{m}})\delta \left(\oint a \right)$& $ ([r], 1^{\mathbb{Z}_3}_{\omega^2})$ \\  \hline
  Non-Lagrangian $D_{[s]}$ & $    ([s], 1^{\Z_2}) $ \\  \hline
   Non-Lagrangian $D_{[s]}\otimes W_s$ & $  ([s], 1_-^{\Z_2})$ \\  \hline
\end{tabular}
\caption{List of QFT operators corresponding to the Drinfeld center of $S_3$. See table \ref{ZS3} for an explanation of the notation.}
\label{identity sector S3}
\end{table}}

\paragraph{Sector:\,$([r], \mathrm{Rep} (\mathbb{Z}_3))$.}
By using the symmetry between $m$ and $\hat{m}$, we can immediately write down the vortex operator
\begin{equation}
 H_{[r]}=  ([r], 1^{\mathbb{Z}_3}) =  (e^{i\frac{2\pi}{3} \oint^\alpha \hat{m}} + e^{i\frac{2\pi}{3} \oint^\alpha 2\hat{m}})\delta \left(\oint a \right) \,,
\end{equation}
 Then, we obtain all the remaining operators in this sector by fusing with the operators in the identity sector. To this end, we notice that fusing with $([e],1_{\text{sign}})$ is trivial due to the Dirac delta factor. Hence, we can only do the fusion with $([e],2^{S_3})$, which produce for us the sum of two gauge invariant operators:

\begin{equation}\label{eq:HWS}
\begin{split}
   H_{[r]} W_s &=
    (e^{i\frac{2\pi}{3}\oint^\alpha m} e^{i\frac{2\pi}{3}\oint^\alpha \hat{m}} + e^{i\frac{2\pi}{3}\oint^\alpha 2m} e^{i\frac{2\pi}{3}\oint^\alpha 2\hat{m}})\delta \left(\oint a \right) \\
   &~~~\,  + (e^{i\frac{2\pi}{3}\oint^\alpha m}e^{i\frac{2\pi}{3}\oint^\alpha 2\hat{m}} + e^{i\frac{2\pi}{3}\oint^\alpha 2m}e^{i\frac{2\pi}{3}\oint^\alpha \hat{m}})\delta \left(\oint a \right) \,.
\end{split}
\end{equation}
corresponding to $ ([r], 1^{\mathbb{Z}_3}) \otimes ([e], 2^{S_3})$. The expression in the first and second line in \eqref{eq:HWS} respectively correspond to $([r], 1^{\mathbb{Z}_3}_\omega)$ and $([r], 1^{\mathbb{Z}_3}_{\omega^2})$. Notice that the operators that we can write down with $\hat{m}$ and $a$ form another copy of $\mathrm{Rep}(S_3)$ inside the Drinfeld center of $S_3$, as expected from the symmetry interchanging $m$ with $\hat{m}$.

As a consistency check for our prescription, we can compute, by multiplying the corresponding entries in table \ref{identity sector S3}, the two fusions:
\begin{equation}
    ([r], 1^{\mathbb{Z}_3}_\omega) \otimes ([r], 1^{\mathbb{Z}_3}_\omega) \,,~ ([r], 1^{\mathbb{Z}_3}_{\omega^2}) \otimes ([r],1^{\mathbb{Z}_3}_{\omega^2}) \,.
\end{equation}
A straightforward computation shows that we get respectively
\begin{equation}
\begin{split}
     1 + e^{i\pi\oint a} + (e^{i\frac{2\pi}{3}\oint^\alpha m} e^{i\frac{2\pi}{3}\oint^\alpha \hat{m}} + e^{i\frac{2\pi}{3}\oint^\alpha 2m} e^{i\frac{2\pi}{3}\oint^\alpha 2\hat{m}})\delta \left(\oint a \right) \,, \\
     1 + e^{i\pi\oint a} + (e^{i\frac{2\pi}{3}\oint^\alpha m}e^{i\frac{2\pi}{3}\oint^\alpha 2\hat{m}} + e^{i\frac{2\pi}{3}\oint^\alpha 2m}e^{i\frac{2\pi}{3}\oint^\alpha \hat{m}})\delta \left(\oint a \right)\,,
\end{split}
\end{equation}
which matches, in the Drinfeld center language, the decomposition in simple objects:
\begin{equation}
\begin{split}
        ([r], 1^{\mathbb{Z}_3}_\omega) \otimes ([r], 1^{\mathbb{Z}_3}_\omega) = ([e], 1^{S_3})\oplus ([e], 1_{\text{sign}}^{S_3})\oplus ([r], 1^{\mathbb{Z}_3}_\omega) \,, \\
        ([r], 1^{\mathbb{Z}_3}_{\omega^2}) \otimes ([r], 1^{\mathbb{Z}_3}_{\omega^2}) = ([e], 1^{S_3})\oplus ([e], 1_{\text{sign}}^{S_3})\oplus ([r], 1^{\mathbb{Z}_3}_{\omega^2}) \,.
\end{split}
\end{equation}
With similar computations one can establish that the Lagrangian line operators given in first six lines in table \ref{identity sector S3} realize the fusion tables  \ref{S3 fusion 1}, \ref{S3 fusion 3a}  and \ref{S3 fusion 3b} under multiplication.

\paragraph{Sector:\,$([s], \mathrm{Rep} (\mathbb{Z}_2))$.} Let us now focus on the non-Lagrangian sector. From the Drinfeld center data, we have the pure magnetic line operators $D_{[s]}$ which corresponds in the Drinfeld center to $([s], 1^{\Z_2})$. According to our general discussion in section \ref{sec: semidirect}, we have the following non-Lagrangian fusions:
\begin{equation}
\begin{split}
    ([s], 1^{\Z_2}) \otimes ([e],1_{\text{sign}}^{S_3}) &= ([s], 1_-^{\Z_2})\,,\\
    ([s], 1^{\Z_2})\otimes ([r], 1^{\mathbb{Z}_3}_{\omega^k}) &= ([s], 1^{\Z_2}) + ([s], 1_-^{\Z_2}) \,,\\
    ([s], 1^{\Z_2}) \otimes([s], 1^{\Z_2}) &= ([e],1^{S_3}) + ([e],2^{S_3}) + ([r],1^{\Z_3}) + ([r],1^{\Z_3}_\omega) + ([r],1^{\Z_3}_{\omega^2})\,,
\end{split}
\end{equation}
where $k=0,1,2$. From these, we can determine for example the following fusion:
\begin{equation}
\begin{split}
         ([s], 1_-^{\Z_2})\otimes([s], 1^{\Z_2}) &=([e],1_{\text{sign}}^{S_3})\otimes[([e],1^{S_3}) + ([e],2^{S_3}) + ([r],1^{\Z_3}) + ([r],1^{\Z_3}_\omega) + ([r],1^{\Z_3}_{\omega^2})] \\
        &=([e],1_{\text{sign}}^{S_3}) + ([e],2^{S_3}) + ([r],1^{\Z_3}) + ([r],1^{\Z_3}_\omega) + ([r],1^{\Z_3}_{\omega^2})\,.
\end{split}
\end{equation}
Notice that despite being a non-Lagrangian fusion, knowing the basic non-Lagrangian fusions discussed in section \ref{sec: semidirect}
would make it very simple to compute.

\subsubsection{Example: $D_{2k+1}$}
The previous discussion on $S_3$ generalizes immediately to the case of dihedral groups $D_n$ with $n=2k +1$ (odd) by noticing the isomorphism of $D_3 \cong S_3$. A presentation for these groups is given as follows:
\begin{equation}
    D_n = \langle r,s \,|\, r^n=s^2=e \,,\, srs = r^{-1}  \rangle \,.
\end{equation}
In particular, this means that we are viewing $D_n = \Z_n \rtimes_\alpha \Z_2$, where $\alpha = \alpha^{\mathrm{op}} = n-1$, and the Lagrangian is then given by
\begin{equation}
    \mathcal{L}_{D_n} = \frac{2\pi i}{n}\hat{m}\cup \delta_\alpha m \,.
\end{equation}
\paragraph{Sector: $([e],\mathrm{Rep}(D_n))$}
We clearly have the one dimensional representations given by
\begin{equation}
\begin{split}
    ([e], 1^{D_n}) &= 1 \,,\\
    ([e], 1_s^{D_n}) &= e^{\pi i \oint a} \,.
\end{split}
\end{equation}
Moreover, on top of these, we have now a bunch of two dimensional representations
\begin{equation}
    ([e], 2^{D_n}_l) = \delta \left(\oint a \right)\Big( e^{\frac{2\pi i}{n}l\oint^\alpha m} +e^{-\frac{2\pi i}{n}l\oint^\alpha m}\Big) \,,
\end{equation}
where $l= 1,\dots, k$.
\paragraph{Sector: $([r^l] , \mathrm{Rep}(\Z_n))$} The pure magnetic line of this sector is given by
\begin{equation}
    ([r^l], 1^{\Z_n}) = \delta \left(\oint a \right)\Big( e^{\frac{2\pi i}{n}l\oint^\alpha \hat m} +e^{-\frac{2\pi i}{n}l\oint^\alpha \hat m}\Big) \,,
\end{equation}
where $l= 1,\dots, k$. The remaining dyons are given by the fusions:
\begin{equation}
\begin{split}
    &([e], 2^{D_n}_l)\otimes ([r^t], 1^{\Z_n}) = ([r^t],1^{Z_n}_l) \oplus ([r^t],1^{Z_n}_{n-l}) =\\ &\delta \left(\oint a \right)\Big( e^{\frac{2\pi i}{n}t\oint^\alpha \hat m} e^{\frac{2\pi i}{n}l\oint^\alpha m}+e^{-\frac{2\pi i}{n}t\oint^\alpha \hat m}e^{-\frac{2\pi i}{n}l\oint^\alpha m}\Big)+ \\&\delta \left(\oint a \right)\Big(e^{\frac{2\pi i}{n}t\oint^\alpha \hat m} e^{-\frac{2\pi i}{n}l\oint^\alpha m}+e^{-\frac{2\pi i}{n}t\oint^\alpha \hat m}e^{\frac{2\pi i}{n}l\oint^\alpha m} \Big)\,.
\end{split}
\end{equation}

\paragraph{Sector: $([s], \mathrm{Rep}(\Z_2))$} Let us now focus on the non-Lagrangian sector. From the Drinfeld center data, we have the pure magnetic line operators $D_{[s]}$ which corresponds in the Drinfeld center to $([s], 1^{\Z_2})$. According to our general discussion in section \ref{sec: semidirect}, we have the following non-Lagrangian fusions:
\begin{equation}
    \begin{split}
    ([s], 1^{\Z_2}) \otimes ([e],1_s^{D_n}) &= ([s], 1_-^{\Z_2})\,,\\
    ([s], 1^{\Z_2})\otimes ([r^l], 1^{\mathbb{Z}_n}_{t}) &= ([s], 1^{\Z_2}) + ([s], 1_-^{\Z_2}) \,,\\
    ([s], 1^{\Z_2}) \otimes([s], 1^{\Z_2}) &= ([e],1^{D_n}) + \sum_{l=1}^{k} ([e],2^{D_n}_l) +  \sum_{l=1}^{k}\sum_{t=0}^{n-1}([r^l],1^{\Z_n}_t) \,.
\end{split}
\end{equation}
With these fusions settled, we can use the Lagrangian approach discussed above to determine all the remaining fusions.

Finally, notice that for the case of even dihedral groups $D_{2k}=\Z_{2k}\rtimes \Z_2$ the $\Z_2$ does not act freely on $\Z_{2k}$ and, in an approach similar to the above, the non-Lagrangian sector requires additional steps to analyze. However, we can consider the pure 2-cocycle extension: $\Z_2\to D_{2k}\to D_{k}$ and see that (eventually iterating this procedure) we can decompose the $D_{2k}$ data into a bunch of $\Z_2$ and $D_{r}$ with $r$ odd. This motivates us to consider 2-cocycle extensions, which we discuss next.

\subsection{Pure 2-Cocycle Extensions}
\label{sec:Pure2Cocycle}

We now turn to the opposite extreme of the previous subsection and consider the case of
a \emph{pure} group cocycle extension. Concretely, we take the short exact sequence
\begin{equation}
1 \to M \to G \to A \to 1
\label{eq:pure-beta-extension-seq}
\end{equation}
to be classified by a non-trivial $2$-cocycle
$\beta \in Z^{2}(A,M)$ with \emph{trivial} action
\begin{equation}
\alpha : A \to \mathrm{Aut}(M),
\qquad \alpha_a = \mathrm{id}_M \quad \forall\,a \in A \, .
\end{equation}

In this case, the exact sequence is non-split and $G$ is a central extension of $A$ by $M$. Any non-Abelianity of $G$ is entirely due to the cocycle $\beta$ rather than a non-trivial
semi-direct product structure. The twisted coboundary
$\delta_\alpha$ thus reduces to the ordinary coboundary $\delta$, and the constraint
\eqref{eq:twisted-m-eom} becomes
\begin{equation}\label{eq:pure-beta-constraint}
\delta m + a^{\ast}\beta \;=\; 0 \, ,
\end{equation}
where $a \in Z^{1}(X,A)$ is the background $A$-bundle and
$m \in C^{1}(X,M)$ is an $M$-valued $1$-cochain.

From the above equation, one easily sees that even though the $A$-connection $a$ is flat, the $M$-valued field $m$ is forced
to have a non-trivial ``curvature'' $\delta m = - a^{\ast}\beta$, whenever the
pullback of the cocycle $\beta$ has support.

The Lagrangian description in the pure $\beta$-extension case can be realized by imposing (\ref{eq:pure-beta-constraint}) via a Lagrangian multiplier $\hat m \in C^{1}(X,M^\vee)$, which can be treated as the background gauge field for the quantum dual $M^\vee=\text{Hom}(M,U(1))$. The resulting action reads
\begin{equation}\label{eq:pure-beta-LM}
S_{\mathrm{LM}} \;=\; 2\pi i \int_X \left\langle \hat m \,,\, \delta m + a^{\ast}\beta \right\rangle_M \, .
\end{equation}
Gauge invariance of this action is guaranteed by taking $\hat m$ to transform
covariantly with the dual action $\alpha^\text{op}: A \rightarrow \text{Aut}(M^\vee)$, which
simplifies to a right action by the identity $\mathrm{id}_M$ in the present case.

When $A$ is an Abelian group, the trivial action of $A$ on $M$ admits an equivalent expression of the Lagrangian by writing all the $A$-connections to be off-shell, imposing their flatness via extra couplings to their quantum dual fields. The resulting fully off-shell action reads
\begin{equation}\label{eq:pure-beta-LM-off}
S_{\mathrm{LM}}
\;=\;
2\pi i \int_X \Big(
\left\langle \hat m \,,\, \delta m + a^{\ast}\beta \right\rangle_M
+
\left\langle \hat a,\,\delta a\right\rangle_A
\Big).
\end{equation}
We will see later in this section via explicit examples that most realizations of non-Abelian group SymTFTs in the recent literature fall into this expression.

An upshot of this Lagrangian description is that it exhibits a particularly transparent realization of non-Abelian symmetry data in terms
of non-genuine extended operators. Let us now construct the topological operators associated with the $M$-sector in this
pure $\beta$-extension. Since $\hat m$ is a $1$-cochain which is also closed, its Wilson line operator
supported on a closed loop $\gamma \subset X$ is simply
\begin{equation}
L_{\hat m}(\gamma) \;=\; \exp\!\left( 2\pi i \int_{\gamma} \langle \hat m , \lambda \rangle_M \right) ,
\label{eq:naive-m-line}
\end{equation}
for some fixed charge $\lambda \in M$. However, because $m$ is only defined up
to shifts \eqref{eq:pure-beta-constraint} that keep $S_{\mathrm{LM}}$ invariant, a similar form as \eqref{eq:naive-m-line} would not by
itself be gauge invariant. Instead, gauge invariance forces us to attach a two-dimensional
surface operator built from the $\beta$-flux.

To see this explicitly, pick a $2$-chain $N$ with $\partial N = \gamma$ and consider the
composite operator
\begin{equation}
\mathcal{U}_{\lambda}(\gamma,N)
\;=\;
\exp\!\left( 2\pi i \int_{\gamma} \langle  m , \lambda \rangle_M \right)
\exp\!\left( 2\pi i \int_{N} \langle \beta(a,a) , \lambda \rangle_M \right) .
\label{eq:beta-line-surface-composite}
\end{equation}
Using the constraint \eqref{eq:pure-beta-constraint}, a deformation of the surface
$N$ that keeps $\gamma = \partial N$ fixed shifts the second factor by
$\exp\!\left( 2\pi i \int_{N} \langle \delta m , \lambda \rangle_M \right)$,
which is precisely cancelled by the variation of the first factor built from
$m$. In other words, the dependence on the choice of spanning surface $N$ is
compensated by the equation of motion, and the composite
$\mathcal{U}_{\lambda}(\gamma,N)$ is well-defined as long as $(\gamma,N)$ bounds a
cycle on which \eqref{eq:pure-beta-constraint} holds. Crucially, the line operator
can never detach completely from the $\beta$-surface: the non-trivial transformation
of $m$ enforced by \eqref{eq:pure-beta-constraint} requires the presence of the
flux sheet $N$.

Moreover, it is also possible to build genuine line operator for $m$, by stacking the naive Wilson line $\exp\!\left( 2\pi i \int_{\gamma} \langle  m , \lambda \rangle_M \right)$ onto a non-invertible line operator or by using the $\beta$-twisted integral. This non-invertible line operator can be viewed as an anomalous 1D theory, whose anomaly inflow theory is exactly the surface operator $\exp\!\left( 2\pi i \int_{N} \langle \beta(a,a) , \lambda \rangle_M \right)$. Its explicit forms can be in principle written down for $M$ and $A$ Abelian, which we will show in the following subsections.

Similarly, one can construct topological operators associated with the $A$-sector. Expressing operators in this sector in a general form as the $M$-sector could be subtle, so we leave their explicit realizations in the examples in the following subsection. Here we just argue their existence qualitatively. Since $a$ is the background field for $A$ which is a closed 1-chain, its Wilson line operator is simply
\begin{equation}
L_{\hat m}(\gamma) \;=\; \exp\!\left( 2\pi i \int_{\gamma} \langle a , \lambda \rangle_A \right).
\label{eq:naive-a-line}
\end{equation}
However, for the quantum dual fields $\hat{a}$, the equation of motion will not be that for a simply closed chain, due to the non-trivial coupling between $a$ and $\hat{m}$ in the first term of \eqref{eq:pure-beta-LM-off}. This will lead to non-genuine operators composed by the naive line operator built from $\hat{a}$ attaching to a surface operator. One can also build genuine non-invertible line operators for $\hat{a}$, whose non-invertible part is a 1D anomalous theory with the anomaly inflow theory as the surface operator in the non-genuine operator.

Remarkably, given these non-genuine operators are built from quantum dual fields for $A$, which is the quotient of $G$, they can be regarded as symmetry generators responsible for the non-Abelian structure of $G$. That is to say, without even referring to the specific electric Lagrangian algebra/topological boundaries realizing the $G$ symmetry in 2D QFTs, the non-Abelian nature of $G$ is already encoded in the bulk.

\subsubsection{Example: $Q_8$}
As a first concrete example, we consider the order 8 quaternion group
\begin{equation}\label{eq:q8 group}
    \left\langle x,y|x^2=y^2=(xy)^2, x^4=1\right\rangle.
\end{equation}
It can be realized as a pure beta-extension of $\mathbb{Z}_2^2$ by $\mathbb{Z}_2$
\begin{equation}\label{eq:q8 short exact sequence}
    1 \rightarrow \mathbb{Z}_2 \rightarrow Q_8 \rightarrow \mathbb{Z}_2^2 \rightarrow 1.
\end{equation}
This implies that the 3D SymTFT for $Q_8$ admits a purely off-shell realization, in terms of $\mathbb{Z}_2^3$ gauge fields, following the general form \eqref{eq:pure-beta-LM-off}:
\begin{equation}\label{eq:Q8 SymTFT}
S_{Q_8}
\;=\;
\pi i \int_X
\hat m \cup \delta m + \hat a \cup \delta a + \hat b \cup \delta b + \hat{m}\cup a \cup b+ \frac{1}{2}\hat{m} \cup (\delta a +\delta b).
\end{equation}
The equation of motion for $\hat{m}$
\begin{equation}\label{eq:q8 beta extension}
    \delta m + a\cup b + \frac{1}{2}(\delta a + \delta b)=0
\end{equation}
exactly falls in the general form \eqref{eq:pure-beta-constraint} for pure beta-extension. Note that the cubic interaction in \eqref{eq:Q8 SymTFT} has been known as a type-III anomaly for $\mathbb{Z}_2^3$ to realize a $Q_8$ gauge theory \cite{deWildPropitius:1995cf}, see also discussion in \cite{Bergman:2024its}.

In order to check that the above off-shell action is well-defined, one needs to show it is gauge-invariant. The gauge transformation ensuring this invariance reads (see, e.g., \cite{He:2016xpi, Kaidi:2023maf,Robbins:2025puq})
\begin{equation}\label{eq:lagrangian for q8 lines}
\begin{aligned}
a &\;\to\; a + \delta \alpha,
&\qquad
\hat a &\;\to\; \hat a + \delta \hat \alpha - (\beta m - \gamma b + \beta \delta \gamma), \\[6pt]
b &\;\to\; b + \delta \beta,
&\qquad
\hat b &\;\to\; \hat b + \delta \hat \beta - (\gamma a - \alpha m + \gamma \delta \alpha), \\[6pt]
m &\;\to\; m + \delta \gamma,
&\qquad
\hat m &\;\to\; \hat m + \delta \hat \gamma - (\alpha b - \beta a + \alpha \delta \beta).
\end{aligned}
\end{equation}

The simple objects in the abstract Drinfeld center $\mathcal{Z}(\mathrm{Vec}_{Q_8})$ can be built explicitly via gauge-invariant line operators under the above gauge transformation. This includes the following electric Wilson and magnetic vortex line operators as building blocks (see, e.g., \cite{Kaidi:2023maf, Robbins:2025puq}),
\begin{align}\label{eq:Lagrangian}
W_{\hat m} &= e^{\pi i \oint \hat m},\qquad
W_a = e^{\pi i \oint a},\qquad
W_b = e^{\pi i \oint b}, \notag \\[6pt]
V_m(M_1)
&=
\sum_{\phi_2,\phi_3\in C^0(M_1,\mathbb Z_2)}
\exp\!\Big(\pi i \oint_{M_1} m\Big)\,
\exp\!\Big(\pi i \oint_{M_1}\big(
\phi_2\,\delta\phi_3-\phi_2 b+\phi_3 a
\big)\Big), \notag \\[6pt]
V_{\hat a}(M_1)
&=
\sum_{\phi_1,\phi_3\in C^0(M_1,\mathbb Z_2)}
\exp\!\Big(\pi i \oint_{M_1} \hat a\Big)\,
\exp\!\Big(\pi i \oint_{M_1}\big(
\phi_3\,\delta\phi_1+\phi_1 b-\phi_3\hat m
\big)\Big), \notag \\[6pt]
V_{\hat b}(M_1)
&=
\sum_{\phi_1,\phi_2\in C^0(M_1,\mathbb Z_2)}
\exp\!\Big(\pi i \oint_{M_1} \hat b\Big)\,
\exp\!\Big(\pi i \oint_{M_1}\big(
\phi_1\,\delta\phi_2-\phi_1 a+\phi_2\hat m
\big)\Big), \notag \\[6pt]
V_{m\hat a}(M_1)
&=
\sum_{\phi_{12},\phi_3\in C^0(M_1,\mathbb Z_2)}
\exp\!\Big(\pi i \oint_{M_1} (m+\hat a)\Big)\,
\exp\!\Big(\pi i \oint_{M_1}\big(
\phi_{12}\,\delta\phi_3+\phi_{12} b-\phi_3(\hat m-a)
\big)\Big), \notag \\[6pt]
V_{m\hat b}(M_1)
&=
\sum_{\phi_{13},\phi_2\in C^0(M_1,\mathbb Z_2)}
\exp\!\Big(\pi i \oint_{M_1} (m+\hat b)\Big)\,
\exp\!\Big(\pi i \oint_{M_1}\big(
\phi_{13}\,\delta\phi_2+\phi_{13} a-\phi_2(b-\hat m)
\big)\Big), \notag \\[6pt]
V_{\hat a\hat b}(M_1)
&=
\sum_{\phi_{23},\phi_1\in C^0(M_1,\mathbb Z_2)}
\exp\!\Big(\pi i \oint_{M_1} (\hat a+\hat b)\Big)\,
\exp\!\Big(\pi i \oint_{M_1}\big(
\phi_{23}\,\delta\phi_1-\phi_{23}\hat m+\phi_1(a-b)
\big)\Big), \notag \\[6pt]
V_{m\hat a\hat b}(M_1)
&=
\sum_{\phi_{12},\phi_{23}\in C^0(M_1,\mathbb Z_2)}
\exp\!\Big(\pi i \oint_{M_1} (m+\hat a+\hat b)\Big)\,
\exp\!\Big(\pi i \oint_{M_1}\big(
\phi_{12}\,\delta\phi_{23}-\phi_{12}(a-b)+\phi_{23}(\hat m-a)
\big)\Big).
\tag{3.79}
\end{align}
together with fusions of these operators resulting in 22 simple line operators. Line operators composed only of Wilson lines $W$ are invertible ones with quantum dimension 1, and enjoy obvious fusion rules. Line operators involved vortex lines $V$ are non-invertible ones with quantum dimension 2, any two of which fuse into a direct sum of dimension-1 Wilson line operators (see e.g., \cite{Kaidi:2023maf} for detailed computations). For example
\begin{equation}
	V_{m} \times V_{m}=1 + W_{a} + W_{b} + W_{ab}.
\end{equation}

The above Lagrangian description of line operators enjoys a nice correspondence with the abstract Drinfeld center $\mathcal{Z}(\mathrm{Vec}_{Q_8})$ simple objects. Recall that there are five conjugacy classes for $Q_8$, under the notation in \eqref{eq:q8 group}, given by
 \begin{equation}
 	[1]=\{ 1 \}, [x^2]= \{ x^2 \}, [x]= \{ x, x^3 \}, [y]= \{ y, y^3 \}, [xy]= \{ xy, x^3 y \}.
 \end{equation}
The irreducible representations of centralizers for $[1]$ and $[x^2]$ coincide with those of $Q_8$ itself, which we denote as\footnote{This notation is also used for simple objects of TY$(\mathbb{Z}_2\times \mathbb{Z}_2)$ fusion rings, of which there are four fusion categories including Rep$(Q_8)$. See \cite{TAMBARA1998692, Perez-Lona:2023djo} for more details.}
\begin{equation}
	1, \eta, \eta', \eta\eta', \mathcal{D}.
\end{equation}
The irreducible representations of the other three conjugacy classes are all Rep$(\mathbb{Z}_4)$, which can be equivalently expressed as $\mathbb{Z}_4$ group elements:
\begin{equation}
	\rho_0, \rho_1, \rho_2, \rho_3.
\end{equation}
In total, this gives us $5+5+4+4+4=22$ simple objects in $\mathcal{Z}(Q_8)$. The dictionary between these simple objects and Lagrangian description for line operators in \eqref{eq:Lagrangian} is collected in table \ref{tab:q8 center and q8 lagrangian} (see, e.g., \cite{Robbins:2025puq}).
\begin{table}[h]
\centering
\renewcommand{\arraystretch}{1.15}
\begin{tabular}{|c|c|}
\hline
\textbf{Simple objects in $\mathcal{Z}(Q_8)$}
& \textbf{Lagrangian description} \\
\hline
$([1],1)$ & $1$ \\
$([1],\eta)$ & $W_{\hat m}$ \\
$([1],\eta')$ & $W_{a}$ \\
$([1],\eta\eta')$ & $W_{\hat m a}$ \\
$([1],\mathcal D)$ & $V_{\hat b}$ \\
\hline
$([x^2],1)$ & $W_{b}$ \\
$([x^2],\eta)$ & $W_{\hat m b}$ \\
$([x^2],\eta')$ & $W_{a b}$ \\
$([x^2],\eta\eta')$ & $W_{\hat m a b}$ \\
$([x^2],\mathcal D)$ & $V_{\hat b}\,W_{b}$ \\
\hline
$([x],\rho_0)$ & $V_{\hat a}$ \\
$([x],\rho_1)$ & $V_{\hat a\hat b}$ \\
$([x],\rho_2)$ & $V_{\hat a}\,W_{a}$ \\
$([x],\rho_3)$ & $V_{\hat a\hat b}\,W_{a} = V_{\hat a\hat b}\,W_{b}$ \\
\hline
$([y],\rho_0)$ & $V_{m}$ \\
$([y],\rho_1)$ & $V_{m\hat b}$ \\
$([y],\rho_2)$ & $V_{m}\,W_{\hat m}$ \\
$([y],\rho_3)$ & $V_{m\hat b}\,W_{\hat m} = V_{m\hat b}\,W_{b}$ \\
\hline
$([xy],\rho_0)$ & $V_{m\hat a}$ \\
$([xy],\rho_1)$ & $V_{m\hat a\hat b}$ \\
$([xy],\rho_2)$ & $V_{m\hat a}\,W_{\hat m} = V_{m\hat a}\,W_{a}$ \\
$([xy],\rho_3)$
& $V_{m\hat a\hat b}\,W_{\hat m}
 = V_{m\hat a\hat b}\,W_{a}
 = V_{m\hat a\hat b}\,W_{b}$ \\
\hline
\end{tabular}
\caption{Simple objects in $\mathcal{Z}(Q_8)$ and their corresponding field expressions.}
\label{tab:q8 center and q8 lagrangian}
\end{table}

\paragraph{Non-genuine operators and non-abelian group multiplication.}
Now let us go one step further beyond simple line operators in the Drinfeld center, and discuss how the SymTFT bulk sees the non-Abelian structure of the $Q_8$ group. Similar to the defining equation \eqref{eq:q8 beta extension} for the pure beta-extension of $Q_8$ as an equation of motion from the SymTFT Lagrangian \eqref{eq:Q8 SymTFT}, we can read equations of motions for $a$ and $b$, which are background fields for the quotient $A=\mathbb{Z}_2^2$ in the short exact sequence \eqref{eq:q8 short exact sequence}:
\begin{equation}
\begin{aligned}
\delta \hat a + \hat m \cup b + \tfrac{1}{2}\,\delta \hat m &= 0,\\
\delta \hat b + \hat m \cup a + \tfrac{1}{2}\,\delta \hat m &= 0 .
\end{aligned}
\end{equation}
It is easy to see that, similar to $m$, $\hat{a}$ and $\hat{b}$ are not closed cochains. This aligns with the fact that they cannot build naive invertible line operators, but are involved in the non-invertible line operators in \eqref{eq:lagrangian for q8 lines}.

However, in addition to the non-invertible genuine line operators, one can also construct non-genuine line operators attaching to surface operators using these equations of motions:
\begin{equation}
\begin{split}
	\mathcal{U}_{\hat{a}}(M_2)=\exp \left( \pi i \int_{\partial M_2}\hat{a} \right) \exp \left( \pi i \int_{M_2} \hat{m} b + \frac{1}{2}\delta \hat{m} \right),\\
	\mathcal{U}_{\hat{b}}(M_2)=\exp \left( \pi i \int_{\partial M_2}\hat{b} \right) \exp \left( \pi i \int_{M_2} \hat{m} a + \frac{1}{2}\delta \hat{m} \right).\\
\end{split}
\end{equation}
We claim the following correspondence between the non-genuine operators and the $Q_8$ group generators in \eqref{eq:q8 group}
\begin{equation}
	\mathcal{U}_{\hat{a}}(M_2) \leftrightarrow x, ~\mathcal{U}_{\hat{b}}(M_2) \leftrightarrow y.
\end{equation}

First, notice that the square of $\mathcal{U}_{\hat{a}}(M_2)$ (similarly for $\mathcal{U}_{\hat{b}}(M_2)$) will degenerate the $\hat{a}$ and $\hat{m}b$ terms due to the $2\pi i $ factor, and the only surviving part is
\begin{equation}
	\left( \mathcal{U}_{\hat{a}}(M_2) \right)^2= \left( \mathcal{U}_{\hat{b}}(M_2) \right)^2 =\exp \left( \pi i \int_{M_2} \delta \hat{m} \right) = \exp \left( \pi i \oint_{\partial M_2} \hat{m}\right)=W_{\hat{m}}.
\end{equation}
This is exactly the generator for the center $M=\mathbb{Z}_2$ of $Q_8$:
\begin{equation}
	W_{\hat{m}} \leftrightarrow x^2=y^2,
\end{equation}
Obviously, $\left( W_{\hat{m}} \right)^2=1$. We thus reproduce the group property $x^2=y^2$ and $(x^2)^2=x^4=1$.

The only remaining group property we need to check is $(xy)^2=y^2$, which is equivalent to
\begin{equation}
\begin{split}
    &(xy)\cdot (xy)=y\cdot x^4 \cdot y=(yx) \cdot x^2 \cdot (xy) \\
    \Rightarrow &~~~~~~~~~~~~~~xy=yx\cdot x^2
\end{split}
\end{equation}
as the non-Abelian multiplication between $x$ and $y$ up to the center $x^2$. Therefore, we need to compute the commutator between $\mathcal{U}_{\hat{a}}(M_2)$ and $\mathcal{U}_{\hat{b}}(M_2)$, and show it produces the $W_{\hat{m}}$ operator.

To derive this commutator, we can characterize the canonical quantization of the phase space for the SymTFT \eqref{eq:Q8 SymTFT}. A convenient choice of configuration variables is
\begin{equation}
	\{ m,\hat{a}, \hat{b} \},
\end{equation}
whose conjugate momenta are simply\footnote{One can certainly choose a different set of configuration variables;
in that case the symplectic structure of the phase space is no longer linear but affine,
leading to shifted canonical momenta.
}
\begin{equation}
	\{ \hat{m}, a, b \}.
\end{equation}
The canonical commutators then read
\begin{equation}\label{eq:q8 phase space commutators}
	[\langle m, \lambda \rangle, \langle \hat{m}, \gamma \rangle]=[\langle a, \lambda \rangle, \langle \hat{a}, \gamma \rangle]=[\langle b, \lambda \rangle, \langle \hat{b}, \gamma \rangle]=\frac{i}{\pi}\langle \lambda, \gamma \rangle,
\end{equation}
where $\mu, \nu$ label spacetime components of the bulk manifold.

\begin{figure}
    \centering
    \scalebox{0.825}{
    \begin{tikzpicture}
	\begin{pgfonlayer}{nodelayer}
		\node [style=SmallCircle] (0) at (-1, 2) {};
		\node [style=SmallCircle] (1) at (2, 4) {};
		\node [style=none] (2) at (-1, -1.5) {};
		\node [style=none] (3) at (2, -1.5) {};
		\node [style=none] (4) at (-1, 5.75) {$\mathcal{U}_{\hat{a}}$};
		\node [style=none] (5) at (2, 5.75) {$\mathcal{U}_{\hat{b}}$};
		\node [style=none] (6) at (-1, 2.625) {$\exp\lb \pi i \int \hat{a}\rb $};
		\node [style=none] (7) at (2, 4.75) {$\exp\lb \pi i \int \hat{b}\rb$};
		\node [style=none] (8) at (-0.375, 1.5) {$\partial M_2$};
		\node [style=none] (9) at (-0.5, 0) {$M_2$};
		\node [style=none] (10) at (1.5, -1) {$M_2'$};
		\node [style=none] (11) at (1.375, 3.5) {$\partial M_2'$};
		\node [style=none] (12) at (4.125, -1) {$\exp\lb\pi i \int \hat{m}b+\frac{1}{2}\delta \hat{m} \rb $};
		\node [style=none] (13) at (-3, 0) {$\exp\lb \pi i \int \hat{m} b +\frac{1}{2}\delta \hat{m}  \rb$};
		\node [style=none] (14) at (0.5, 2) {};
		\node [style=none] (15) at (3.5, 2) {};
		\node [style=none] (16) at (1.875, 2) {};
		\node [style=none] (17) at (2.125, 2) {};
		\node [style=none] (18) at (0.5, -2.25) {(i)};
		\node [style=SmallCircle] (19) at (11, 2) {};
		\node [style=SmallCircle] (20) at (8, 4) {};
		\node [style=none] (21) at (11, -1.5) {};
		\node [style=none] (22) at (8, -1.5) {};
		\node [style=none] (23) at (11, 5.75) {$\mathcal{U}_{\hat{a}}$};
		\node [style=none] (24) at (8, 5.75) {$\mathcal{U}_{\hat{b}}$};
		\node [style=none] (25) at (11, 2.625) {$\exp\lb \pi i \int \hat{a}\rb$};
		\node [style=none] (26) at (8, 4.75) {$\exp\lb \pi i \int \hat{b}\rb$};
		\node [style=none] (27) at (10.375, 1.5) {$\partial M_2$};
		\node [style=none] (28) at (10.5, -0.5) {$M_2$};
		\node [style=none] (29) at (8.5, 0.25) {$M_2'$};
		\node [style=none] (30) at (8.625, 3.5) {$\partial M_2'$};
		\node [style=none] (31) at (6, 0.25) {$\exp\lb \pi i \int \hat{m}a+\frac{1}{2}\delta \hat{m} \rb $};
		\node [style=none] (32) at (13., -0.5) {$\exp\lb \pi i \int \hat{m}b+\frac{1}{2}\delta \hat{m} \rb $};
		\node [style=none] (37) at (9.5, -2.25) {(ii)};
		\node [style=none] (38) at (8, 3) {};
		\node [style=none] (39) at (8.5, 2.5) {};
		\node [style=none] (40) at (8.5, 1.5) {};
		\node [style=none] (41) at (8, 1) {};
		\node [style=none] (42) at (11, 3.5) {};
		\node [style=none] (43) at (11, 0.5) {};
		\node [style=none] (44) at (12.5, 2) {};
		\node [style=SmallCircle] (45) at (2, -7.75) {};
		\node [style=SmallCircle] (46) at (-1, -5.75) {};
		\node [style=none] (47) at (2, -11.25) {};
		\node [style=none] (48) at (-1, -11.25) {};
		\node [style=none] (50) at (-1, -4) {$\mathcal{U}_{\hat{b}}$};
		\node [style=none] (51) at (2, -7.125) {$\exp\lb \pi i \int \hat{a}\rb$};
		\node [style=none] (52) at (-1, -5) {$\exp\lb \pi i \int \hat{b}\rb$};
		\node [style=none] (53) at (1.375, -8.25) {$\partial M_2$};
		\node [style=none] (54) at (1.5, -10.75) {$M_2$};
		\node [style=none] (55) at (-0.5, -9.75) {$M_2'$};
		\node [style=none] (56) at (-0.375, -6.25) {$\partial M_2'$};
		\node [style=none] (57) at (-3.125, -9.75) {$\exp\lb \pi i \int \hat{m}a+\frac{1}{2}\delta \hat{m} \rb $};
		\node [style=none] (58) at (4.125, -10.75) {$\exp\lb \pi i\int  \hat{m} b+\frac{1}{2}\delta \hat{m} \rb $};
		\node [style=none] (59) at (0.5, -12) {(iii)};
		\node [style=none] (60) at (-1, -6.75) {};
		\node [style=none] (63) at (-1, -8.75) {};
		\node [style=none] (64) at (2, -6.25) {};
		\node [style=none] (65) at (2, -9.25) {};
		\node [style=none] (66) at (3.5, -7.75) {};
		\node [style=none] (67) at (0.5, -7.75) {};
		\node [style=SmallCircle] (68) at (11, -7.75) {};
		\node [style=SmallCircle] (69) at (8, -5.75) {};
		\node [style=none] (70) at (11, -11.25) {};
		\node [style=none] (71) at (8, -11.25) {};
		\node [style=none] (72) at (11, -4) {$\mathcal{U}_{\hat{a}}W_{\hat{m}}$};
		\node [style=none] (73) at (8, -4) {$\mathcal{U}_{\hat{b}}$};
		\node [style=none] (74) at (11.5, -7.125) {$\exp\lb \pi i \int \hat{a}+\pi i  \int \hat m\rb $};
		\node [style=none] (75) at (8, -5) {$\exp\lb \pi i \int \hat{b}\rb$};
		\node [style=none] (76) at (10.375, -8.25) {$\partial M_2$};
		\node [style=none] (77) at (10.5, -9.75) {$M_2$};
		\node [style=none] (78) at (8.5, -9.75) {$M_2'$};
		\node [style=none] (79) at (8.625, -6.25) {$\partial M_2'$};
		\node [style=none] (80) at (6, -9.75) {$\exp\lb \pi i\int  \hat{m}a+\frac{1}{2}\delta \hat{m} \rb$};
		\node [style=none] (81) at (13, -9.75) {$\exp\lb \pi i \int \hat{m} b+\frac{1}{2}\delta \hat{m}  \rb$};
		\node [style=none] (82) at (9.5, -12) {(iv)};
		\node [style=none] (83) at (8, -6.75) {};
		\node [style=none] (84) at (8, -8.75) {};
    \node [style=none] (85) at (2, -5.75) {$\exp\lb \pi i \int \hat{m}a \rb $};
    \node [style=none] (86) at (3.5, -6.5) {$\Sigma_2$};
	\end{pgfonlayer}
	\begin{pgfonlayer}{edgelayer}
		\draw [style=ThickLine] (0) to (2.center);
		\draw [style=ThickLine] (1) to (3.center);
		\draw [style=ArrowLineRight] (17.center) to (15.center);
		\draw [style=ThickLine] (14.center) to (16.center);
		\draw [style=ThickLine] (19) to (21.center);
		\draw [style=ThickLine] (22.center) to (41.center);
		\draw [style=ThickLine, bend left=45] (41.center) to (40.center);
		\draw [style=ThickLine] (38.center) to (20);
		\draw [style=ThickLine, bend right=45] (38.center) to (39.center);
		\draw [style=ThickLine, in=180, out=0] (39.center) to (42.center);
		\draw [style=ThickLine, in=90, out=0] (42.center) to (44.center);
		\draw [style=ThickLine, in=0, out=-90] (44.center) to (43.center);
		\draw [style=ThickLine, in=0, out=180] (43.center) to (40.center);
		\draw [style=ThickLine] (45) to (47.center);
		\draw [style=ThickLine] (48.center) to (63.center);
		\draw [style=ThickLine] (60.center) to (46);
		\draw [style=ThickLine, in=90, out=0] (64.center) to (66.center);
		\draw [style=ThickLine, in=0, out=-90] (66.center) to (65.center);
		\draw [style=ThickLine] (60.center) to (63.center);
		\draw [style=ThickLine, in=90, out=180] (64.center) to (67.center);
		\draw [style=ThickLine, in=180, out=-90] (67.center) to (65.center);
		\draw [style=ThickLine] (68) to (70.center);
		\draw [style=ThickLine] (71.center) to (84.center);
		\draw [style=ThickLine] (83.center) to (69);
		\draw [style=ThickLine] (83.center) to (84.center);
	\end{pgfonlayer}
\end{tikzpicture}
    }
    \caption{When interchanging the location of $\mathcal{U}_{\hat{a}}$ and $\mathcal{U}_{\hat{b}}$ one of their end points is dragged through the SPT supporting surface of the other. We sketch four steps. (i): Initial configuration. The dragging is indicated by the arrow. (ii): we can deform $M_2'$ as indicated and realize a setup trivially identical to (i). The position of $\mathcal{U}_{\hat{a}}$ and $\mathcal{U}_{\hat{b}}$ have been interchanged. (iii): We reconnect the $M_2'$ shown in (ii) to obtain the $M_2'$ shown here and a disjoint surface $\Sigma_2$. (iv): we shrink $\Sigma_2$ onto $\partial M_2$. The non-trivial commutators \eqref{eq:q8 phase space commutators} dress  $\partial M_2$ with a genuine line $W_{\hat{m}}$.}
    \label{fig:LoopyLoop}
\end{figure}
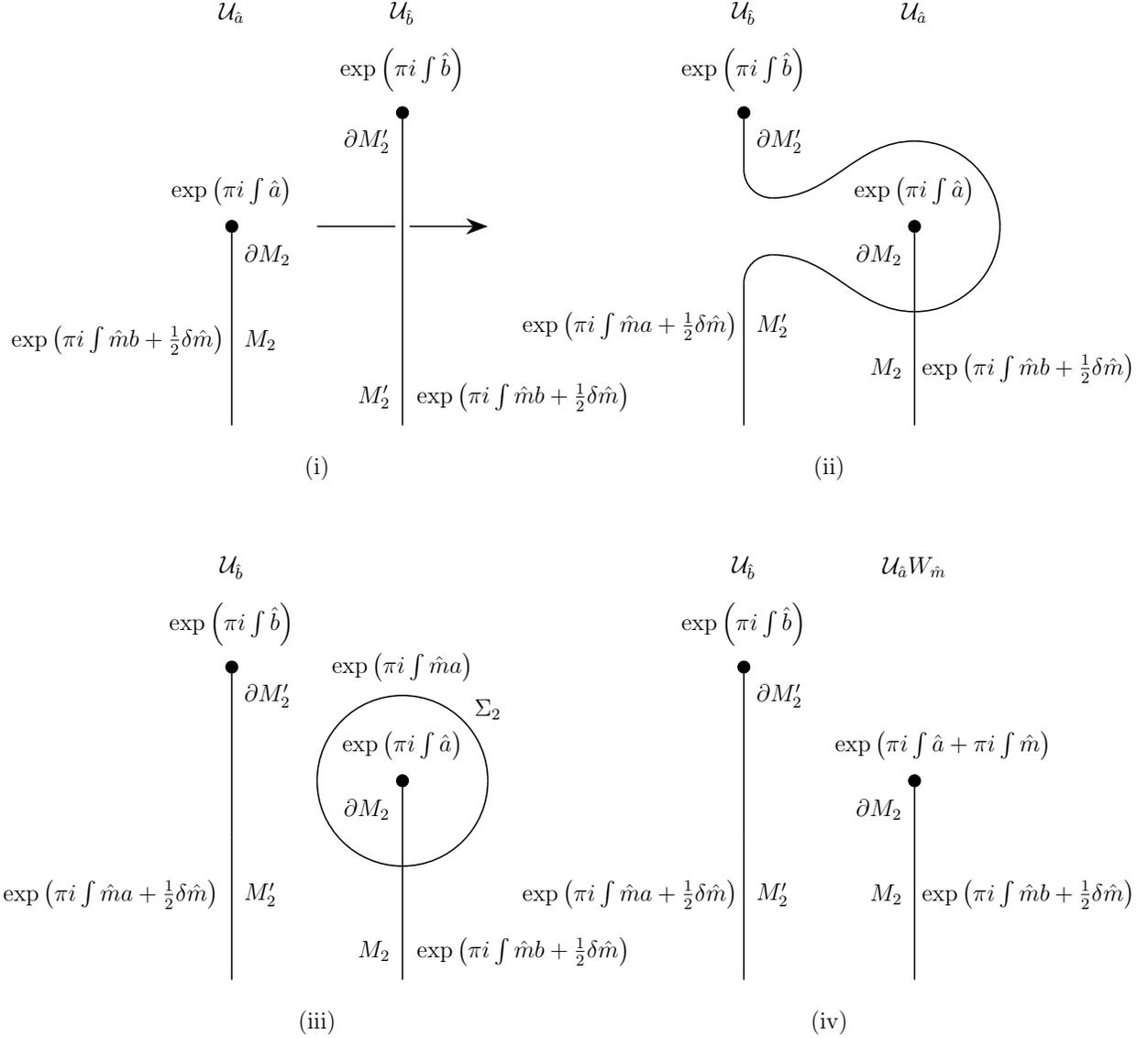

As illustrated in Figure \ref{fig:LoopyLoop}, the nontrivial result for commuting $\mathcal{U}_{\hat{a}}(M_2)$ and $\mathcal{U}_{\hat{b}}(M_2)$ is produced by how the surface operator inside $\mathcal{U}_{\hat{b}}(M_2)$ acts on the line operator inside $\mathcal{U}_{\hat{a}}(M_2)$. Using the commutators \eqref{eq:q8 phase space commutators}, one can compute it explicitly:
\begin{equation}\label{eq:commutator of q8}
\begin{split}
	&\exp \left( \pi i \int \hat{m} a + \frac{1}{2}\delta \hat{m} \right) \exp \left( \pi i \int \hat{a} \right) \exp \left( \pi i \int \hat{m} a + \frac{1}{2}\delta \hat{m} \right)\\
	= &\exp \left( \pi i \int \hat{m} a+ \frac{1}{2}\delta \hat{m} +\pi i \int \hat{a} + \frac{1}{2}(\pi i)^2 \left[ \int \hat{m} a, \int \hat{a} \right] \right) \exp \left( \pi i \int \hat{m} a + \frac{1}{2}\delta \hat{m} \right)\\
	= &\exp \left( \pi i \int \hat{m} a+ \frac{1}{2}\delta \hat{m} +\pi i \int \hat{a} + \frac{\pi i}{2} \int \hat{m} \right) \exp \left( \pi i \int \hat{m} a + \frac{1}{2}\delta \hat{m} \right)\\
	= &\exp \left( \pi i \int \delta \hat{m} + \frac{\pi i}{2} \int \hat{m} +\pi i \int \hat{a} + \frac{1}{2}(\pi i)^2\left[ \int \hat{a}, \int \hat{m}a \right] \right)\\
	= &\exp \left( \pi i \int \delta \hat{m}\right) \exp \left(  \pi i \int \hat{a} \right)\\
	= & \exp\left( \pi i \int \hat{a} \right) W_{\hat{m}}.
\end{split}
\end{equation}
The resulting commutation relation for $\mathcal{U}_{\hat{a}}(M_2)$ and $\mathcal{U}_{\hat{b}}(M_2)$ are
\begin{equation}
	\mathcal{U}_{\hat{a}}\mathcal{U}_{\hat{b}}=\mathcal{U}_{\hat{b}} \mathcal{U}_{\hat{a}} W_{\hat{m}},
\end{equation}
which reproduces the group law $xy=yx\cdot x^2$ exactly.

We conclude this subsection by summarizing the following complete dictionary between the $Q_8$ group elements and (non-genuine) operators in the SymTFT in table \ref{tab:q8 elements and non-genuine operators}.
\begin{table}[t]
\centering
\renewcommand{\arraystretch}{1.2}
\begin{tabular}{|c|c|}
\hline
\textbf{$Q_8$ element} & \textbf{SymTFT operator} \\
\hline
$1$      & $1$ \\

$x^2$     & $W_{\hat m}$ \\

$x$      & $\mathcal U_{\hat a}$ \\
$x^3$     & $\mathcal U_{\hat a}\, W_{\hat m}$ \\

$y$      & $\mathcal U_{\hat b}$ \\
$y^3$     & $\mathcal U_{\hat b}\, W_{\hat m}$ \\

$xy$      & $\mathcal U_{\hat a}\mathcal U_{\hat b}$ \\
$x^3y$     & $\mathcal U_{\hat a}\mathcal U_{\hat b}\, W_{\hat m}$ \\
\hline
\end{tabular}
\caption{Identification of $Q_8$ group elements with SymTFT operators.}
\label{tab:q8 elements and non-genuine operators}
\end{table}

\subsection{Relating Different Lagrangians: The $D_4$ Example}
\label{sec:MultipleSES}

The decomposition of solvable groups into Abelian building blocks is not unique.
For example, the dihedral group $D_4$ fits into the two sequences\footnote{Here, we use ``$1$'' to denote the trivial group of one element.}
\begin{equation}\ba \label{eq:2sequences}
	    1\to\Z_4 \to D_4 \to \Z_2\to 1 \,,\\
    1 \to \Z_2 \to D_4 \to \Z_2^2 \to 1\,.
\ea\end{equation}
The first sequence has non-trivial $\alpha$ and trivial $\beta$ while for the second sequence the converse holds. Our goal in this section will be to relate the Lagrangians for each, focussing on the example $G=D_4$, as they derive from our discussion in section \ref{sec: semidirect} and \ref{sec:Pure2Cocycle}.

The two sequences \eqref{eq:2sequences} are related using the sequence
\be\label{eq:242}
1\rightarrow \Z_2\rightarrow \Z_4\rightarrow \Z_2\rightarrow 1\,.
\ee
We will ``substitute'' this sequence into the first sequence in \eqref{eq:2sequences} (this works as the subgroup $\Z_2\subset \Z_4$ is normal in $D_4$). The sequence \eqref{eq:242} does not split and the rewriting of $\Z_4$ variables as a pair of $\Z_2$ variables will depend on a choice of section. In this example, there are two sections, and we need to apply the decomposition to two sets of $\Z_4$ valued fields. Therefore we will find four equivalent Lagrangians associated to the second sequence in \eqref{eq:2sequences}, related by a suitable change of variables.

With our expectations laid out we begin by considering the Lagrangian
\begin{equation}
    \mathcal{ L}_{D_4=\Z_4\rtimes \Z_2} = \frac{2\pi i}{4} \hat m \cup \delta_\alpha m \,,
\end{equation}
which makes reference to the semi-direct product presentation of $D_4$. Recall, here $\hat m,m$ are 1-cochains taking values in $\Z_4$. First, following \eqref{eq:242} we introduce four new fields $m^F,m^B$ and $\hat m^F,\hat m^B$  valued in $\Z_2$ by writing
\be\label{eq:section}
m=2m^F+s(m^B)\,, \qquad \hat m=2\hat m^F+\hat s(\hat m^B)\,.
\ee
with two local sections $s,\hat s:\Z_2\rightarrow \Z_4$ of \eqref{eq:242}, evaluated on the component functions of $ m^B, \hat m^B$. These sections need not exist globally. There are two choices of sections $s_1,s_2$ locally, distinguished by $s_1(1)=1,s_2(1)=3$ and with $s_1(0)=s_2(0)=0$, and $s,\hat s$ independent. With this one has
\be\ba
(\delta_\alpha m)_{ijk}&=(2m^F+s(m^B))_{ij}-(2m^F+s(m^B))_{ik}+(-1)^{a_{ij}}(2m^F+s(m^B))_{jk} \\
&=2m^F_{ij}-2m^F_{ik}+2m^F_{jk} + s(m^B)_{ij}-s(m^B)_{ik}+(-1)^{a_{ij}}s(m^B)_{jk}\\
&=2(\delta m^F)_{ijk} + s(m^B)_{ij}-s(m^B)_{ik}+(-1)^{a_{ij}}s(m^B)_{jk}\\
&=2(\delta m^F)_{ijk} + s(m^B)_{ij}-s(m^B)_{ik}+s(m^B)_{jk}+2a_{ij}s(m^B)_{jk}\\
&=2(\delta m^F)_{ijk} + (\delta( s(m^B)))_{ijk}+2a_{ij}m^B_{jk}\\
&=2(\delta m^F+a \cup m^B)_{ijk} + (\delta(s( m^B)))_{ijk}\,,\\
\ea\ee
where in going to the second line we used that $2m^F$ is in the kernel of $\alpha$, i.e., $(-1)^{a_{ij}} 2m^F_{jk}=2m^F_{jk}$. In going to the third line we recognized the component expression for the differential $\delta m^F$ and in going to the fourth line we used that this equation is mod 4 and therefore $-s(m^B)_{jk}=s(m^B)_{jk}+2s(m^B)_{jk}$. In going to the fifth line we recognized the differential $\delta (s(m^B))_{ijk}$ and in going to the sixth line we identify the cup product and reorganize.

Next, we substitute this expression into the semi-direct product Lagrangian
\be\label{eq:D4action}\ba
\mathcal{ L}_{D_4} &= \frac{2\pi i}{4} (2\hat m^F +\hat s(\hat m^B)) \cup (2\delta m^F+2a \cup s(m^B)+ \delta(s( m^B))) \\
 &= \frac{2\pi i}{2} \hat m^B\cup (\delta m^F+a \cup m^B) +  \frac{2\pi i}{2} \hat m^F\cup \delta m^B+  \frac{2\pi i}{4}  \hat{s}(\hat m^B) \cup \delta (s(m^B))\,,\\
\ea\ee
 The first two terms are independent of the sections $s,\hat s$. The difference between two sections is a local $\Z_2$-valued function, i.e., $\Delta:\Z_2\rightarrow \Z_2$ such that $s(m^B)=s'(m^B)+2\Delta(m^B)$ in $\Z_4$, and similarly for $\hat{s}$ with function $\hat\Delta$ . Changes in section therefore lead to redefinitions $m^F\rightarrow m^F+\Delta(m^B)$ and $\hat m^F\rightarrow \hat  m^F+\hat \Delta(\hat m^B) $.

 Let us explicitly consider these different choices. We have
\be
\delta (s(m^B))=2\text{Bock}(m^B)\,, \qquad \delta( \hat s(\hat m^B))=2\text{Bock}(\hat m^B) \,,
\ee
in $\Z_4$, and substituting this into the action results in
\be\label{eq:D4action2}\ba
\mathcal{ L}_{D_4}^{(1)} &= \frac{2\pi i}{2} \hat m^B\cup (\delta m^F+a \cup m^B) +  \frac{2\pi i}{2} \hat m^F\cup \delta m^B+  \frac{2\pi i}{2}  \hat m^B \cup \text{Bock}(m^B)\,.\\
\ea\ee
Note next, that changing the section $s$ results in
 \be \ba
\frac{2\pi i}{2} \hat m^B\cup \delta m^F\rightarrow &\frac{2\pi i}{2} \hat m^B\cup \delta m^F + \frac{2\pi i}{2} \hat m^B\cup \delta (\Delta(m^B)) \\ &
\frac{2\pi i}{2} \hat m^B\cup \delta m^F + \frac{2\pi i}{2} \hat m^B\cup \text{Bock}(m^B)\,,
\ea  \ee
and consequently \eqref{eq:D4action2} is equivalent to the Lagrangian
\be\label{eq:D4action3}\ba
\mathcal{ L}_{D_4}^{(2)} &= \frac{2\pi i}{2} \lb \hat m^B\cup \delta m^F+\hat m^B\cup a \cup m^B +   \hat m^F\cup \delta m^B  \rb\,.
\ea\ee
Considering all possible other choice of sections results in two more Lagrangians
\be\label{eq:D4action45}\ba
\mathcal{ L}_{D_4}^{(3)} &= \frac{2\pi i}{2} \lb \hat m^B\cup \delta m^F+\hat m^B\cup a \cup m^B +   \hat m^F\cup \delta m^B  \rb+  \frac{2\pi i}{2}   m^B \cup \text{Bock}(\hat m^B)\,,\\[0.5em]
\mathcal{ L}_{D_4}^{(4)} &= \frac{2\pi i}{2} \lb \hat m^B\cup \delta m^F+\hat m^B\cup a \cup m^B +   \hat m^F\cup \delta m^B  \rb \\&~~~\,+  \frac{2\pi i}{2}  \hat m^B \cup \text{Bock}(m^B)+   \frac{2\pi i}{2}   m^B \cup \text{Bock}(\hat m^B)\,.
\ea\ee

For the Lagrangian $\mathcal{L}^{(2)}_{D_4}$ let us redefine variables, integrate by parts and reorder using that the above is mod 2, then $(m^B,\hat m^B,m^F,\hat m^F)\rightarrow (c, b,\hat b,\hat c)$ gives
\be\label{eq:D4action4}\ba
\mathcal{ L}_{D_4}^{(2)}  &= \frac{2\pi i}{2} \lb \hat b\cup \delta  b+ a\cup b \cup c +   \hat c\cup \delta c  \rb\,,
\ea\ee
where we recall, due to having started from the semi-direct product description, that $a$ is on-shell. Taking $a$ off-shell by introducing a Lagrange multiplier $\hat a$ then gives the Lagrangian
\be\label{eq:D4action5}\ba
\mathcal{ L}_{D_4}^{(5)} &= \frac{2\pi i}{2} \lb \hat a\cup \delta  a+\hat b\cup \delta  b+   \hat c\cup \delta c+ a\cup b \cup c   \rb\,,
\ea\ee
 which is well-known (see, e.g., \cite{He:2016xpi, Kaidi:2023maf, Yu:2023nyn, Bergman:2024its}). Now that all fields in the Lagrangian are off-shell, one can construct explicitly all its genuine line operators in Drinfeld center, as well as surface-attaching non-genuine line operators associated to $D_4$ group elements. The derivation is similar to that of $Q_8$ in section \ref{sec:Pure2Cocycle}, which we summarize in Appendix \ref{app:d4 non-genuine operators}.

 The fields in the Lagrangian \eqref{eq:D4action5} are all $\Z_2$ valued and due to the absence of discrete operations, we can write an analogous Lagrangian using differential forms. This results in a sixth Lagrangian $\mathcal{L}^{(6)}_{D_4}$ associated with the short exact sequence $1 \rightarrow \Z_2\rightarrow D_4\rightarrow \Z_2\times \Z_2 \rightarrow 1$. Finally, for completeness, let us note that in the next section, we will give a seventh Lagrangian for $D_4$ DW theory using differential forms which corresponds to $1\rightarrow \Z_4\rightarrow D_4\rightarrow \Z_2 \rightarrow 1$, see \eqref{eq:D4Lagrangian2}.

\section{Lagrangian Approach with Differential Forms}
\label{sec:DiffApproach}

In the previous section we constructed a class of ``intrinsically discrete'' BF-like theories by packaging the field content in terms of discrete cochains. Now, in the case of BF-theories for finite Abelian symmetries, it is well-known that there are different ways to package the field content. For example, in the case of a $\mathbb{Z}_p$ gauge theory, we can either work in terms of discrete cochains with action:
\begin{equation}
\frac{2 \pi}{p} \int b \cup \delta a,
\end{equation}
or instead in terms of $U(1)$ valued differential forms with action:
\begin{equation}
\frac{p}{2 \pi} \int b' \wedge d a',
\end{equation}
in the obvious notation. One can formally pass between the two presentations by rescaling the fields of one presentation by a suitable power of $2 \pi / p$.

Motivated by this, it is natural to ask whether there is a similar treatment available for finite non-Abelian groups $G$ which sit
in the short exact sequence:
\begin{equation}\label{eq:extendo}
1 \rightarrow M \rightarrow G \rightarrow A \rightarrow 1,
\end{equation}
with $M$ an Abelian finite group and $A$ a finite group.
Said differently, the question becomes whether there is a Lagrangian formulation in terms of continuous differential forms.

In this section we show that for groups $G$ which are constructed from an extension of $M$ and $A$ as in line (\ref{eq:extendo}),
additional restrictions need to be imposed in order to have a consistent Lagrangian treatment. This issue persists even if we take the fields valued in $A$ ``off-shell'' as we did in the discrete cochain approach. The main result of our analysis is that for groups obtained from a central extension, i.e., of the form:
\begin{equation}
1 \rightarrow [G,G] \rightarrow G \rightarrow \mathrm{Ab}(G) \rightarrow 1,
\end{equation}
it is indeed possible to give a presentation in terms of continuous differential forms. We also show that when this condition is not satisfied, that there are generic obstructions to such a presentation. For earlier discussions of SymTFTs for finite non-Abelian groups in terms of a BF-like theory, see e.g., references \cite{deWildPropitius:1995cf,Tachikawa:2017gyf, Franco:2024mxa,Bergman:2024its, Robbins:2025puq}.

To keep our discussion concrete, we primarily focus on $G$ given as a semi-direct product $\mathbb{Z}_{n} \rtimes \mathbb{Z}_k$, though we also sketch how our discussion generalizes. For these finite groups, there is a well-known formulation in terms of $U(1)$-valued differential forms. In $D+1$ spacetime dimensions:
\be \ba
S_{\:\! \Z_n}\equiv S_{\:\! \Z_n}[ m_1,\hat m_{D-1}]&=\frac{n}{2\pi}\int  m_1\wedge d \hat m_{D-1}\,,\\[0.25em]
S_{\:\! \Z_k}\equiv   S_{\:\! \Z_k}[ a_1,\hat a_{D-1}]&=\frac{k}{2\pi}\int  a_1\wedge d \hat a_{D-1}\,.
\ea \ee
From this, we can construct a BF-theory for $G = \mathbb{Z}_n \times \mathbb{Z}_k$ with action $S_{\:\! \Z_n\times \Z_k}=S_{\:\! \Z_n}+S_{\:\! \Z_k}$. The equations of motion are
\be \ba
 m_1\,: \quad 0&= \frac{n}{2\pi} d \hat m_{D-1}\,,\quad \qquad &&\hat m_{D-1}\,: \quad 0= \frac{n}{2\pi} d  m_{1}\,, \\[0.3em]
 a_1\,: \quad 0&= \frac{k}{2\pi} d \hat a_{D-1}\,,\quad \qquad && ~\hat a_{D-1}\,: \quad\:\! 0= \frac{k}{2\pi} d  a_{1}\,, \\[0.3em]
\ea \ee
and the genuine line and surface operators
\be\label{eq:exponentiateEOM} \ba
\hat U &=\exp\lb i\int { \hat m_{D-1}}\rb\,,\qquad \quad &&  U =\exp\lb i\int { m_{1}}\rb\,, \\[0.4em]
\hat V&=\exp\Bigg( i\int {\hat a_{D-1}}\Bigg)\,, \qquad \quad &&  V =\exp\Bigg( i\int { a_{1}}\Bigg) \,,
\ea \ee
are invariant under the gauge symmetry transformations
\be\label{eq:gaugetrafo} \ba
 m_1 ~&\rightarrow~   m_1  +d \xi_0\\\
 \hat m_{D-1}~ &\rightarrow~  \hat m_{D-1}+d\hat  \xi_{D-2} \\
 a_1 ~&\rightarrow~  a_1 + d \zeta_{0} \\
\hat a_{D-1}~ &\rightarrow~ \hat a_{D-1}+ d \hat \zeta_{D-2} \,. \\
\ea \ee

Let us now ask whether we can mimic the discrete cochain construction by switching on suitable interaction terms between these two Abelian factors. To this end, select an automorphism $\alpha$ such that $\Z_n\times \Z_k$ becomes $\Z_n\rtimes \Z_k $. For this the $\Z_n$ fields must pick up charge with respect to the $\Z_k$ fields and we therefore covariantize the differential acting on $ m_1,\hat m_{D-1}$ following
\be\label{eq:covariantderive}
d~\rightarrow~  \mathscr{D}_{\:\! a_1}^{\:\!(q)}=d+  \frac{q  a_1}{2\pi}\,, \qquad  \mathscr{D}_{\:\! a_1}^{\:\!(q)}\mathscr{D}_{\:\! a_1}^{\:\!(q)}=\frac{q}{2\pi} d a_1\,,
\ee
where $q$ is the charge to be associated with $\alpha\in \text{Hom}(\Z_k,\text{Aut}(\Z_n))$. The resulting action is
\be\label{eq:SemiAction}\ba
&\text{Candidate \, Action \, for \,} G = \mathbb{Z}_n \rtimes \mathbb{Z}_k: \\
S_{\:\!\Z_n\rtimes \Z_k}& = \frac{n}{2\pi}\int  m_{1}\wedge \mathscr{D}_{\:\! a_1}^{\:\!(q)}  \hat m_{D-1}+\frac{k}{2\pi}\int  a_{1}\wedge d  \hat a_{D-1}\\[0.25em]
&=\frac{n}{2\pi}\int  m_{1}\wedge d \hat m_{D-1}+\frac{nq}{(2\pi)^2}\int  m_{1}\wedge  {a}_1\wedge \hat m_{D-1}+\frac{k}{2\pi}\int  a_{1}\wedge d  \hat a_{D-1}\,.
\ea\ee
As we show, this candidate action leads to a self-consistent treatment when $G$ is a central extension. Otherwise, we find that the gauge symmetry is pathological in the sense that the gauge symmetry is not really $G$, or even a group.\footnote{Of course, the action is still well-posed, it is simply that it does not accomplish our goal of realizing a SymTFT for the zero-form symmetry of a finite non-Abelian group $G$.}
In more technical terms, we will be able to associate a group to this action when $q/k\in \Z$ and $q=0$ mod $n$ and
$(q/k)^2=0$ mod $n$ and $\text{gcd}(n,1+q/k)=1$.\footnote{Further note that the added $  m_{1}  {a}_1 \hat m_{D-1}$  term does not fit into the type I, II, III cases of anomalies for DW theory distinguished for example in \cite{He:2016xpi}. The $  m_{1}  {a}_1  \hat m_{D-1}$  contains two variables which are conjugate, and many non-trivial features of the discussion below can be traced back to this feature.}

\subsection{Formal Gauge Transformations}\label{ssec:FormalGauge}

We now show that there are a class of formal gauge transformations of this action which never completely truncate unless $G$ is a central extension. We begin by noting that the gauge transformations \eqref{eq:gaugetrafo} leave this action invariant up to terms which are quadratic in the gauge variables. These unwanted terms can be cancelled be altering \eqref{eq:gaugetrafo} and the modified gauge transformations leave the above action invariant up to terms which are cubic in the gauge variables. The new unwanted terms can be used to modify the previously modified gauge transformations further and so on, with formal result:
\be \ba
 m_1 ~&\rightarrow~ m_1+\sum_{\ell=0}^{\infty} \Big( \frac{q  \zeta_0}{2\pi }\Big)^{\!\ell} \Big( d \xi_0- \frac{q}{2\pi} \xi_0 a_1\Big)+  \sum_{\ell=1}^{\infty} \Big( \frac{q  \zeta_0}{2\pi }\Big)^{\!\ell}  m_1\\[0.6em]
\hat m_{D-1}~ &\rightarrow~ \hat m_{D-1}+d \hat \xi_{D-2}+ \frac{q}{2\pi}    a_1 \wedge\hat  \xi_{D-2}-\frac{q}{2\pi }  \zeta_0d \hat \xi_{D-2}- \Big( \frac{ q}{2\pi}\Big)^{\!2}   \zeta_0 a_1 \wedge \hat\xi_{D-2} - \frac{q}{2\pi}  \zeta_0 \hat m_{D-1} \\[0.7em]
 a_1 ~&\rightarrow~  a_1 + \sum_{\ell=0}^{\infty} \Big( \frac{q  \zeta_0}{2\pi }\Big)^{\!\ell} d \zeta_0 \\[0.7em]
\hat a_{D-1}~ &\rightarrow~ \hat a_{D-1}+ d \hat \zeta_{D-2} +\frac{n}{k} \frac{q}{2\pi } \xi_0 d \hat \xi_{D-2} +\frac{n}{k}  \Big( \frac{ q}{2\pi}\Big)^{\!2}    \xi_0  a_1\wedge  \hat \xi_{D-2}   \\[0.25em]&~~~~~ - \frac{n}{k}\frac{ q}{2\pi }    m_1 \wedge \hat \xi_{D-2} +\frac{n}{k} \frac{ q}{2\pi } \xi_0 \hat m_{D-1}  \,.\\
\ea \ee
In more compact notation, making some of the occurring combinations explicit, we have:
\be \label{eq:gaugevariation} \ba
 m_1 ~&\rightarrow~\sum_{\ell=0}^{\infty} \Big( \frac{q  \zeta_0}{2\pi }\Big)^{\!\ell} \Big( \mathscr{D}^{(-q)}_{ a_1} \xi_0+ m_1\Big)\\[0.6em]
\hat m_{D-1}~ &\rightarrow~  \lb1-\frac{q}{2\pi} \zeta_0 \rb\lb \mathscr{D}^{(q)}_{ a_1}\hat \xi_{D-2} +\hat m_{D-1}\rb\\[0.7em]
 a_1 ~&\rightarrow~  a_1 + \sum_{\ell=0}^{\infty} \Big( \frac{q  \zeta_0}{2\pi }\Big)^{\!\ell} d \zeta_0 \\[0.55em]
\hat a_{D-1}~ &\rightarrow~ \hat a_{D-1}+ d \hat \zeta_{D-2}+\frac{n}{k} \frac{ q}{2\pi } \lb  \xi_0\mathscr{D}^{(q)}_{ a_1} \hat \xi_{D-2}  -     m_1 \wedge\hat  \xi_{D-2} + \xi_0 \hat m_{D-1} \rb \,. \\
\ea \ee
The action is invariant under these gauge transformations to all orders in the gauge variables assuming the telescoping of the infinite sums, e.g.,
\be
(1-\frac{q \zeta_0}{2\pi}) \sum_{\ell=0}^{\infty} \Big( \frac{q  \zeta_0}{2\pi }\Big)^{\!\ell}=1\,.
\ee
The constraint for the real gauge parameter $\zeta_0$ to be within the radius of convergence radius is:
\be\label{eq:fakeboy}
\Big|  \frac{q  \zeta_0}{2\pi } \Big| < 1\,.
\ee
Motivated by this, we rewrite the gauge transformations of $ a_1$ into a standard form by making the change of variable:
\be
\frac{q}{2\pi}\zeta_0= 1-\exp\lb -\frac{q}{2\pi }  \eta_0 \rb\,,
\ee
resulting in the gauge transformations
\be \label{eq:gaugevariation2} \ba
 m_1 ~&\rightarrow~ \exp\lb \frac{q}{2\pi }  \eta_0 \rb \Big( \mathscr{D}^{(-q)}_{ a_1} \xi_0+ m_1\Big)\\[0.6em]
\hat m_{D-1}~ &\rightarrow~  \exp\lb -\frac{q}{2\pi }  \eta_0 \rb \lb \mathscr{D}^{(q)}_{ a_1} \hat \xi_{D-2} +\hat m_{D-1}\rb\\[0.7em]
 a_1 ~&\rightarrow~  a_1 +d\eta_0 \\[0.7em]
\hat a_{D-1}~ &\rightarrow~ \hat a_{D-1}+ d \hat \zeta_{D-2}+\frac{n}{k} \frac{ q}{2\pi } \lb  \xi_0\mathscr{D}^{(q)}_{ a_1} \hat \xi_{D-2}  -     m_1 \wedge \hat \xi_{D-2} + \xi_0 \hat m_{D-1} \rb \,, \\
\ea \ee
which vanishes identically to all orders when substituted into the action $S_{\:\!\Z_n\rtimes \Z_k}$ given in \eqref{eq:SemiAction} on closed manifolds. In particular, we now no longer require any restrictions such as \eqref{eq:fakeboy}.

The equations of motion derived from $S_{\Z_n\rtimes \Z_k}$ are
\be \ba
 m_1\,: \qquad 0&= \frac{n}{2\pi} \mathscr{D}_{\:\! a_1}^{\:\!(q)} \hat m_{D-1}=\frac{n}{2\pi} d  \hat m_{D-1}+ \frac{qn}{(2\pi)^2}  a_1 \wedge \hat m_{D-1} \\[0.3em]
 a_1\,: \qquad 0&= \frac{k}{2\pi} d \hat a_{D-1}-\frac{qn}{(2\pi)^2}  m_1\wedge \hat m_{D-1} \\[0.3em]
 \hat m_{D-1}\,: \qquad 0&= \frac{n}{2\pi}  \mathscr{D}_{\:\! a_1}^{\:\!(-q)}  m_{1} = \frac{n}{2\pi}  \mathscr{D}_{- a_1}^{\:\!(q)}  m_{1}= \frac{n}{2\pi} d m_1-\frac{qn}{(2\pi)^2} a_1\wedge  m_1\\[0.3em]
\hat a_{D-1}\,: \qquad 0&= \frac{k}{2\pi} d  a_{1} \,.
\ea \ee
They are not generically gauge invariant, or gauge covariant, e.g.,
\be\ba
 \frac{n}{2\pi} \mathscr{D}_{\:\! a_1}^{\:\!(q)} \hat m_{D-1}~&\rightarrow~ \frac{n}{2\pi}   \exp\lb -\frac{q}{2\pi}  \eta_0\rb \lb  \mathscr{D}_{\:\! a_1}^{\:\!(q)} \hat m_{D-1} +\frac{q}{2\pi} d a_1 \wedge \hat \xi_{D-2}\rb\,, \\[0.3em]
 \frac{k}{2\pi} d \hat a_{D-1}-\frac{qn}{(2\pi)^2}  m_1\wedge \hat m_{D-1}~&\rightarrow~ \frac{k}{2\pi} d \hat a_{D-1}-\frac{qn}{(2\pi)^2}  m_1\wedge \hat m_{D-1} \\&~~~~~\,+\frac{n}{k}\frac{q}{2\pi} \xi_0 \mathscr{D}_{ a_1}^{(q)} \hat m_{D-1}-\frac{n}{k}\frac{q}{2\pi} \hat \xi_{D-2} \mathscr{D}_{ a_1}^{(-q)} m_{1}\,.
\ea \ee
Importantly however, the first equation is gauge covariant on-shell $(q/2\pi)d a_1=0$, similarly the second equation is invariant when the $m$'s are on-shell. Overall, we can therefore not simply exponentiate the equations of motion, as in the direct product case \eqref{eq:exponentiateEOM}. However, we can achieve gauge covariant operators as follows. Integrating the equations of motion over a $D$-dimensional manifold $\Sigma$ with boundary $\partial \Sigma$ (or over a $2$-dimensional manifold $\sigma$ with boundary $\partial \sigma$) we obtain the expressions
\be \label{eq:Operators} \ba
\hat U'[\Sigma]&=\exp\lb i\int_{\partial \Sigma} { \hat m_{D-1}} \rb\exp \lb \frac{ ie}{2\pi}\int_{\Sigma\,} {  a_1}\wedge { \hat m_{D-1}}  \rb \\[0.4em]
\hat V'[\Sigma]&=\exp\lb i\int_{\partial \Sigma} { \hat a_{D-1}}\rb\exp \lb -\frac{ien}{2\pi k}\int_{\Sigma\,} {  m_1} \wedge { \hat m_{D-1}}  \rb \\[0.4em]
 U'[\sigma]&=\exp\lb i\int_{\partial \sigma} { m_{1}} \rb\exp \lb - \frac{ie}{2\pi}\int_{\sigma\,} {  a_1} \wedge {  m_{1}}  \rb \\[0.4em]
 V[\partial \sigma]&=\exp\lb i\int_{\partial \sigma} { a_{1}}\rb \,,
\ea \ee
of which only $\hat V$ is gauge invariant, let alone covariant. Gauge covariant operators are derived from $U',\hat U'$ by setting $(q/2\pi)d a_1=0$ on $\Sigma$ and $\sigma$ respectively. They are
\be\ba
\hat U(\Sigma)&=\hat U'(\Sigma)\prod_{\Gamma_2}\delta \lb \frac{q}{2\pi}\int_{\Gamma_2} d a_1 \rb\,,\\
 U(\sigma)&= U'(\sigma)\prod_{\Gamma_2}\delta \lb \frac{q}{2\pi}\int_{\Gamma_2} d a_1 \rb\,,
\ea \ee
where the product runs over generators of $H_2(\Sigma,\partial \Sigma)$ and $H_2(\sigma,\partial \sigma)$ respectively, and
\be
\hat V(\Sigma)=\hat V'(\Sigma) \prod_{\Gamma_2}\delta\lb \frac{n}{k}\frac{q}{2\pi} \int_{\Gamma_2} \mathscr{D}_{ a_1}^{(-q)} m_{1} \rb  \prod_{\Gamma_{D-1}}\delta\lb \frac{n}{k}\frac{q}{2\pi} \int_{\Gamma_{D-1}} \mathscr{D}_{ a_1}^{(q)} \hat m_{D-2} \rb \,,
\ee
where the products run over generators of $H_2(\Sigma,\partial \Sigma)$ and $H_{D-1}(\Sigma,\partial \Sigma)$ respectively. The operators $\hat U, U,\hat V, V$ now transform covariantly under gauge transformations:
\be\label{eq:gaugetrafo2}
(\hat U, U, \hat V, V) \rightarrow (\hat U^\rho, U^{-\rho}, \hat V, V)\,, \qquad \rho= \exp\lb \frac{q}{2\pi}  \eta_0\rb\,.
\ee
To make the operators $U,\hat U$ gauge invariant we then have to further impose Dirichlet boundary conditions $\eta_0=0$ along their support, i.e., restrict the set of permissible gauge transformations, resulting in gauge invariant operators which we denote $U_D,\hat U_D$.

Note, the operators $U_D,\hat U_D$ realize an example of the ``simplest" deformation of a symmetry operators labelled by $(1,0)\in \Z_n\rtimes \Z_k$ from an appropriate topological boundary into the bulk via the folding trick discussed in section \ref{ssec:NGen} and displayed in figure \ref{fig:ThxKantaro} (strictly matching the figure only when we pick $\Sigma,\sigma$ to connect to the boundary). Here, we can see that they are explicitly ``simpler" than a general Dirichlet boundary condition where both $ \eta_0, \xi_0$ are set to zero (which follows from the folding trick directly when Dirichlet boundary conditions for $ m_1, a_1$ are imposed there), rather than just $ \eta_0$ (i.e., we have condensed the $\Z_n$ symmetry operators which commute with the $U$'s).

We next fix the parameter $q$ and match the above operators to $\Z_n$ and $\Z_k$ group elements. We begin with the low-codimension operators $\hat U, \hat V$ and their fusion algebra. Canonical quantization of the action \eqref{eq:SemiAction} together with the Baker-Campbell-Hausdorff formula results in the (gauge covariant) commutator
\be
\hat V \hat U \hat V^{-1}=\hat U^{1+q/k}\,.
\ee
This computes from the contour construction similar to figure \ref{fig:LoopyLoop}. For example, the action on the part of $\hat U$ supported on the boundary of $\Sigma$ is
\be\ba
\,&~~~\,\exp\lb -\frac{iqn}{2\pi k}\int {  m_1}\wedge {\hat m_{D-1}} \rb\exp\lb i\int_{\partial \Sigma} {\hat m_{D-1}} \rb  \exp\lb \frac{iqn}{2\pi k}\int  {  m_1} \wedge {\hat m_{D-1}}  \rb\\[0.3em]
&= \exp\lb  i\int_{\partial \Sigma} {\hat m_{D-1}} +\lbb  -\frac{iqn}{2\pi k}\int   {m}_1 \wedge {\hat m_{D-1}}  , i\!\int_{\partial \Sigma} {\hat m_{D-1}} \rbb \rb \\[0.3em]
&= \exp\lb  i\int_{\partial \Sigma} {\hat m_{D-1}} +\frac{iq}{k} \int_{\partial \Sigma} {\hat m_{D-1}} \rb\,,
\ea \ee
and a similar computation holds away from the boundary, i.e., here we have written out the codimension-2 part of the operator $\hat U$ and identical computation can be made for the codimension-1 part. The above computation generalizes, with arbitrary integers $r,s$, to
\be \label{eq:bummer}
\hat V^r \hat U^s \hat V^{-r}=\hat U^{s(1+rq/k)}\,.
\ee
This is NOT a group action in general, the above relation leads to offenders such as
\be \ba
\hat V^2\hat U\hat V^{-2}&=\hat U^{1+2q/k}\\
\hat V(\hat V\hat U\hat V^{-1})\hat V^{-1}&=\hat U^{(1+q/k)^2}\,.
\ea \ee
We have, by the equation of motion $\hat U^n=1$ and $\hat V^k=1$ and the relation \eqref{eq:bummer} closes when $q/k\in \Z$ and describes a group action when
\be\label{eq:linearizablecondition}
1+\frac{2q}{k}=\lb 1+\frac{q}{k}\rb^2\qquad \text{mod}\,n\,,
\ee
i.e., when $(q/k)^2=0$ mod $n$. In this case we further require $\text{gcd}(n,1+q/k)=1$ other wise conjugation has a kernel. In other words, the action induced by conjugation on the $\Z_n$ variable $\hat U$ is required to linearize in the exponent, essentially due to canonical quantization being linear.

The condition \eqref{eq:linearizablecondition} allows for a straightforward characterization of groups
$G$ which we may realize using differential forms. Assume \eqref{eq:linearizablecondition} holds. Then with $g=\text{gcd}(n,q/k)$ we have
\be
[G,G]=\langle \hat U^g \rangle\cong \Z_{n/g}\,,
\ee
and the canonical central extension
\be
1\rightarrow [G,G] \rightarrow G\rightarrow G/[G,G]=\text{Ab}(G)\rightarrow 1\,.
\ee
We have a fully off-shell description precisely when the semi-direct product may be recast as a pure 2-cocycle extension.

There are also some straightforward generalizations of the above computation. Consider for example the semi-direct product $\Z_n^p\rtimes \Z_k$. Starting from the BF-theory for $\Z_n^p\times \Z_k$ one then covariantizes the derivative as in \eqref{eq:covariantderive} but now with a matrix of charges:
\be\label{eq:covariantderive2}
\delta_{ab}d~\rightarrow~  \lb \mathscr{D}_{\:\! a_1}^{\:\!(q)}\rb_{ab}=\delta_{ab}d+  \frac{q_{ab}  a_1}{2\pi}\,.
\ee
Here $a,b=1,\dots,p$ and $\delta_{ab}d$ is associated with the diagonal single derivative BF-terms for the $\Z_n^p$ factor. The conditions for this covariant derivative to be associated with a group, following analogous steps as above, are
\be
q_{ab}/k \in \Z\,, \qquad \sum_{c=1}^p q_{ac}q_{cb}/k^2=0 ~~~\text{mod}\,n\,,
\ee
and that the smallest matrix power for which $\delta_{ab}+(q_{ab}/k)$ vanishes mod $n$ is $n$. In generalizing to $\Z_n^{p_1}\times \Z_k^{p_2}$ one introduces $p_2$ many charge matrices $q_{ab}^{(i)}$. Generalizing to arbitrary semi-direct products of Abelian groups works in principle analogously, although writing out the conditions explicitly is cumbersome. In any case, the conditions on the problematic integers again descends  from requiring closure of the conjugations action, linearity (to achieve a group structure), and maximality, i.e.,  the conjugation action should not have a non-trivial kernel.

One can also ask: when \eqref{eq:linearizablecondition} fails to hold, what algebraic structure, if any, can be associated to the action  \eqref{eq:SemiAction}? Equivalently, in the general case, what is the interpretation of \eqref{eq:bummer}? There is not much to be said, consider $\phi_r(s):=(1+r(q/k))s$ as extracted from \eqref{eq:bummer}. The failure to form a group action is characterized by
\be
\psi_{r_1,r_2}(s)=(\phi_{r_1}\circ\phi_{r_2}-\phi_{r_1+r_2})(s)=sr_1r_2(q/k)^2\,.
\ee
Because the $\phi$'s are linear we have associativity $(\phi_{r_1}\phi_{r_2})\phi_{r_3}=\phi_{r_1}(\phi_{r_2}\phi_{r_3})$.
Overall, $\Z_n$ simply realizes a pseudo-representation of $\Z_k$ with error term $\psi$.

The general claim follows from similar considerations, namely when $G$ is a central extension arising from:
\begin{equation}
1 \rightarrow [G,G] \rightarrow G \rightarrow \mathrm{Ab}(G) \rightarrow 1,
\end{equation}
a BF-like theory treatment is available, and otherwise, is destined to fail.

\subsection{Illustrative Examples}

Let us close this section with a few illustrative examples.

Consider $D_4=\Z_4\rtimes \Z_2$ we have the fully off-shell differential form action
\be\label{eq:D4Lagrangian2}\ba
S_{D_4}&=\frac{4}{2\pi}\int  m_{1}\wedge \mathscr{D}_{\:\! a_1}^{\:\!(-4)}  \hat m_{D-1}+\frac{2}{2\pi}\int  a_{1}\wedge d  \hat a_{D-1}\\[0.25em]
&=\frac{4}{2\pi}\int  m_{1}\wedge d \hat m_{D-1}-\frac{16}{(2\pi)^2}\int  m_{1}\wedge  {a}_1\wedge  \hat m_{D-1}+\frac{2}{2\pi}\int  a_{1}\wedge d  \hat a_{D-1}\,.
\ea\ee
On the other hand $S_3=D_3$ \textit{cannot} be presented using differential forms. However, note that while $D_4$ is the only dihedral group satisfying \eqref{eq:linearizablecondition}. Nonetheless, more generally, $D_N$ DW theory has a differential form presentation when $N=2^r$ for some integer $r$. This follows from repeatedly applying the short exact sequence $\Z_2\rightarrow D_N\rightarrow D_{N/2}$ which then represents $D_{N}$ as $\Z_2^r$ theory with $r-1$ group 2-cocycles which are of type III following \cite{He:2016xpi}. This example shows that concretely for the case of $\Z_n\rtimes \Z_k$ the failure of the conditions listed below \eqref{eq:SemiAction} only excludes the existence of an action with 4 separate differential form fields (here $ m_1, a_1,\hat m_{D-1}, \hat a_{D-1}$), but not more.

\section{Defects Beyond the Drinfeld Center}
\label{sec:X-check}
In the previous section, we analyzed several SymTFTs for non-Abelian group symmetries,
including $G=Q_8$ and $G=D_4$, and explicitly constructed line operators
implementing the action of group elements.
A salient feature of these examples is that, while central group elements are realized
as genuine bulk line operators, non-central elements nevertheless admit a concrete
realization as non-genuine line operators, i.e., those attaching to surface operators in the theory.
These operators play an essential role in reproducing the full non-Abelian group
structure, yet they do not fit into the standard description in terms of the
Drinfeld center.

From the viewpoint of 3D TFT, the Drinfeld center
$\mathcal Z(\mathcal C)$ of a given fusion category $\mathcal{C}$ captures the genuine bulk line operators, namely those
topological lines that can freely move and braid in the three-dimensional bulk.
Equivalently, they can be regarded as line operators living on the trivial surface defect.
As such, the Drinfeld center provides a complete characterization of bulk line defects,
but does not capture line defects that are intrinsically attached to nontrivial
surface defects.

The additional operators appearing in the non-Abelian examples are necessarily
surface-attached line defects.
They cannot be detached from the surfaces on which they live, and therefore do not
admit an interpretation as genuine bulk line defects.
Consequently, they are invisible to the Drinfeld center construction, despite being physically meaningful operators in the SymTFT.
This observation indicates that a proper understanding of these non-genuine line
operators requires a framework that goes beyond the Drinfeld center and incorporates
surface defects and their junctions with line operators on an equal footing.

In the remainder of this section, we adopt the fully extended perspective on
3D TFTs, in which defects of all codimensions are
systematically organized.
This framework will allow us to precisely characterize surface defects and
surface-attached non-genuine line operators, and to clarify how non-central group elements are
realized in group-theoretical SymTFTs.

\subsection{Defect Hierarchy and Fully Extended 3D SymTFTs}

To systematically incorporate surface-attached non-genuine line operators, it is natural to
adopt the fully extended perspective on 3D TFTs.
In this framework, a 3D SymTFT is not merely a theory assigning numbers or vector spaces
to closed manifolds, but rather a functor defined on manifolds with defects of all
codimensions.
Mathematically, defects in a fully extended 3D TFT are organized into an
$(\infty,3)$-category,\footnote{
The notation ``$(\infty,3)$'' indicates that morphisms exist in \emph{all} higher
degrees (hence the $\infty$), but only those up to degree $3$ are allowed to be
non-invertible; all $k$-morphisms for $k>3$ are invertible.
Thus an $(\infty,3)$-category has objects, 1-morphisms, 2-morphisms, and 3-morphisms,
together with higher coherence data that are invertible.
In the context of fully extended TFTs, this language conveniently organizes defects of
codimension $0,1,2,$ and $3$, corresponding respectively to bulk phases, surface defects,
line defects, and local junction operators.
We will not require a specific model of $(\infty,3)$-categories in what follows.}
as emphasized by the cobordism hypothesis and its applications
to TFTs \cite{Lurie:2009keu}.

In this framework, a 3D TFT is determined by its value on a
point, which defines an object in the target $(\infty,3)$-category. The requirement of 3D TFT being fully extended imposes strong finiteness and
duality conditions on this object. In particular, the object must be fully dualizable in the Morita $(\infty,3)$-category and it is a nontrivial result that fully dualizable objects in this setting are precisely
(spherical) fusion categories \cite{Douglas:2013aea}. Consequently, the bulk data of a fully extended 3D SymTFT is naturally encoded by a
fusion category $\mathcal{C}$. This fusion category should be regarded as the algebraic avatar of the 3D bulk phase,
rather than as a category of bulk excitations.
A concrete realization of such a fully extended theory is provided, for example, by
the Turaev-Viro(-Barrett-Westbury) TFT constructed from $\mathcal{C}$ \cite{Turaev:1992hq,Barrett:1993ab}.\footnote{
For comparison, a more frequently used construction in the physics literature for 3D SymTFTs is the Reshetikhin-Turaev construction from the Drinfeld
center $\mathcal{Z}(\mathcal{C})$ .
This theory describes the same bulk anyon content as the Turaev-Viro(-Barrett-Westbury) TFT. However, it has not been realized as
a fully extended TFT until recently \cite{freed2026fullylocalreshetikhinturaevtheories}.}

\begin{figure}
\centering
\scalebox{0.9}{
\begin{tikzpicture}
	\begin{pgfonlayer}{nodelayer}
		\node [style=none] (0) at (-5, 1.5) {};
		\node [style=none] (1) at (-3, 1.5) {};
		\node [style=none] (2) at (-1, 2) {};
		\node [style=none] (3) at (1, 1) {};
		\node [style=none] (4) at (3, 1.5) {};
		\node [style=none] (5) at (5, 1.5) {};
		\node [style=none] (6) at (-1, 0.925) {};
		\node [style=none] (7) at (-5, -1.5) {};
		\node [style=none] (8) at (-3, -1.5) {};
		\node [style=none] (9) at (-1, -1) {};
		\node [style=none] (10) at (1, -2) {};
		\node [style=none] (11) at (3, -1.5) {};
		\node [style=none] (12) at (5, -1.5) {};
		\node [style=none] (13) at (-4, 0) {$\mathcal{M}_1$};
		\node [style=none] (14) at (4, 0) {$\mathcal{M}_2$};
		\node [style=none] (15) at (-1.5, -2.5) {$\mathcal{F}_1$};
		\node [style=none] (16) at (1.5, 2.5) {$\mathcal{F}_2$};
		\node [style=none] (17) at (2.25, -2.5) {$\mathcal{F}_3$};
		\node [style=none] (18) at (-2.25, 2.5) {$\mathcal{F}_4$};
		\node [style=none] (19) at (0, 0) {$\mathcal{M}_3$};
		\node [style=none] (20) at (-1, 2.5) {$\eta_1$};
		\node [style=none] (21) at (1, -2.5) {$\eta_2$};
		\node [style=none] (22) at (-1.5, -0.5) {$\mathcal{M}_4$};
		\node [style=none] (23) at (1.5, 0.5) {$\eta_3$};
		\node [style=none] (24) at (-3.25, -2) {$\eta_4$};
		\node [style=none] (25) at (3.25, -2) {$\eta_5$};
        \node [style=none] (26) at (0, -3.25) {};
	\end{pgfonlayer}
	\begin{pgfonlayer}{edgelayer}
		\draw [style=ThickLine] (0.center) to (1.center);
		\draw [style=ThickLine, in=-180, out=30, looseness=0.75] (1.center) to (2.center);
		\draw [style=ThickLine, in=-165, out=-30, looseness=0.75] (1.center) to (3.center);
		\draw [style=ThickLine, in=135, out=-45] (2.center) to (3.center);
		\draw [style=ThickLine, in=150, out=15, looseness=0.75] (2.center) to (4.center);
		\draw [style=ThickLine, in=0, out=-150] (4.center) to (3.center);
		\draw [style=ThickLine] (4.center) to (5.center);
		\draw [style=ThickLine] (7.center) to (8.center);
		\draw [style=ThickLine, in=-165, out=-30, looseness=0.75] (8.center) to (10.center);
		\draw [style=ThickLine, in=0, out=-150] (11.center) to (10.center);
		\draw [style=ThickLine] (11.center) to (12.center);
		\draw [style=DottedLine, in=-180, out=30] (8.center) to (9.center);
		\draw [style=DottedLine, in=135, out=-45] (9.center) to (10.center);
		\draw [style=DottedLine, in=150, out=15, looseness=0.75] (9.center) to (11.center);
		\draw [style=ThickLine] (1.center) to (8.center);
		\draw [style=ThickLine] (3.center) to (10.center);
		\draw [style=ThickLine] (4.center) to (11.center);
		\draw [style=DottedLine] (6.center) to (9.center);
		\draw [style=ThickLine] (2.center) to (6.center);
	\end{pgfonlayer}
\end{tikzpicture}
}
\caption{
Defects in a fully extended 3D SymTFT with fixed bulk phase $\mathcal C$.
The ambient 3D bulk (not explicitly labeled) is characterized by a fusion
category $\mathcal C$.
Codimension-one surface defects are labeled by $\mathcal C$--$\mathcal C$ bimodule
categories $\mathcal M_i$.
Codimension-two line defects separating surfaces are labeled by bimodule functors
$\mathcal F_i$,
and codimension-three junction operators are given by natural transformations $\eta_i$.
The figure illustrates the defect hierarchy underlying the 2-category
$\mathrm{Bimod}_2(\mathcal C)$.
}
\label{fig:defects-morphisms}
\end{figure}
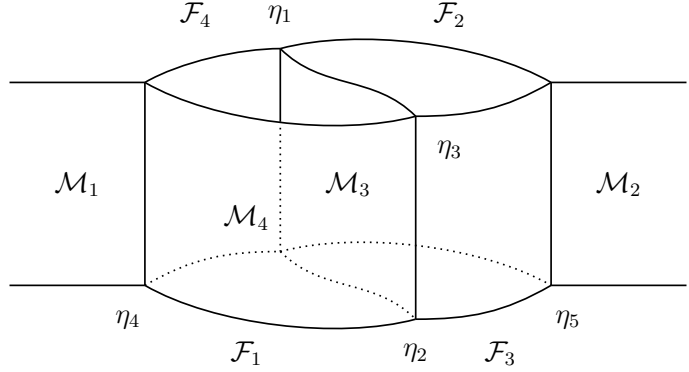

Once the bulk SymTFT as an object associated to a fusion category $\mathcal{C}$ is fixed, surface defects as 1-morphisms are then described by $\mathcal{C}-\mathcal{C}$ bimodule categories. Physically, they are topological interfaces separating identical bulk
phases. Line defects, as 2-morphisms of the $(\infty, 3)$-category, separating two surface defects are then described by bimodule functors
between the corresponding bimodule categories, while local operators at junctions of
line defects are given by natural transformations between such functors.
We denote by $\mathrm{Fun}_{\mathcal{C} | \mathcal{C}}(\mathcal{M}, \mathcal{N})$ the category of  $\mathcal{C} - \mathcal{C}$ bimodule functors between bimodule categories $\mathcal{M}$ and $\mathcal{N}$, respecting both left and right actions of $\mathcal{C}$. See Figure \ref{fig:defects-morphisms} for a rough depiction.

For our purposes, the relevant defect data can be captured by restricting attention to
the endomorphism 2-category
\begin{equation}
    \mathrm{Bimod}_2(\mathcal C),
\end{equation}
organized by surface and line defects. The objects, 1-morphisms, and 2-morphisms of this 2-category are $\mathcal{C}-\mathcal{C}$ bimodule categories, bimodule functors, and natural transformations, respectively.

Within this framework, the Drinfeld center admits a natural reinterpretation.
The trivial surface defect is described by the regular $\mathcal{C}-\mathcal{C}$
bimodule category, i.e., $\mathcal C$ itself.
Line defects living entirely on this trivial surface thus correspond to bimodule
endofunctors of $\mathcal C$.
As a result, the category of bulk genuine line operators is identified with the
endomorphism category of the trivial surface defect \cite{Schommer-Pries:2011rhj},
\begin{equation}
    \mathcal Z(\mathcal C)\;\simeq\;
    \mathrm{End}_{\mathrm{Bimod}_2(\mathcal C)}(\mathcal C).
\end{equation}
This formulation makes manifest that the Drinfeld center captures precisely those
line defects that can be inserted in the bulk without being attached to any nontrivial
surface defect.

More general line defects in a 3D SymTFT are now described by a
1-morphism between surface defects in $\mathrm{Bimod}_2(\mathcal C)$, namely a
bimodule functor between two (possibly not regular) $\mathcal C$--$\mathcal C$ bimodule categories.
In particular, the non-genuine line operators constructed in the previous section are
precisely of one special type: they correspond to bimodule functors from the trivial surface
defect (regular bimodule category) to a nontrivial surface defect (or equivalently, from a nontrivial surface
defect to the trivial one).

To be more precise, consider a line defect separating the trivial surface defect, described by the
regular bimodule category $\mathcal C$, and a nontrivial surface defect described by
a $\mathcal C$--$\mathcal C$ bimodule category $\mathcal M$.
Such a line defect is encoded by a bimodule functor
\begin{equation}
    F:\ \mathcal C \to \mathcal M
    \qquad \text{or equivalently} \qquad
    F^{\mathrm{op}}:\ \mathcal M \to \mathcal C .
\end{equation}
However, because $\mathcal C$ is the regular bimodule category, bimodule functors out
of $\mathcal C$ are in one-to-one correspondence with objects of $\mathcal M$. Schematically, one can understand this correspondence due to that for every object $m\in \mathcal{M}$, it is a module acted by objects $L \in \mathcal{C}$:
\begin{equation}
	 L \,\triangleright\, m = m',
\end{equation}
thus constructing a functor $F: \mathcal{C} \to \mathcal{M}$. Formally, this one-to-one correspondence can be written as \cite{Ostrik:2001xnt,Etingof:2002vpd}
\begin{equation}
    \mathrm{Fun}_{\mathcal C|\mathcal C}(\mathcal C,\mathcal M)
    \;\simeq\;
    \mathcal M .
\end{equation}
Therefore, non-genuine line operators connecting the trivial surface to a nontrivial
surface defect are naturally labeled by objects in the corresponding bimodule category.

We conclude the correspondence between physical defects and their categorical counterparts objects in table \ref{tab:defect-dictionary}.
\begin{table}[t]
\centering
\begin{tabular}{|c|c|}
\hline
Defect in 3D SymTFT & Categorical description \\
\hline
surface defect
& $\mathcal C$--$\mathcal C$ bimodule category $\mathcal M$ \\[0.5em]

trivial surface defect
& regular bimodule category $\mathcal C$ \\[0.5em]

nontrivial surface defect
& $\mathcal C$--$\mathcal C$ bimodule category $\mathcal M \neq \mathcal C$ \\[0.5em]

generic line defect
& bimodule functor $F:\mathcal N \to \mathcal M$ \\[0.5em]

non-genuine line defect
& object $m \in \mathcal M \;\simeq\;
\mathrm{Fun}_{\mathcal C|\mathcal C}(\mathcal C,\mathcal M)$ \\
\hline
\end{tabular}
\caption{Dictionary between defect operators in a fully extended 3D SymTFT and categorical data in the Morita 2-category $\mathrm{Bimod}_2(\mathcal C)$.}
\label{tab:defect-dictionary}
\end{table}

\subsection{Non-Genuine Defects in SymTFTs for $\mathrm{Vec}_G$}

We now specialize to the case $\mathcal C=\mathrm{Vec}_G$, where $G$ is a finite group.
Recall that $\mathrm{Vec}_G$ denotes the fusion category of $G$-graded vector spaces,
with tensor product given by the group multiplication.
As explained in the previous subsection, surface defects in the corresponding 3D
SymTFT are described by $\mathrm{Vec}_G-\mathrm{Vec}_G$ bimodule categories.

\subsubsection{Surface Defects and Group Automorphisms}
The classification of $\mathrm{Vec}_G-\mathrm{Vec}_G$ bimodule categories is well
understood \cite{Ostrik:2001xnt,Etingof:2002vpd}.
In particular, a special class of these bimodule categories are characterized by automorphisms of $G$, which will capture the essential features needed to realize non-Abelian symmetry
generators.

Given an automorphism $\varphi \in \mathrm{Aut}(G)$, one can associate a
$\mathrm{Vec}_G-\mathrm{Vec}_G$ bimodule category $\mathcal M_{\varphi}$.
The simple objects in $\mathcal M_{\varphi}$ are in one-to-one correspondence with those of $\mathrm{Vec}_G$, and are therefore labeled by elements of $G$ \cite{etingof2015tensor}.
The left and right actions of $\mathrm{Vec}_G$ are, however, twisted relative to one
another by the automorphism $\varphi$.
Concretely, the left action is given by ordinary group multiplication, while the
right action is composed with $\varphi$. This twisting distinguishes $\mathcal M_{\varphi}$ from the regular bimodule category
and endows it with a nontrivial surface defect interpretation.

Among all these surface defects, those constructed via inner automorphisms are of interest to us.
For each group element $g\in G$, the inner automorphism
\begin{equation}
    \varphi_g = \mathrm{Ad}_g : h \mapsto g h g^{-1}
\end{equation}
defines a $\mathrm{Vec}_G-\mathrm{Vec}_G$ bimodule category $\mathcal M_{\varphi_g}$.
For simplicity, we will denote these bimodule categories by $\mathcal{M}_g$. The bimodule structure encodes a nontrivial action of the group
element $g$ on objects of $\mathcal{M}_g$ which are also labeled by group elements. To be more specific, denote $m_k$ as a simple object of $\mathcal M_g$ labeled by $k\in G$, then the left and right actions of $\mathrm{Vec}_G$ are given explicitly by
\begin{equation}
    h \,\triangleright\, m_k \,\triangleleft\, h'
    \;=\;
    m_{\,h\,k\,\varphi_g(h')}
    \;=\;
    m_{\,h\,k\,g h' g^{-1}} ,
\end{equation}
for $h,h'\in G$.
This bimodule category $\mathcal{M}_g$ thus encodes the presence of a nontrivial surface defect labeled by
the group element $g$.

In fact, given the above surface defects are labeled by group elements and are codimension-1 defects in the SymTFT, we already see how the non-Abelian structure is present in the bulk. To match and generalize our results in previous sections illustrated with examples, we will see shortly that there are indeed non-genuine line operators attaching to these surface defects.

\subsubsection{Non-Genuine Line Operators and Group Elements}

We now turn to the line defects associated to the inner-automorphism surface defects
$\mathcal M_g$.
As previously explained in this section, a line defect separating the trivial surface defect,
described by the regular bimodule category $\mathcal C=\mathrm{Vec}_G$, and a
nontrivial surface defect described by a bimodule category $\mathcal M$ is encoded by
a bimodule functor
\begin{equation}
    F:\ \mathrm{Vec}_G \to \mathcal M .
\end{equation}
When $\mathcal M=\mathcal M_g$, this describes a line operator attached to the surface
defect labeled by the group element $g$. Recall further that because $\mathrm{Vec}_G$ is the regular bimodule category, such bimodule functors are
canonically classified by objects of $\mathcal M_g$.
We therefore conclude that non-genuine line operators associated to the surface defect
$\mathcal M_g$ are naturally labeled by the objects of $\mathcal M_g$.

Since the simple objects of $\mathcal M_g$ are labeled by group elements, we obtain a
canonical family of surface-attached line operators $\{\mathcal{U}_{g,h}\}$, indexed by
$h\in G$.
Among these, there is a special choice corresponding to the object $m_e \in \mathcal{M}_g$ labeled
by the identity element, which we will denote as $\mathcal{U}_g$ for simplicity.
Physically, $\mathcal{U}_g$ should be viewed as the non-genuine line operator implementing the action of the
group element $g$, attached to the non-trivial surface defect $\mathcal M_g$, as illustrated by examples in the previous section.

The fusion of non-genuine line operators is naturally inherited from the tensor
structure of the bimodule category.
In particular, consider two such operators $\mathcal{U}_g$ and $\mathcal{U}_h$, associated to surface
defects $\mathcal{M}_g$ and $\mathcal{M}_h$, respectively. The fusion of such defects is described by the relative
tensor product of bimodule categories,
\begin{equation}
    \mathcal M_g \boxtimes_{\mathrm{Vec}_G} \mathcal M_h \;\simeq\; \mathcal M_{gh}.
\end{equation}
More concretely,
let $m^{(g)}_e\in\mathcal M_g$ and $m^{(h)}_e\in\mathcal M_h$ denote the objects labeled
by the identity element.
In the relative tensor product one identifies
\begin{equation}
    \bigl(m^{(g)}_e \triangleleft x\bigr)\boxtimes m^{(h)}_e
    \;\sim\;
    m^{(g)}_e \boxtimes \bigl(x\triangleright m^{(h)}_e\bigr),
    \qquad x\in G,
\end{equation}
where $\triangleleft$ and $\triangleright$ denote the right and left actions.
Using the twisted right action in $\mathcal M_g$,
$m^{(g)}_e \triangleleft x = m^{(g)}_{\varphi_g(x)}$, one finds that the composed
bimodule structure is twisted by $\varphi_{gh}$, hence
$\mathcal M_g \boxtimes_{\mathrm{Vec}_G} \mathcal M_h \simeq \mathcal M_{gh}$.
Choosing the special non-genuine line $U_g$ (resp.\ $U_h$) to correspond to
$m^{(g)}_e$ (resp.\ $m^{(h)}_e$), this induces the fusion rule
\begin{equation}
    \mathcal{U}_g \,\otimes\, \mathcal{U}_h \;\simeq\; \mathcal{U}_{gh} ,
\end{equation}
reproducing the group multiplication law.\footnote{Recall that these $\mathcal{U}$'s label surface operators. The line operators have a more intricate fusion rule.}

We conclude this section by summarizing defects in SymTFT for $\mathrm{Vec}_G$ and their categorical counterparts in table \ref{tab:defect-dictionary-vecG}.
\begin{table}[t]
\centering
\begin{tabular}{|c|c|}
\hline
Defect in $\mathrm{Vec}_G$ SymTFT
& Categorical description \\
\hline
surface defect
& $\mathrm{Vec}_G$--$\mathrm{Vec}_G$ bimodule category $\mathcal M$ \\[0.5em]

trivial surface defect
& regular bimodule category $\mathrm{Vec}_G$ \\[0.5em]

surface defect labeled by $g\in G$
& twisted bimodule category $\mathcal M_g$
\quad ($\mathcal M_g \boxtimes_{\mathrm{Vec}_G} \mathcal M_h \simeq \mathcal M_{gh}$) \\[0.5em]

generic line defect
& bimodule functor $F:\mathcal M_g \to \mathcal M_h$ \\[0.5em]

non-genuine line defect
& object $m_k \in \mathcal M_g \;\simeq\;
\mathrm{Fun}_{\mathrm{Vec}_G|\mathrm{Vec}_G}(\mathrm{Vec}_G,\mathcal M_g)$ \\[0.5em]

special non-genuine line
& $\mathcal U_g \;\equiv\; m_e \in \mathcal M_g  $ \\
\hline
\end{tabular}
\caption{Defect dictionary for $\mathrm{Vec}_G$ SymTFTs.
The group multiplication law is realized defect-theoretically through the relative
tensor product of surface defects and the fusion of surface-attached line operators.}
\label{tab:defect-dictionary-vecG}
\end{table}

\section{Higher-Dimensional Generalization} \label{sec:HIGHER}
So far we restricted ourselves to the study of DW theory in dimension 3. However, most of the things we discussed have a straightforward generalization to higher dimension.

Let us now take $X$ to be a $d$-dimensional manifold, with $\omega \in H^d(G,U(1))$, where $G$ is the usual finite gauge group of DW theory. Then, up to an overall normalization factor, the path-integral of DW theory is given by
\begin{equation}
    Z(X)=\sum_{g\in Z^1(X,G)} \int_X g^*\omega \,.
\end{equation}
Assume now that $G$ fits into a short exact sequence of the following form:
\begin{equation}
    1\to M \to G \xrightarrow{\pi} A \to 1\,,
\end{equation}
where $M$ is Abelian, and assume that $\omega$ factors as follows:
\[\begin{tikzcd}
	{G^d} & {A^d} \\
	& {U(1)}
	\arrow["{\pi^d}", from=1-1, to=1-2]
	\arrow["\omega"', from=1-1, to=2-2]
	\arrow["{\tilde{\omega}}", from=1-2, to=2-2]
\end{tikzcd}\]
where $\tilde\omega \in H^d(A,U(1))$. Then, we can introduce the dual field $\hat{m}\in C^{d-2}(X,M)$, and recast the path-integral of DW theory in the following form:
\begin{equation}
    Z(X) = \sum_{a\in H^1(X,A)}\sum_{m\in C^1(X,M)}\sum_{\hat m\in C^{d-2}(X,M)} e^{2\pi i \int_X \hat{m}\mathbf{Q}\cup(\delta_\alpha m +a^*\beta)+ 2\pi i \int_X a^*\tilde\omega}\,,
\end{equation}
where $\alpha$ and $\beta$ are respectively the action and the group 2-cocycle classifying the above short exact sequence. The discussion for the pair $(a,m)$ is the same as in the 3-dimensional case. Namely, this pair (on-shell) describes the field $g\in Z^1(X,G)$. Simple line operators associated to this field are labeled by the irreducible representations of $G$.

On the other hand, the pair $(a,\hat m)$ is now describing the gauge field associated to a higher group \cite{Kapustin:2013uxa}. To be precise, we have the data of a gauge field for a $(d-2)$-group with trivial Postnikov class, but non-trivial action of the 0-form symmetry to the $(d-3)$-form symmetry, given by $\alpha^{\mathrm{op}}$. Hence, the simple line operators of this theory are labeled by irreducible $(d-2)$-representations for a $(d-2)$-group. The construction is going to be the analogous to the 3-dimensional case: First, we need to write down the holonomies for the gauge field of the $(d-2)$-group, and then we need to construct the most general gauge invariant operators with these holonomies \cite{attal2002combinatoricsnonAbeliangerbesconnection , Pfeiffer_2003 , Girelli_2004 , baez2006highergaugetheory}. We will discuss this construction in the next section in the case of DW theory in dimension 4.

Connecting to section \ref{sec:X-check}, from a mathematical point of view this is telling us the following two facts:
\begin{itemize}
    \item Let us denote by $D$-Vect the target $D$-category (whatever it is) of $D$-dimensional Dijkgraaf-Witten theories in arbitrary dimension $d$. Then, the operators supported on manifolds of dimension $(D-2)$ are given by the objects of the $(D-2)$-category associated to the endomorphism of the identity 1-morphism of the DW functor $\mathcal{Z}_{DW}^{G,\omega}$. Such category must contain the $(D-2)$-representations for the $(D-2)$-group $M^{D-3}\rtimes_{\alpha^{\mathrm{op}}}A^0$ as subcategory.
    \item Let us now consider the identity $(D-2)$-morphism of $\mathcal{Z}_{DW}^{G,\omega}$. Then the 1-category of endomorphisms of the $(D-2)$-identity must contain $\mathrm{Rep}(G)$ as a subcategory.
\end{itemize}

\subsection{4D Dijkgraaf-Witten SymTFTs}
We now discuss a bit more in depth DW theory in dimension 4. In particular, we want to understand how to construct the holonomy of the 2-group gauge field appearing in the theory, which is described by the pair of fields $(a,\hat m)$. As mentioned above, this 2-group is characterized by having trivial Postnikov class, but non-trivial action of the group $A$ on the 1-form symmetry $M$.

In order to construct the holonomy operator we need a bit of mathematical background first. We will mainly follow \cite{Pfeiffer_2003}.

A (strict) 2-group consists in the following data: A group of objects $G_0$, a group of arrows $G_1$, two group homomorphisms $s,t: G_1 \to G_0$ (source and target), and identity morphism $id: G_0 \to G_1$ and a multiplication map $\circ: G_1\times_{G_0}G_1 \to G_1$, where $G_1\times_{G_0}G_1 = \{ (g,h)\in G_1\times G_1 : t(g)=s(h) \}$ is the set of composable morphisms. These data have to satisfy the following axioms:
\begin{equation}
\begin{split}
    s(id_x)=t(id_x) = x \,\,\, \forall x\in G_0 \,, \\
    s(g\circ h) = g \, \,, \, t(g\circ h) =h \, \, \forall(g,h)\in G_1\times_{G_0}G_1 \,,\\
    id_{s(g)}\circ g = g\circ id_{t(g)} = g \, \, \forall g\in G_1 \,, \\
    (g_1\circ g_2)\circ g_3 = g_1\circ(g_2\circ g_3) \,,
\end{split}
\end{equation}
where $g_1$, $g_2$ and $g_3$ are composable morphisms. In the case we are studying, $G_0$ can be identified with the group $A$ while $G_1$ can be identified with the semidirect product $M\rtimes_{\alpha^\mathrm{op}} A$. Source and target homomorphisms coincide and are given by the projection on $A$. The identity morphism $id:A \to M\rtimes_{\alpha^\mathrm{op}} A$ is given by $a\to id_a = (1,a)$, and the composition is given as follows:
\begin{equation}
    (m,a) \circ (m', a) = (mm', a) \,.
\end{equation}
W\[\begin{tikzcd}
	\bullet && \bullet & {=} & \bullet && \bullet
	\arrow[""{name=0, anchor=center, inner sep=0}, "a"', curve={height=18pt}, from=1-1, to=1-3]
	\arrow[""{name=1, anchor=center, inner sep=0}, "a", curve={height=-18pt}, from=1-1, to=1-3]
	\arrow[""{name=2, anchor=center, inner sep=0}, "a", curve={height=-30pt}, from=1-5, to=1-7]
	\arrow[""{name=3, anchor=center, inner sep=0}, "a"', curve={height=30pt}, from=1-5, to=1-7]
	\arrow[""{name=4, anchor=center, inner sep=0}, "a"{pos=0.7}, from=1-5, to=1-7]
	\arrow["{mm'}", between={0.2}{0.8}, Rightarrow, from=1, to=0]
	\arrow["{m'}", between={0.2}{0.8}, Rightarrow, from=4, to=3]
	\arrow["m", between={0.2}{0.8}, Rightarrow, from=2, to=4]
\end{tikzcd}\]
On top of this, we can also define vertical composition as follows:
\[\begin{tikzcd}
	\bullet & \bullet & \bullet & {=} & \bullet &&& \bullet
	\arrow[""{name=0, anchor=center, inner sep=0}, "a"', curve={height=18pt}, from=1-1, to=1-2]
	\arrow[""{name=1, anchor=center, inner sep=0}, "a", curve={height=-18pt}, from=1-1, to=1-2]
	\arrow[""{name=2, anchor=center, inner sep=0}, "b", curve={height=-18pt}, from=1-2, to=1-3]
	\arrow[""{name=3, anchor=center, inner sep=0}, "b"', curve={height=18pt}, from=1-2, to=1-3]
	\arrow[""{name=4, anchor=center, inner sep=0}, "ab", curve={height=-18pt}, from=1-5, to=1-8]
	\arrow[""{name=5, anchor=center, inner sep=0}, "ab"', curve={height=18pt}, from=1-5, to=1-8]
	\arrow["m", between={0.2}{0.8}, Rightarrow, from=1, to=0]
	\arrow["n", between={0.2}{0.8}, Rightarrow, from=2, to=3]
	\arrow["{m+\alpha_a(n)}", between={0.2}{0.8}, Rightarrow, from=4, to=5]
\end{tikzcd}\]
Concretely, we have defined a tensor product over the category defining the 2-group, using the multiplication in the semi-direct product group for morphisms.

Let us recall that the fundamental 2-groupoid of a manifold $X$ is the 2-category whose objects are point of the manifold, 1-morphisms are paths between points, and 2-morphisms are (homotopy classes of) surfaces connecting two 1-morphisms. Realizing the fundamental 2-groupoid of $X$ in terms of a triangulation allows us to furnish the datum of a 2-functor between the fundamental 2-groupoid of a space $X$ and the 2-group by diagrams of the following form:
\[\begin{tikzcd}
	& j \\
	i && k
	\arrow["{a_{jk}}", from=1-2, to=2-3]
	\arrow["{a_{ij}}", from=2-1, to=1-2]
	\arrow[""{name=0, anchor=center, inner sep=0}, "{a_{ik}}"', from=2-1, to=2-3]
	\arrow["{\hat m_{ijk}}", between={0}{0.8}, Rightarrow, from=1-2, to=0]
\end{tikzcd}\]
where $\delta a = 0$ and $\delta_{\alpha^{\mathrm{op}}} \hat m = 0$.e can visualize all this data diagramatically as follows:\footnote{To be precise, what we are actually representing is not the category defined by the 2-group, but the 2-category defined by its delooping. We will be sloppy and we will identity the two, as it will be not important in our discussion to keep the two notions separate.}

Now we have all the ingredients ready to describe the holonomy map. Indeed, with this description a closed surface $\Sigma$ is simply a 2-endomorphism of the constant loop over the base point. Under the above 2-functor, the surface is mapped into the flux of $\hat m\in H^2(X,M_{\alpha^\mathrm{op}})$ through this surface:
\begin{equation}
    \oint_\Sigma^{\alpha^\mathrm{op}} \hat m \in M \,.
\end{equation}
We take the above as the definition of twisted integral in this general setting. Notice however that this definition makes sense only if the $A$-bundle over $\Sigma$, defined by restricting $a$, is trivial. 

The next step to build the genuine surface operators of this theory is by considering how the flux of the $\hat m$ field transforms under gauge transformations. In this formalism, gauge transformations are given by pseudo-natural transformations \cite{Pfeiffer_2003}. Unpacking their combinatorics result in the following set of gauge transformation for the twisted integral (provided the triviality of the $A$-bundle over the surface):
\begin{equation}
\begin{split}
    \oint_\Sigma^{\alpha^\mathrm{op}} \hat m \to \alpha_{\eta_{i_0}} \oint_\Sigma^{\alpha^\mathrm{op}} \hat m \, , 
\end{split}
\end{equation}
where $i_0$ is the base point of our space. From this discussion, we can construct the genuine surface operators of this theory as follows:
\begin{equation}
    H_{[\hat\xi_i]}(\Sigma) = \prod_{\gamma\in\mathrm{Gen}[\pi_1(\Sigma)]}\delta \left(\oint_\gamma a \right) \sum_{\hat\zeta\in[\hat\xi_i]} e^{2\pi i \oint_\gamma^{\alpha^\mathrm{op}}\hat m^T \mathbf{Q}\hat\zeta}\,,
\end{equation}
where $[\hat{\xi}_i]$ denotes, as in the 3D case, the $\alpha$-orbits in the semi-direct product $M\rtimes_\alpha A$, and $\xi_i \in M$ is a representative of this orbit. $\mathrm{Gen}[\pi_1(\Sigma)]$ denotes a finite set of generators for the fundamental group of $\Sigma$. The Dirac delta is ensuring that the $A$-bundle, when restricted to $\Sigma$ is trivial.

\section{Conclusions} \label{sec:CONC}

In this paper we have constructed 3D discrete BF-like SymTFTs for 2D QFTs
with a finite non-Abelian zero-form symmetry group $G$.
In particular, we have shown that when $G$ sits in the short exact sequence:
\begin{equation}\label{eq:extendoagain}
1 \rightarrow M \rightarrow G \rightarrow A \rightarrow 1
\end{equation}
with $M$ a finite Abelian group, there is a corresponding treatment in terms of a BF-theory with fields valued in $M$, and extended by possibly on-shell fields valued in $A$. We have also shown precisely when all of the fields in $A$ can be taken off-shell, and have also explained that a presentation in terms of continuous differential forms is available when $G$ is a central extension, i.e., when $M = [G,G]$. This formulation makes manifest the non-Abelian structure of the corresponding symmetry category, and provides a streamlined way to read off many details of the corresponding fusion rules. This approach provides access to both genuine and non-genuine operators of the symmetry category, with the latter realized as surfaces (i.e., flux tubes) terminated by non-genuine line operators. Our method applies to a broad class of finite symmetries, and we have also sketched how the same technique generalizes to higher-dimensional systems. In the remainder of this section we discuss some potential avenues of future investigation.

The main focus in this work has been on a class of finite non-Abelian symmetries $G$ which are realized as a finite extension, as in line (\ref{eq:extendoagain}). It is natural to ask whether we can interchange the position of the Abelian factor. Along these lines, a particularly promising class of examples are those associated with solvable groups, namely those groups which admit a filtration by normal subgroups:
\begin{equation} \label{eq:NormalSeries}
\{ e \}\equiv G_0 \subset G_1 \subset \dots \subset G_{n-1}\subset G_n\equiv G\,,
\end{equation}
such that the factor groups $G_{i+1}/G_i$ are Abelian. It would be interesting to see whether an iterative / ``partially off-shell'' formulation of BF-like theories for this class of symmetry groups is also available.

A common feature of $D = 2$ theories is the further enrichment of the symmetry category beyond a grouplike structure, e.g., as in the case of Tambara-Yamagami categories \cite{TAMBARA1998692}.
It would be interesting to formulate a more direct Lagrangian formalism for these cases as well.

We have also seen that especially in $D > 2$ QFT$_D$'s, the additional structure provided by having a BF-like theory presentation provides insights into structures not directly accessible in the Drinfeld center. Giving a more precise identification of the relevant categorical structures would be instructive.

Along these lines, we have already seen the utility of twisted integrals in the construction of appropriate gauge invariant operators.
We suspect that this will lead to additional applications of interest in their own right.

It is also natural to ask whether similar considerations hold for more involved discrete groups. For example, the group $\mathrm{SL}(2,\mathbb{Z})$ is an amalgamated product: $\mathbb{Z}_4 \ast_{\mathbb{Z}_2} \mathbb{Z}_{6}$, so it is tempting to speculate that
the simpler cyclic building blocks appearing in this expression can be used to construct explicit SymTFTs for such situations as well.

String theory provides a wealth of examples where finite non-Abelian symmetries naturally arise from either higher-dimensional Chern-Simons couplings or metric isometries of an extra-dimensional geometry. The former case leads to pure beta-extension constructions (see Appendix \ref{app: top-down beta extension}). It would be interesting to see whether there is a ``top down'' derivation for the latter case, where SymTFTs beyond pure beta-extensions can be directly obtained via dimensional reduction of
the higher-dimensional theory.

\newpage

\section*{Acknowledgements}

We thank V. Chakrabhavi and S.N. Meynet for comments on an earlier draft. We thank V. Chakrabhavi, J. Kaidi, H.-T. Lam, J. McNamara, S.N. Meynet, K. Ohmori, B. Rayhaun,  Z. Sun, and E. Torres for helpful discussions. JJH thanks the Kavli IPMU for hospitality during part of this work.
JJH thanks the organizers and administrative staff for the KITP meeting
``Generalized Symmetries in Quantum Field Theory: High Energy Physics, Condensed Matter, and Quantum Gravity'' for rejecting his application. JJH, MH, XY and HYZ thank the Simons Collaboration on Global Categorical Symmetries Annual Meeting 2025 for the hospitality during part of this work. JJH, MH, XY and HYZ thank the Simons Physics Summer Workshop for the hospitality during part of this work. XY thanks the Kavli IPMU, CMSA and Jefferson Physical Laboratory at Harvard, and Department of Physics at UMass Amherst for their hospitality during part of this work.
The work of OB and JJH is supported by BSF grant 2022100. The work of OB is also supported in part by the Israel Science Foundation under grant No. 1254/22T. The work of JJH is also supported by DOE (HEP) Award DE-SC0013528 as well as a University Research Foundation grant at the University of Pennsylvania. MH acknowledges support from the VR Centre for Geometry and Physics (VR grant No. 2022-06593).
The work of XY is partially supported by the NSF grant PHY2310588. The work of DM has been funded partly by the Simons Foundation International (Simons Collaboration on Global Categorical Symmetries), by the VR project grant number 2023-06593 and by the European Research Council (ERC grant agreement No. 101171852). DM also thanks the Liljewalch Travel Scholarship for funding a trip to the US during the final stages of writing this paper.


\appendix


\section{Groups and Bundles}
\label{group and bundles}

In this Appendix we discuss finite non-Abelian groups which admit a presentation in terms of an extension of other finite groups. We then characterize the bundle structure induced from this algebraic extension data.

\subsection{Group Extensions}
Consider a finite non-Abelian group $G$ which fits into a short exact sequence of (possibly non-Abelian) finite groups
\begin{equation}
  1 \to  M \to G \to A \to 1 \, .
\end{equation}
As a set, $G$ can be regarded as the product $M\times A$. The group structure is then given by defining the following product \cite{brown2011nonabelian}:
\begin{equation}
    \begin{bmatrix}
        m_1 \\
        a_1
    \end{bmatrix}\begin{bmatrix}
        m_2 \\
        a_2
    \end{bmatrix} = \begin{bmatrix}
       m_1\alpha_{a_1}(m_2)\beta(a_1,a_2) \\
        a_1a_2
    \end{bmatrix} \,,
\end{equation}
where $\beta:A\times A \to M$ and $\alpha:A\to\mathrm{aut}(M)$ must satisfy the identities\footnote{Notice that $\alpha$ at this level may not be a group homomorphism.}
\begin{equation}
\begin{split}
\label{conditions: alpha and beta}
    \alpha_a(\beta(b,c))\beta(a,bc) = \beta(a,b)\beta(ab,c) \\
    \alpha_a\alpha_b = \mathrm{Ad}^{M}_{\beta(a,b)}\alpha_{ab} \, ,
\end{split}
\end{equation}
where $\mathrm{Ad}^{M}_{\beta(a,b)}$ denotes the adjoint action in $M$ by the group element $\beta(a,b)$.
Special cases of the above characterization are given by:
\begin{itemize}
    \item $\alpha$ and $\beta$ trivial: Then we have a direct product,
    \item $\alpha$ trivial and $\beta$ nontrivial: Then we have the analogue of a central extension,
    \item $\alpha$ nontrivial and $\beta$ trivial: Then we have a semidirect product.
\end{itemize}
If we specialize to the case in which $M$ is Abelian, then we clearly see that the first condition in \eqref{conditions: alpha and beta} is a cohomological constraint on $\beta$. Indeed, we can write this condition in additive notation as
\begin{equation}
\label{cohomological beta}
    \beta(a,bc) + \alpha_a(\beta(b,c)) - \beta(ab,c) - \beta(a,b) = 0 \, ,
\end{equation}
which coincide with the differential in group cohomology, meaning that $\beta$ defines an element in group cohomology with values in $M$, seen as an $A$-module via the action of $\alpha$: $\beta \in H^2(A, M_\alpha)$. On the other hand, the second condition in \eqref{conditions: alpha and beta} implies that $\alpha$ is a group homomorphism, as the adjoint action in this case is trivial. In additive notation we can rewrite the group multiplication as
\begin{equation}
\label{group operation}
    \begin{bmatrix}
        m_1 \\
        a_1
    \end{bmatrix}\begin{bmatrix}
        m_2 \\
        a_2
    \end{bmatrix} = \begin{bmatrix}
       m_1 + \alpha_{a_1}(m_2) + \beta(a_1,a_2)\\
        a_1a_2
    \end{bmatrix} \,.
\end{equation}
Let us briefly discuss some useful formulas deriving from this: First, for the inverse we have
\begin{equation}
\begin{bmatrix}
    m\\ a
\end{bmatrix}^{-1} = \begin{bmatrix}
    -\alpha_a^{-1}(m+\beta(a,a^{-1})) \\ a^{-1}
\end{bmatrix}\,,
\end{equation}
and by imposing that this is both the left and right inverse we get that
\begin{equation}
    \beta(a,a^{-1}) = \alpha_a\beta(a^{-1},a)\,.
\end{equation}
Moreover, by requiring that $(0,1)$ is the identity, we further obtain the condition that $\beta(a,1)=0$. Let now compute the adjoint action:
\begin{equation}
    \begin{bmatrix}
        \epsilon\\ \kappa
    \end{bmatrix} \begin{bmatrix}
        m \\ a
    \end{bmatrix} \begin{bmatrix}
        \epsilon \\ \kappa
    \end{bmatrix}^{-1} = \begin{bmatrix}
        \epsilon + \alpha_\kappa m + \beta(\kappa,a) - \alpha_{\kappa a\kappa^{-1}}(\epsilon +\beta(\kappa,\kappa^{-1}) ) +\beta(\kappa a, \kappa^{-1}) \\ \kappa a \kappa^{-1}
    \end{bmatrix} \,.
\end{equation}
In particular, if $a=1$, the top component simplifies to $\alpha_\kappa m$.

\subsection{Bundle Decomposition}
Let us now consider the data of a $G$ bundle over a manifold $X$, which fits into the short exact sequence of the previous section. These data consist of transition functions $g_{ij}\in G$ which have to satisfy the following cocycle condition:
\begin{equation}
    g_{ij}g_{jk}=g_{ik} \, ,
\end{equation}
and are determined up to gauge transformations $g_{ij}\sim \lambda_i g_{ij} \lambda_j^{-1}$. According to the previous section, we can decompose the bundle data, yielding
\begin{equation}
 \begin{bmatrix}
       m_{ij}\alpha_{a_{ij}}(m_{jk})\beta(a_{ij},a_{jk}) \\
        a_{ij}a_{jk}
    \end{bmatrix}= \begin{bmatrix}
        m_{ik} \\
        a_{ik}
    \end{bmatrix} \, ,
\end{equation}
where $m_{ij}$ and $a_{ij}$ are determined up to gauge transformations induced by $g_{ij}$. Consider now the case in which the module $M$ is Abelian. Then we can rewrite the cocycle condition as follows:
\begin{equation}
\begin{split}
\label{bundle decomposition}
   \delta_\alpha m_{ijk}  &= m_{ij} - m_{ik} +\alpha_{a_{ij}}m_{jk} = - \beta(a_{ij},a_{jk}) = -a^*\beta_{ijk} \, ,\\
   a_{ij}a_{jk} &= a_{ik}\,.
\end{split}
\end{equation}
Let us discuss these equations further. The latter is straightforward: the data for a $G$ bundle automatically defines also data  for an $A$ bundle, where the transition functions are simply given by the projection of $G$ into $A$. Now, we notice that $\beta: H^1(X,A)\to H^2(X,M_\alpha)$, i.e., it is a cohomological operation from the first cohomology group of $X$ with coefficient in $A$ to the second cohomology group of $X$ with coefficients in $M$, twisted by the action $\alpha$. This can be checked via a straightforward computation:
\begin{equation}
    \delta_\alpha (a^*\beta)_{ijkl} = \beta(a_{ij}, a_{jk}) - \beta(a_{ij}, a_{jl}) + \beta(a_{ik}, a_{kl}) - \alpha_{a_{ij}}(\beta(a_{jk}, a_{kl})) \,.
\end{equation}
If we use the equations: $a_{jk} a_{kl} = a_{jl}$, $a_{ij}a_{jk} = a_{ik}$ and \eqref{cohomological beta}, we obtain the following:
\begin{equation}
    \delta_\alpha (a^*\beta)_{ijkl} = \beta(a_{ij}, a_{jk}) - \beta(a_{ij}, a_{jk}a_{kl}) + \beta(a_{ij} a_{jk}, a_{kl}) - \alpha_{a_{ij}}(\beta(a_{jk}, a_{kl})) = 0 \,.
\end{equation}
Hence, $a^*\beta$ is an element in the second twisted cohomology of $X$. In particular, the first equation in \eqref{bundle decomposition} tells us that this element must be exact (i.e., trivial in cohomology).

Finally, let us discuss the gauge transformations for the fields $m$ and $a$. These are inherited by the gauge transformations of the starting background field $g\in H^1(X,G)$, which are given by $g_{ij}\to \lambda_i g_{ij} \lambda_j^{-1}$. Looking at the action of these on $M$ and $A$ we obtain the following:
\begin{equation}
\begin{split}
    a_{ij} &\to \kappa_i a_{ij}\kappa_j^{-1} \,,\\
    m_{ij} &\to \alpha_{\kappa_i}(m_{ij}) + (\epsilon_i - \alpha_{\kappa_ia_{ij}\kappa_j^{-1}}(\epsilon_j)) - \alpha_{\kappa_i a_{ij}\kappa_j^{-1}}(\beta(\kappa_j,\kappa_j^{-1}))\,.
\end{split}
\end{equation}
In particular, the differential is not gauge invariant but transforms as follows:
\begin{equation}
    \delta_\alpha m_{ijk} \to \alpha_{\kappa_i}(\delta_\alpha m_{ijk}) + \alpha_{\kappa_ia_{ij}\kappa_j^{-1}}(\beta(\kappa_j,\kappa_j^{-1})) \,.
\end{equation}
However, the following combination will transform in a covariant way:
\begin{equation}
    \delta_\alpha m_{ijk} + a^*\beta_{ijk} \to \alpha_{\kappa_i}(\delta_\alpha m_{ijk} + a^*\beta_{ijk})\,.
\end{equation}
Notice that in the case of a pure cocycle extension, the differential transforms in a covariant way.

\section{Representation Theory for Semi-Direct Products}
\label{app:RepTheory}

In this Appendix we summarize the irreducible representations of the semi-direct product $G=\Z_n\rtimes \Z_k$ with $M=\Z_n$ and $A=\Z_k$ following Serre \cite{serre1977representations}. Every such representation will be constructed as an induced representation of a subgroup $G_i=M\rtimes A_i$ of $G$, where $A_i$ is a subgroup of $A$ explained below.




In order to construct all representation of $G = M \rtimes A$, we start by examining representations of $M$ (same as characters for $M$ Abelian) and how they are acted upon by $A$-conjugation. The character group $\text{Char}(M)= \text{Hom}(M,U(1))$ are acted on by $A$ via $\chi\mapsto a\cdot\chi$ with $(a\cdot\chi)(m)=\chi(a^{-1}ma)$. Under such $A$-conjugation of characters/representations of $M$, we pick a set of representatives $\chi_i$ with $i\in (\mathrm{char}{M})^A$ labeling the coset of such $A$-actions. Then, one introduces the subgroups $\text{Fix}_A(\chi_i)\equiv A_i\subset A$ of $A$ fixed by $\chi_i$ and further introduces $G_i=M\rtimes A_i=M\rtimes \text{Fix}_A(\chi_i)$. The character $\chi_i$ of $M$ extends naturally to a character of $G_i$ by defining $\tilde\chi_i(ma)=\chi_i(m)$, i.e., projecting out the $A$ part. Similarly, any irreducible representation $\rho$ of $A_i$ extends to an irreducible representation $\tilde\rho$ of $G_i$ by the composition with projection, i.e.,  $\tilde\rho: G_i\xrightarrow{\pi_i} A_i \xrightarrow{\rho} V$, where $V$ is the representation space.

Now, Serre showed in \cite{serre1977representations} that the character $\tilde\chi_i\otimes \tilde \rho$ of $G_i$ induces an irreducible representation $\sigma_{i,\rho}$ of $G$, that two such representations $\sigma_{i,\rho},\sigma_{i',\rho'}$ are isomorphic iff $i=i'$ and $\rho\cong \rho'$, and that every irreducible representation of $G$ is isomorphic to one such induced representation.

With Serre's construction one can, for example, count the number of irreducible representations of $G = \Z_n \rtimes \Z_k$ with group presentation
\be
G=\langle m,a\,|\, m^n=1\,,~a^k=1\,, ~ama^{-1}=m^{1+q}\rangle\,.
\ee
The above discussion instructs us to first count orbits of $\Z_n$ with respect to the $\Z_k$ action which is determined by $q$. Let $[i]_q$ be such an orbit, then $A_i\cong \Z_{k/|[i]_q|}$ and there are $k/|[i]_q|$ representations $\sigma_{i,\rho}$ for fixed $i$. The overall number of representations summed over all the $i$'s is:
\be
|\text{Irreps}(\Z_n\rtimes_q \Z_k)|=\sum_{[i]_q\,\subset\,  \Z_n}\frac{k}{|[i]_q|}\,.
\ee
Further, the $\chi_i$'s are in 1:1 with all subgroups $A_i$ of $A$.

Next, we turn to discuss the representation ring of $G=\Z_n\rtimes_q \Z_k$. By the above discussion irreducible representations are uniquely labeled by orbits in $\Z_n$, denoted $[i]_q$, together with a representation $\rho$ of $A_i$. The fusion of two representations $([i]_q,\rho)\otimes([i^{\prime}]_q,\rho^{\prime})$ is computed by decomposing the set
\be
\{j+j'\in \Z_n\,|\,j\in [i]_q, \,j'\in [i']_q\}
\ee
into $\Z_k$ orbits. Orbits with stabilizer $A_{jj'}\subset A$ appear with multiplicity $|A_{jj'}|$. These degenerate orbits are uniquely distinguished by their label $\rho\in \text{Irrep}(A_{jj'})$ where each value is taken exactly once. Of course the representations $\rho$ here are 1-dimensional and can Pontryagin dually be labeled by elements of $A_{jj'}$.

As an example, consider $D_4\cong \Z_4\rtimes \Z_2$. The irreducible representations are
\be
\{[1]_{2},0\}=\{[3]_{2},0\}\,, \qquad \{[0]_{2},0\}\,,  \{[0]_{2},1\}\,, \{[2]_{2},0\}\,,  \{[2]_{2},1\}\,,
\ee
which are one 2-dimensional and four 1-dimensional representations. The fusions with the 2-dimensional representation are
\be\ba
\{[1]_{2},0\} \otimes \{[1]_{2},0\}&=  \{[0]_{2},0\}\oplus  \{[0]_{2},1\}\oplus \{[2]_{2},0\}\oplus  \{[2]_{2},1\}\,,\\
\{[1]_{2},0\}\otimes \Pi &=\{[1]_{2},0\} \,,
\ea\ee
where $\Pi$ is any of the 1-dimensional representations. The fusion ring for the 1-dimensional representations
is modeled on $\Z_2\oplus \Z_2$ with $ \{[0]_{-1},1\}$ the identity element.

\section{Drinfeld Center Fusion Table: $S_3$}
\label{App: S3 fusion}

{\renewcommand*{\arraystretch}{1.2}

In tables \ref{S3 fusion 1}, \ref{S3 fusion 2a}, \ref{S3 fusion 2b}, \ref{S3 fusion 3a}  and \ref{S3 fusion 3b} we list out the Drinfeld center fusion table for the symmetric group $S_3$ (see for example \cite{Chen:2024xkw}). Our notational conventions are as in section \ref{ssec:S3Example}.

\begin{table}[]
\centering
\[
\begin{array}{c|ccc}
\otimes
& ([e],{\bf{1}}^{S_3}) & ([e],{\bf{1}}_{\text{sign}}^{S_3}) & ([e],{\bf{2}}^{S_3}) \\ \hline

([e],{\bf{1}}^{S_3}) & ([e],{\bf{1}}^{S_3}) & ([e],{\bf{1}}_{\text{sign}}^{S_3}) & ([e],{\bf{2}}^{S_3}) \\

([e],{\bf{1}}_{\text{sign}}^{S_3}) & ([e],{\bf{1}}_{\text{sign}}^{S_3}) & ([e],{\bf{1}}^{S_3}) & ([e],{\bf{2}}^{S_3}) \\

([e],{\bf{2}}^{S_3}) & ([e],{\bf{2}}^{S_3}) & ([e],{\bf{2}}^{S_3}) &
  ([e],{\bf{1}}^{S_3})\oplus([e],{\bf{1}}_{\text{sign}}^{S_3})\oplus([e],{\bf{2}}^{S_3}) \\

([s],{\bf 1}^{\mathbb{Z}_2}) & ([s],{\bf 1}^{\mathbb{Z}_2}) & ([s],{\bf 1}_-^{\mathbb{Z}_2}) &
  ([s],{\bf 1}^{\mathbb{Z}_2})\oplus([s],{\bf 1}_-^{\mathbb{Z}_2}) \\

([s],{\bf 1}_-^{\mathbb{Z}_2}) & ([s],{\bf 1}_-^{\mathbb{Z}_2}) & ([s],{\bf 1}^{\mathbb{Z}_2}) &
  ([s],{\bf 1}_-^{\mathbb{Z}_2})\oplus([s],{\bf 1}^{\mathbb{Z}_2}) \\

([r],{\bf 1}^{\mathbb{Z}_3}) & ([r],{\bf 1}^{\mathbb{Z}_3}) & ([r],{\bf 1}^{\mathbb{Z}_3}) &
  ([r],{\bf 1}_{\omega^2}^{\mathbb{Z}_3})\oplus([r],{\bf 1}_{\omega}^{\mathbb{Z}_3}) \\

([r],{\bf 1}_{\omega}^{\mathbb{Z}_3}) & ([r],{\bf 1}_{\omega}^{\mathbb{Z}_3}) & ([r],{\bf 1}_{\omega}^{\mathbb{Z}_3}) &
  ([r],{\bf 1}^{\mathbb{Z}_3})\oplus([r],{\bf 1}_{\omega^2}^{\mathbb{Z}_3}) \\

([r],{\bf 1}_{\omega^2}^{\mathbb{Z}_3}) & ([r],{\bf 1}_{\omega^2}^{\mathbb{Z}_3}) & ([r],{\bf 1}_{\omega^2}^{\mathbb{Z}_3}) &
  ([r],{\bf 1}_{\omega}^{\mathbb{Z}_3})\oplus([r],{\bf 1}^{\mathbb{Z}_3})
\end{array}
\]
\caption{Fusions with the sector $(\{e\}, \mathrm{Rep}(S_3))$.}
\label{S3 fusion 1}
\end{table}
\begin{table}[]
\centering
\[
\begin{array}{c|c}
\otimes & ([s],{\bf 1}^{\mathbb{Z}_2}) \\ \hline

([e],{\bf{1}}^{S_3}) & ([s],{\bf 1}^{\mathbb{Z}_2}) \\

([e],{\bf{1}}_{\text{sign}}^{S_3}) & ([s],{\bf 1}_-^{\mathbb{Z}_2}) \\

([e],{\bf{2}}^{S_3}) & ([s],{\bf 1}^{\mathbb{Z}_2}) \oplus ([s],{\bf 1}_-^{\mathbb{Z}_2}) \\

([s],{\bf 1}^{\mathbb{Z}_2}) & ([e],{\bf{1}}^{S_3}) \oplus ([e],{\bf{2}}^{S_3}) \oplus ([r],{\bf 1}^{\mathbb{Z}_3}) \oplus ([r],{\bf 1}_{\omega}^{\mathbb{Z}_3}) \oplus ([r],{\bf 1}_{\omega^2}^{\mathbb{Z}_3}) \\

([s],{\bf 1}_-^{\mathbb{Z}_2}) & ([e],{\bf{1}}_{\text{sign}}^{S_3}) \oplus ([e],{\bf{2}}^{S_3}) \oplus ([r],{\bf 1}^{\mathbb{Z}_3}) \oplus ([r],{\bf 1}_{\omega}^{\mathbb{Z}_3}) \oplus ([r],{\bf 1}_{\omega^2}^{\mathbb{Z}_3}) \\

([r],{\bf 1}^{\mathbb{Z}_3}) & ([s],{\bf 1}^{\mathbb{Z}_2}) \oplus ([s],{\bf 1}_-^{\mathbb{Z}_2}) \\

([r],{\bf 1}_{\omega}^{\mathbb{Z}_3}) & ([s],{\bf 1}^{\mathbb{Z}_2}) \oplus ([s],{\bf 1}_-^{\mathbb{Z}_2}) \\

([r],{\bf 1}_{\omega^2}^{\mathbb{Z}_3}) & ([s],{\bf 1}^{\mathbb{Z}_2}) \oplus ([s],{\bf 1}_-^{\mathbb{Z}_2})
\end{array}
\]
\caption{Fusions with $([s],{\bf 1}^{\mathbb{Z}_2}))$.}
\label{S3 fusion 2a}
\end{table}

\begin{table}[]
\centering
\[
\begin{array}{c|c}
\otimes & ([s],{\bf 1}_-^{\mathbb{Z}_2}) \\ \hline

([e],{\bf{1}}^{S_3}) & ([s],{\bf 1}_-^{\mathbb{Z}_2}) \\

([e],{\bf{1}}_{\text{sign}}^{S_3}) & ([s],{\bf 1}^{\mathbb{Z}_2}) \\

([e],{\bf{2}}^{S_3}) & ([s],{\bf 1}_-^{\mathbb{Z}_2}) \oplus ([s],{\bf 1}^{\mathbb{Z}_2}) \\

([s],{\bf 1}^{\mathbb{Z}_2}) & ([e],{\bf{1}}_{\text{sign}}^{S_3}) \oplus ([e],{\bf{2}}^{S_3}) \oplus ([r],{\bf 1}^{\mathbb{Z}_3}) \oplus ([r],{\bf 1}_{\omega}^{\mathbb{Z}_3}) \oplus ([r],{\bf 1}_{\omega^2}^{\mathbb{Z}_3}) \\

([s],{\bf 1}_-^{\mathbb{Z}_2}) & ([e],{\bf{1}}^{S_3}) \oplus ([e],{\bf{2}}^{S_3}) \oplus ([r],{\bf 1}^{\mathbb{Z}_3}) \oplus ([r],{\bf 1}_{\omega}^{\mathbb{Z}_3}) \oplus ([r],{\bf 1}_{\omega^2}^{\mathbb{Z}_3}) \\

([r],{\bf 1}^{\mathbb{Z}_3}) & ([s],{\bf 1}^{\mathbb{Z}_2}) \oplus ([s],{\bf 1}_-^{\mathbb{Z}_2}) \\

([r],{\bf 1}_{\omega}^{\mathbb{Z}_3}) & ([s],{\bf 1}^{\mathbb{Z}_2}) \oplus ([s],{\bf 1}_-^{\mathbb{Z}_2}) \\

([r],{\bf 1}_{\omega^2}^{\mathbb{Z}_3}) & ([s],{\bf 1}^{\mathbb{Z}_2}) \oplus ([s],{\bf 1}_-^{\mathbb{Z}_2})
\end{array}
\]
\caption{Fusions with $([s],{\bf 1}_-^{\mathbb{Z}_2})$.}
\label{S3 fusion 2b}
\end{table}

\begin{table}[]
\centering
\[
\begin{array}{c|c}
\otimes & ([r],{\bf 1}^{\mathbb{Z}_3}) \\ \hline

([e],{\bf{1}}^{S_3}) & ([r],{\bf 1}^{\mathbb{Z}_3}) \\

([e],{\bf{1}}_{\text{sign}}^{S_3}) & ([r],{\bf 1}^{\mathbb{Z}_3}) \\

([e],{\bf{2}}^{S_3}) & ([r],{\bf 1}_{\omega^2}^{\mathbb{Z}_3}) \oplus ([r],{\bf 1}_{\omega}^{\mathbb{Z}_3}) \\

([s],{\bf 1}^{\mathbb{Z}_2}) & ([s],{\bf 1}^{\mathbb{Z}_2}) \oplus ([s],{\bf 1}_-^{\mathbb{Z}_2}) \\

([s],{\bf 1}_-^{\mathbb{Z}_2}) & ([s],{\bf 1}^{\mathbb{Z}_2}) \oplus ([s],{\bf 1}_-^{\mathbb{Z}_2}) \\

([r],{\bf 1}^{\mathbb{Z}_3}) & ([e],{\bf{1}}^{S_3}) \oplus ([e],{\bf{1}}_{\text{sign}}^{S_3}) \oplus ([r],{\bf 1}^{\mathbb{Z}_3}) \\

([r],{\bf 1}_{\omega}^{\mathbb{Z}_3}) & ([e],{\bf{2}}^{S_3}) \oplus ([r],{\bf 1}_{\omega^2}^{\mathbb{Z}_3}) \\

([r],{\bf 1}_{\omega^2}^{\mathbb{Z}_3}) & ([e],{\bf{2}}^{S_3}) \oplus ([r],{\bf 1}_{\omega}^{\mathbb{Z}_3})
\end{array}
\]
\caption{Fusions with $([r],{\bf 1}^{\mathbb{Z}_3})$.}
\label{S3 fusion 3a}
\end{table}

\begin{table}[]
\centering
\[
\begin{array}{c|cc}
\otimes & ([r],{\bf 1}_{\omega}^{\mathbb{Z}_3}) & ([r],{\bf 1}_{\omega^2}^{\mathbb{Z}_3}) \\ \hline

([e],{\bf{1}}^{S_3}) & ([r],{\bf 1}_{\omega}^{\mathbb{Z}_3}) & ([r],{\bf 1}_{\omega^2}^{\mathbb{Z}_3}) \\

([e],{\bf{1}}_{\text{sign}}^{S_3}) & ([r],{\bf 1}_{\omega}^{\mathbb{Z}_3}) & ([r],{\bf 1}_{\omega^2}^{\mathbb{Z}_3}) \\

([e],{\bf{2}}^{S_3}) & ([r],{\bf 1}^{\mathbb{Z}_3}) \oplus ([r],{\bf 1}_{\omega^2}^{\mathbb{Z}_3}) &
([r],{\bf 1}_{\omega}^{\mathbb{Z}_3}) \oplus ([r],{\bf 1}^{\mathbb{Z}_3}) \\

([s],{\bf 1}^{\mathbb{Z}_2}) & ([s],{\bf 1}^{\mathbb{Z}_2}) \oplus ([s],{\bf 1}_-^{\mathbb{Z}_2}) &
([s],{\bf 1}^{\mathbb{Z}_2}) \oplus ([s],{\bf 1}_-^{\mathbb{Z}_2}) \\

([s],{\bf 1}_-^{\mathbb{Z}_2}) & ([s],{\bf 1}^{\mathbb{Z}_2}) \oplus ([s],{\bf 1}_-^{\mathbb{Z}_2}) &
([s],{\bf 1}^{\mathbb{Z}_2}) \oplus ([s],{\bf 1}_-^{\mathbb{Z}_2}) \\

([r],{\bf 1}^{\mathbb{Z}_3}) & ([e],{\bf{2}}^{S_3}) \oplus ([r],{\bf 1}_{\omega^2}^{\mathbb{Z}_3}) &
([e],{\bf{2}}^{S_3}) \oplus ([r],{\bf 1}_{\omega}^{\mathbb{Z}_3}) \\

([r],{\bf 1}_{\omega}^{\mathbb{Z}_3}) & ([e],{\bf{1}}^{S_3}) \oplus ([e],{\bf{1}}_{\text{sign}}^{S_3}) \oplus ([r],{\bf 1}_{\omega}^{\mathbb{Z}_3}) &
([e],{\bf{2}}^{S_3}) \oplus ([r],{\bf 1}^{\mathbb{Z}_3}) \\

([r],{\bf 1}_{\omega^2}^{\mathbb{Z}_3}) & ([e],{\bf{2}}^{S_3}) \oplus ([r],{\bf 1}^{\mathbb{Z}_3}) &
([e],{\bf{1}}^{S_3}) \oplus ([e],{\bf{1}}_{\text{sign}}^{S_3}) \oplus ([r],{\bf 1}_{\omega^2}^{\mathbb{Z}_3})
\end{array}
\]
\caption{Fusions with $([r],{\bf 1}_{\omega}^{\mathbb{Z}_3}))$ and $([r],{\bf 1}_{\omega^2}^{\mathbb{Z}_3})$.}
\label{S3 fusion 3b}
\end{table}
}

\clearpage

\newpage

\section{Some Topological Boundary Conditions}
\label{app:BC}

In this Appendix we discuss choices of topological boundary conditions for the SymTFTs of some finite symmetry groups.
At the level of categorical data, this can be helpful in determining which topological operators are genuine, and which are non-genuine.

We primarily focus on a few representative examples where the group can be presented as:
\begin{equation}\label{eq:extendoagainagain}
1 \rightarrow M \rightarrow G \rightarrow A \rightarrow 1,
\end{equation}
with $M$ Abelian. In particular, we focus on the groups $S_3$, $D_4$, $A_4$ and $\Delta_{3 N^2}$ which are realized as the group extensions:
\begin{align}
1 & \rightarrow \mathbb{Z}_3 \rightarrow S_3 \rightarrow \mathbb{Z}_2 \rightarrow 1 \\
1 & \rightarrow \mathbb{Z}_{4} \rightarrow D_4 \rightarrow \mathbb{Z}_2 \rightarrow 1 \\
1 & \rightarrow \mathbb{Z}_{2} \times \mathbb{Z}_2 \rightarrow A_4 \rightarrow \mathbb{Z}_3 \rightarrow 1 \\
1 & \rightarrow \mathbb{Z}_{N} \times \mathbb{Z}_N \rightarrow \Delta_{3 N^2} \rightarrow \mathbb{Z}_3 \rightarrow 1,
\end{align}
We remark that for $D_4$ and $\Delta_{3N^2}$ with $N$ divisible by $3$ (i.e., $3 \vert N$) there is an alternative presentation in terms of a central extension:
\begin{align}
1 & \rightarrow \mathbb{Z}_{2} \rightarrow D_4 \rightarrow \mathbb{Z}_2 \times \mathbb{Z}_2 \rightarrow 1 \label{eq:D4central}\\
1 & \rightarrow \mathbb{Z}_3 \rightarrow \Delta_{3 N^2} \rightarrow \mathbb{Z}_{N} \times \mathbb{Z}_N \rightarrow 1 \label{eq:DELTAcentral}.
\end{align}

In all of these cases, the group $M$ of line appearing in line (\ref{eq:extendoagainagain}) is Abelian
so we can apply our general methodology to construct a corresponding SymTFT.
Moreover, in the presentation of lines (\ref{eq:D4central}) and (\ref{eq:DELTAcentral}) we have a central extension and so a formulation in terms of continuous differentials in a BF-like theory is also available, as per our discussion in section \ref{sec:DiffApproach}.

Identifying the Lagrangian algebras of the Drinfeld center $\mathcal{Z}(\mathcal{C})$ for this category then tells us the collection of genuine topological operators which can be fully detached from the boundary. The relevant Lagrangian algebras are summarized in tables \ref{tab:lagrangian algebras of s3}, \ref{tab:D4_lagrangian_algebra}, \ref{tab:L_Z_A4}, and \ref{tab:Delta_27_lagrangian_algebra} for the respective cases $S_3$, $D_4$, $A_4$, and $\Delta_{27}$, i.e. $\Delta_{3N^2}$ with $N = 3$. See, e.g., \cite{DAVYDOV2017149} for a general treatment, as well as explicit tables for $S_3$ and $D_4$.

\begin{table}[]
\centering
\renewcommand{\arraystretch}{1.25}
\begin{tabular}{|c|c|}
\hline
Gauging in $\mathrm{Vec}_{S_3}$ & Lagrangian Algebra Object of $\mathcal{Z}(\mathrm{Vec}_{S_3})$  \\
\hline
$e$ &
$([e],\mathbf{1}^{S_3})+([e],\mathbf{1}_{\mathrm{sign}}^{S_3})+2([e],\mathbf{2}^{S_3})$
\\ \hline
$\Z_2$ &
$([e],\mathbf{1}^{S_3})+([e],\mathbf{2}^{S_3})+([s],\mathbf{1}^{\Z_2})$
\\ \hline
$\Z_3$ &
$([e],\mathbf{1}^{S_3})+([e],\mathbf{1}_{\mathrm{sign}}^{S_3})+2([r],\mathbf{1}^{\Z_3})$
\\ \hline
$S_3$ &
$([e],\mathbf{1}^{S_3})+([s],\mathbf{1}^{\Z_2})+([r],\mathbf{1}^{\Z_3})$
\\ \hline
\end{tabular}
\caption{Lagrangian algebras of $\mathcal{Z}(\mathrm{Vec}_{S_3})$ and their corresponding gaugings in $S_3$.}
\label{tab:lagrangian algebras of s3}
\end{table}

\begin{table}[]
    \centering
    \begin{tabular}{|c|c|} \hline
        Gauging in $\mathrm{Vec}_{D_4}$ & Lagrangian Algebra Object of $\mathcal{Z}(D_4)$ \\ \hline
        $(e, 1)$ & $\sum_{\rho \in \{\mathbf{1}^{D_4}, \mathbf{1}^{D_4}_x, \mathbf{1}^{D_4}_y, \mathbf{1}^{D_4}_z, \mathbf{2}^{D_4}\}} \mathrm{dim}(\rho)([e], \rho)$ \\ \hline
        $(\langle r^2 \rangle, 1)$ & $\sum_{\rho \in \{\mathbf{1}^{D_4}, \mathbf{1}^{D_4}_x, \mathbf{1}^{D_4}_y, \mathbf{1}^{D_4}_z\}} (([e], \rho) + ([r^2], \rho))$ \\ \hline
        $(\langle r \rangle, 1)$ & $\sum_{\rho \in \{\mathbf{1}^{D_4}, \mathbf{1}^{D_4}_x\}} (([e], \rho) +  ([r^2], \rho)) + 2([r], i)$ \\ \hline
        $(\langle r^2, s \rangle, 1)$ & $\sum_{\rho \in \{\mathbf{1}^{D_4}, \mathbf{1}^{D_4}_y\}} (([e], \rho) +([r^2], \rho)) + 2([s], \mathbf{1}^{\Z_2^2})$ \\ \hline
        $(\langle r^2, s \rangle, \alpha)$ & $([e], \mathbf{1}^{D_4}) + ([e], \mathbf{1}^{D_4}_y) + ([r^2], \mathbf{1}^{D_4}_x) + ([r^2], \mathbf{1}^{D_4}_z) + 2([s], \mathbf{1}^{\Z_2^2}_b)$ \\ \hline
        $(\langle r^2, rs \rangle, 1)$ & $\sum_{\rho \in \{\mathbf{1}^{D_4}, \mathbf{1}^{D_4}_z\}} (([e], \rho) + ([r^2], \rho)) + 2([rs], \mathbf{1})$ \\ \hline
        $(\langle r^2, rs \rangle, \beta)$ & $([e], \mathbf{1}^{D_4}) + ([e], \mathbf{1}_z^{D_4}) + ([r^2], \mathbf{1}^{D_4}_x) + ([r^2], \mathbf{1}^{D_4}_y) + 2([rs], \mathbf{1}^{\Z_2^2}_{ab})$ \\ \hline
        $(D_4, 1)$ & $([e], \mathbf{1}^{D_4}) + ([r^2], \mathbf{1}^{D_4}) + ([r], \mathbf{1}^{\Z_4}) + ([s], \mathbf{1}^{\Z_2^2}) + ([rs], \mathbf{1}^{\Z_2^2})$ \\ \hline
        $(D_4, \gamma)$ & $([e], \mathbf{1}^{D_4}) + ([r^2], \mathbf{1}^{D_4}_x) + ([r], \mathbf{1}^{\Z_4}) + ([s], \mathbf{1}^{\Z_2^2}_b) + ([rs], \mathbf{1}^{\Z_2^2}_{ab})$ \\ \hline
        $(\langle s \rangle, 1)$ & $\sum_{r \in \mathbf{1}^{D_4}, \mathbf{1}^{D_4}_y, \mathbf{2}^{D_4}}([e], r) + \sum_{s \in \mathbf{1}^{\Z_2^2}, \mathbf{1}^{\Z_2^2}_b} ([s], s)$\\ \hline
        $(\langle rs \rangle, 1)$ & $\sum_{r \in \mathbf{1}^{D_4}, \mathbf{1}^{D_4}_z, \mathbf{2}^{D_4}}([e], r) + \sum_{s \in \mathbf{1}^{\Z_2^2}, \mathbf{1}^{\Z_2^2}_{ab}} ([rs], s)$ \\ \hline
    \end{tabular}
    \caption{Lagrangian Algebras in $\mathcal{Z}(D_4)$. Here $\alpha, \beta, \gamma \in H^2(H, U(1))$ are the non-trivial generators of the discrete torsion, where $H$ is either $D_4$ or $\Z_2^2$.}
    \label{tab:D4_lagrangian_algebra}
\end{table}

\begin{table}[]
    \centering
    \begin{tabular}{|c|c|} \hline
        Gauging in $\mathrm{Vec}^{A_4}$ & Lagrangian Algebra Object of $\mathcal{Z}(A_4)$ \\ \hline
        $(e, 1)$ & $\sum_{\rho \in \{\mathbf{1}^{A_4}, \mathbf{1}^{A_4}_a, \mathbf{1}^{A_4}_{a^2}, \mathbf{3}^{A_4}\}} \mathrm{dim}(\rho)([e], \rho)$ \\ \hline
        $(\langle u \rangle, 1)$ & $\sum_{\rho \in \{\mathbf{1}^{A_4}, \mathbf{1}^{A_4}_a, \mathbf{1}^{A_4}_{a^2}, \mathbf{3}^{A_4}\}} ([e], \rho) + \sum_{\rho \in \{\mathbf{1}^{\Z_2^2}, \mathrm{1}^{\Z_2^2}_v\}} ([u], \rho)$ \\ \hline
        $(\langle t \rangle, 1)$ & $([e], \mathbf{1}^{A_4}) + ([e], \mathbf{3}^{A_4}) + ([t], \mathbf{1}^{\Z_3}) + ([t^2], \mathbf{1}^{\Z_3})$ \\ \hline
        $(\mathbb{Z}_2^2 = \langle t^2ut, u \rangle, 1)$ & $\sum_{\rho \in \{\mathbf{1}^{A_4}, \mathbf{1}^{A_4}_a, \mathbf{1}^{A_4}_{a^2}\}} ([e], \rho) + 3([u], \mathbf{1}^{\Z_3})$ \\ \hline
        $(\mathbb{Z}_2^2 = \langle t^2ut, u \rangle, \alpha)$ & $\sum_{\rho \in \{\mathbf{1}^{A_4}, \mathbf{1}^{A_4}_a, \mathbf{1}^{A_4}_{a^2}\}} ([e], \rho) + 3([u], \mathbf{1}^{\Z_3}_v)$ \\ \hline
        $(A_4, 1)$ & $([e], \mathbf{1}^{A_4}) + ([t^2], \mathbf{1}^{\Z_3}) + ([t], \mathbf{1}^{\Z_3}) + ([u], \mathbf{1}^{\Z_2^2})$ \\ \hline
        $(A_4, \beta)$ & $([e], \mathbf{1}^{A_4}) + ([t^2], \mathbf{1}^{\Z_3}_x) + ([t], \mathbf{1}^{\Z_3}) + ([u], \mathbf{1}^{\Z_2^2}_v)$ \\ \hline
    \end{tabular}
    \caption{Lagrangian algebras of $\mathcal{Z}(A_4)$.}
    \label{tab:L_Z_A4}
\end{table}

\begin{table}[]
    \centering
    \begin{tabular}{|c|c|} \hline
        Gauging in $\mathrm{Vec}_{\Delta_{27}}$ & Lagrangian Algebra Object of $\mathcal{Z}(\Delta_{27})$ \\ \hline
        $(e, 1)$ & $\sum_{\rho \in \mathrm{Rep}(\Delta_{27})}([e], \rho)$ \\ \hline
        $(\langle a \rangle, 1)$ & $\sum_{\rho \in \{{\bf 1}^{\Delta_{27}}, {\bf 1}^{\Delta_{27}}_y, {\bf 1}^{\Delta_{27}}_{y^2}, {\bf 3}^{\Delta_{27}}_p, {\bf 3}^{\Delta_{27}}_q\}}([e], \rho) + \sum_{\ell = \pm 1} \sum_{k = 0, 1, 2} ([a^\ell], {\bf 1}^{\Z_3^2}_{z^k})$ \\ \hline
        $(\langle b \rangle, 1)$ & $\sum_{\rho \in \{{\bf 1}^{\Delta_{27}}, {\bf 1}^{\Delta_{27}}_x, {\bf 1}^{\Delta_{27}}_{x^2}, {\bf 3}^{\Delta_{27}}_p, {\bf 3}^{\Delta_{27}}_q\}}([e], \rho) + \sum_{\ell = \pm 1} \sum_{k = 0, 1, 2} ([b^\ell], {\bf 1}^{\Z_3^2}_{z^k})$ \\ \hline
        $(\langle ab \rangle, 1)$ & $\sum_{\rho \in \{{\bf 1}^{\Delta_{27}}, {\bf 1}^{\Delta_{27}}_{x^2 y}, {\bf 1}^{\Delta_{27}}_{x y^2}, {\bf 3}^{\Delta_{27}}_p, {\bf 3}^{\Delta_{27}}_q\}}([e], \rho) + + \sum_{\ell = \pm 1} \sum_{k = 0, 1, 2} ([(ab)^\ell], {\bf 1}^{\Z_3^2}_{z^k})$ \\ \hline
        $(\langle a^{-1}b \rangle, 1)$ & $\sum_{\rho \in \{{\bf 1}^{\Delta_{27}}, {\bf 1}^{\Delta_{27}}_{x y}, {\bf 1}^{\Delta_{27}}_{x^2 y^2}, {\bf 3}^{\Delta_{27}}_p, {\bf 3}^{\Delta_{27}}_q\}}([e], \rho) + + \sum_{\ell = \pm 1} \sum_{k = 0, 1, 2} ([(ab^{-1})^\ell], {\bf 1}^{\Z_3^2}_{z^k})$ \\ \hline
        $(\langle c \rangle, 1)$ & $\sum_{j = 0, 1, 2} \sum_{k = 0, 1, 2} \sum_{\ell = 0, 1, 2} ([c^j], {\bf 1}^{\Delta_{27}}_{x^k y^\ell})$ \\ \hline
        $(\langle a, c \rangle, \alpha^s)$ & $\sum_{j = 0, 1, 2} \sum_{k = 0, 1, 2} ([c^j], {\bf 1}^{\Delta_{27}}_{x^{-sj} y^k}) + \sum_{\ell = \pm 1} 3 ([a^\ell], {\bf 1}^{\Z_3^2}_{z^{s\ell} })$ \\ \hline
        $(\langle b, c \rangle, \alpha^s)$ & $\sum_{j = 0, 1, 2} \sum_{k = 0, 1, 2} ([c^j], {\bf 1}^{\Delta_{27}}_{x^{k} y^{-sj}}) + \sum_{\ell = \pm 1} 3 ([b^\ell], {\bf 1}^{\Z_3^2}_{z^{s\ell}})$ \\ \hline
        $(\langle ab, c \rangle, \alpha^s)$ & $\sum_{j = 0, 1, 2} \sum_{k = 0, 1, 2} ([c^j], {\bf 1}^{\Delta_{27}}_{y^{-sj} (xy^{-1})^k}) + \sum_{\ell = \pm 1} 3 ([(ab)^\ell], {\bf 1}^{\Z_3^2}_{z^{s\ell} })$ \\ \hline
        $(\langle ab^{-1}, c \rangle, \alpha^s)$ & $\sum_{j = 0, 1, 2} \sum_{k = 0, 1, 2} ([c^j], {\bf 1}^{\Delta_{27}}_{x^{sj} (xy)^k}) + \sum_{\ell = \pm 1} 3 ([(ab^{-1})^\ell], {\bf 1}^{\Z_3^2}_{z^{s\ell}})$ \\ \hline
        $(\langle a, b, c \rangle, \alpha^s\beta^t)$ & $\sum_{k \in 0, 1, 2} ([c^k], {\bf 1}_{x^{sk} y^{tk}}) + \sum_{j, \ell = 0, 1, 2; (j, \ell) \neq (0, 0)} ([a^j b^\ell], {\bf 1}_{z^{-sj-t\ell}})$ \\ \hline
    \end{tabular}
    \caption{Lagrangian Algebras in $\mathcal{Z}(\Delta_{27})$ for all $A(H, \sigma)$. Here when $|H| = 9$, $\sigma = \alpha^s$ is such that $\alpha$ generates $H^2(\Z_3 \times \Z_3, U(1)) = \Z_3$ and $s = 0, \pm 1$. We define $\Delta_{27} = \{a^3 = b^3 = c^3 = 1, ac = ca, bc = cb, aba^{-1}b^{-1} = c\}$. The representations of $\Delta_{27}$ are denoted by ${\bf 1}_{x^k y^l}$ with $k, l = 0, 1, 2$ and $3_p, 3_q$, all with superscript $\Delta_{27}$. The convention is fixed by $\chi_{{\bf 1}_x}(a) = \chi_{{\bf 1}_y}(b) = e^{2\pi i/3}$ and $\chi_{{\bf 1}_x}(b) = \chi_{{\bf 1}_y}(a) = 1$. On the other hand, for any element $d$ not in the center of $\Delta_{27}$, its centralizer $\Z_3 \times \Z_3$ is always given by $\langle c, d\rangle$ ($d = a, b, ab, ac$, etc.). We denote the nine one-dimensional representations of $\Z_3 \times \Z_3$ as ${\bf 1}^{\Z_3^2}_{z^k w^l}$ fixed by $\chi_{{\bf 1}_z}(c) = \chi_{{\bf 1}_w}(d) = e^{2\pi i/3}$ and $\chi_{{\bf 1}_z}(d) = \chi_{{\bf 1}_w}(c) = 1$. In the end, we have $H^2(\Delta_{27}, U(1)) = \Z_3 \times \Z_3$, which can be computed by the Lyndon-Hochschild-Serre spectral sequence associated with the central extension of $1 \rightarrow \Z_3 \rightarrow \Delta_{27} \rightarrow \Z_3^2 \rightarrow 1$. We express the element of $H^{2}(\Delta_{27}, U(1))$ as $\alpha^s \beta^t, s, t = 0, 1, 2$, where $\alpha = \exp^{2\pi i(xz'-x'z)/3}, \beta =  \exp^{2\pi i(yz'-y'z)/3}$ and $(x, y, z)$ is the matrix entries in the $\Z_3$-valued upper triangular matrix presentation of $\Delta_{27}$ so that $x, y$ are on the first off-diagonal and $z$ is on the second off-diagonal.}
    \label{tab:Delta_27_lagrangian_algebra}
\end{table}

\clearpage

\newpage

\section{Non-Genuine Operators in $D_4$ SymTFT}
\label{app:d4 non-genuine operators}
In this Appendix we summarize the non-genuine operators in the SymTFT associated to group elements of $D_4$. We set the notation for $D_4$ as
\begin{equation}\label{eq:d4 group}
    \left\langle x,y|x^4=y^2=1, xy=yx^3\right\rangle.
\end{equation}

For the ease of reading, we reproduce the $D_4$ SymTFT with all fields off-shell
\be\ba
\mathcal{ L}_{D_4}^{(5)} &= \frac{2\pi i}{2} \lb \hat a\cup \delta  a+\hat b\cup \delta  b+   \hat c\cup \delta c+ a\cup b \cup c   \rb\,,
\ea\ee
This Lagrangian and its genuine line operators have been investigated in the literature, e.g., \cite{He:2016xpi, Kaidi:2023maf, Yu:2023nyn}. The set of genuine line operators, associated to objects in the Drinfeld center, is the same as that of $Q_8$ (\ref{eq:Lagrangian})

We are instead interested in surface-attaching non-genuine line operators. Following the discussion in section \ref{sec:Pure2Cocycle}, the equations of motion
\begin{equation}
\begin{split}
    &\delta \hat{b} + a\cup c=0,\\
    &\delta \hat{c} + a \cup b=0
\end{split}
\end{equation}
tell us there exist the following two operators
\begin{equation}
\begin{split}
	\mathcal{U}_{\hat{b}}(M_2)=\exp \left( \pi i \int_{\partial M_2}\hat{b} \right) \exp \left( \pi i \int_{M_2} a c   \right),\\
	\mathcal{U}_{\hat{c}}(M_2)=\exp \left( \pi i \int_{\partial M_2}\hat{c} \right) \exp \left( \pi i \int_{M_2} a b  \right).\\
\end{split}
\end{equation}
The commutation relation of these two operators can be derived similarly to (\ref{eq:commutator of q8}) as follows:
\begin{equation}
\begin{split}
    &\exp \left( \pi i \int ab \right)\exp \left(\pi i \int \hat{b} \right)\exp\left(\pi i \int ab \right)\\
    =&\exp \left( \pi i\int ab+\pi i\int \hat{b}+\frac{1}{2}(\pi i)^2\left[\int ab, \int \hat{b}\right] \right)\exp \left({\pi i\int ab}\right)\\
    =&\exp \left(\pi i \int ab +\pi i \int \hat{b}+\frac{\pi i}{2}\int a \right)\exp \left({\pi i\int ab}\right)\\
    =&\exp\left( \pi i\int \hat{b}+ \frac{\pi i}{2}\int a+\frac{1}{2}(\pi i)^2\left[ \int \hat{b}, \int ab \right] \right)\\
    =&\exp \left( \pi i \int \hat{b} \right) \exp\left( \pi i \int a \right),
\end{split}
\end{equation}
where in the second and the fourth line, we have used commutators of the canonical quantization for the SymTFT
\begin{equation}\label{eq:d4 phase space commutators}
	[\langle a, \lambda \rangle, \langle \hat{a}, \gamma \rangle]=[\langle b, \lambda \rangle, \langle \hat{b}, \gamma \rangle]=[\langle c, \lambda \rangle, \langle \hat{c}, \gamma \rangle]=\frac{i}{\pi}\langle \lambda, \gamma \rangle.
\end{equation}

From the above computation, we see that commuting $\mathcal{U}_{\hat{b}}$ and $\mathcal{U}_{\hat{c}}$ produces invertible genuine line operator $W_a\equiv \exp \left( \pi i \int a  \right)$:
\begin{equation}
	\mathcal{U}_{\hat{b}}\mathcal{U}_{\hat{c}}=\mathcal{U}_{\hat{c}} \mathcal{U}_{\hat{b}} W_{a}.
\end{equation}
Furthermore, the $\pi i $ coefficient in $\mathcal{U}_{\hat{b}}$ and $\mathcal{U}_{\hat{c}}$ simply implies
\begin{equation}
   \left( \mathcal{U}_{\hat{b}} \right)^2=\left( \mathcal{U}_{\hat{c}} \right)^2 = 1.
\end{equation}
We can then summarize the dictionary between the $D_4$ group elements and (non-genuine) operators in the SymTFT in table \ref{tab:d4 elements and non-genuine operators}.
\begin{table}[]
\centering
\renewcommand{\arraystretch}{1.2}
\begin{tabular}{|c|c|}
\hline
\textbf{$D_4$ element} & \textbf{SymTFT operator} \\
\hline
$1$      & $1$ \\

$x$     & $\mathcal{U}_{\hat{b}}\mathcal{U}_{\hat{c}}$ \\

$x^2$      & $W_a$ \\
$x^3$     & $\mathcal{U}_{\hat{b}}\mathcal{U}_{\hat{c}}W_a$ \\

$y$      & $\mathcal U_{\hat b}$ \\
$yx$     & $\mathcal U_{\hat c}$ \\

$yx^2$      & $\mathcal U_{\hat b} W_a$ \\
$yx^3$     & $\mathcal U_{\hat c}W_a$ \\
\hline
\end{tabular}
\caption{Identification of $D_4$ group elements with SymTFT operators.}
\label{tab:d4 elements and non-genuine operators}
\end{table}

\section{Top-Down Approach to Pure 2-Cocycle Extensions}
\label{app: top-down beta extension}
In this Appendix we illustrate that SymTFTs for non-Abelian group symmetries, constructed as pure beta-extensions, arise very naturally in string theory. Instead of exhausting all possible realizations, we show that there exists an infinite class of examples in which the SymTFT contains cubic couplings that generate non-Abelian group extensions in a systematic way.

We consider four-dimensional quantum field theories realized on stacks of D3-branes
probing a singular Calabi--Yau threefold of the form
\begin{equation}
X \;=\; \mathrm{Cone}(\partial X),
\end{equation}
where $\partial X$ is a compact five-dimensional manifold, commonly referred to as
the \emph{link} of the singularity. The topology of $\partial X$ encodes the discrete
data governing global symmetries of the D3-brane worldvolume theory.

For a Calabi--Yau cone, $\partial X$ is a Sasaki--Einstein manifold (or a mild
generalization thereof), and its cohomology takes the schematic form
\begin{equation}
\label{eq:link-cohomology}
\begin{aligned}
H^0(\partial X,\mathbb Z) &\cong \mathbb Z,\\
H^1(\partial X,\mathbb Z) &\cong 0,\\
H^2(\partial X,\mathbb Z) &\cong \mathbb Z^{b_2}\oplus \mathrm{Tor}\,H^2(\partial X,\mathbb Z),\\
H^3(\partial X,\mathbb Z) &\cong \mathbb Z^{b_3},\\
H^4(\partial X,\mathbb Z) &\cong \mathrm{Tor}\,H^4(\partial X,\mathbb Z),\\
H^5(\partial X,\mathbb Z) &\cong \mathbb Z.
\end{aligned}
\end{equation}
Here the free parts, characterized by the Betti numbers $b_2$ and $b_3$, are associated
with continuous background fields and marginal couplings, while the torsion subgroups
$\mathrm{Tor}\,H^2(\partial X)$ and $\mathrm{Tor}\,H^4(\partial X)$ give rise to
\emph{discrete} background gauge fields.

From the perspective of type IIB string theory, the cohomology of $\partial X$
controls the possible topological backgrounds for the bulk $p$-form gauge fields
reduced on the internal space. In particular, torsion cohomology classes leading to discrete degrees of freedom in the SymTFT, capturing finite symmetries associated with the 4D theory.

Since our goal in this Appendix is to
exhibit the mechanism by which non-Abelian finite global symmetries arise from
string theory, we will work in a
minimal torsion subsector $\mathrm{Tor}\,H^2(\partial X)=\mathrm{Tor}\,H^4(\partial X)=\mathbb{Z}_N$. Concretely, we restrict attention to a set of discrete
background fields valued in a common cyclic group $\mathbb Z_N$, which suffices to
illustrate the universal features of the resulting SymTFT.

Following the discussion in, e.g.\cite{Apruzzi:2021nmk,Yu:2023nyn,Baume:2023kkf,GarciaEtxebarria:2024fuk,Franco:2024mxa}, the SymTFTs come from the dimensional reduction of both kinetic terms and CS couplings in type IIB string theory. In particular, the ten-dimensional
topological action contains the term
\begin{equation}
S_{\text{CS}}^{\text{IIB}}
=
\frac{2\pi i}{2}
\int_{M_{10}}
C_4 \wedge H_3 \wedge F_3 ,
\label{eq:IIB-CS-explicit}
\end{equation}
where $C_4$ is the Ramond--Ramond four-form potential, and
$H_3=dB_2$, $F_3=dC_2$ are the NS-NS and R-R three-form field strengths.\footnote{
We suppress numerical factors and subtleties associated with the self-duality of
$F_5$, which are not essential for the present discussion.
}A precise
treatment of this reduction is naturally formulated in terms of differential
cohomology. Here we will only record the resulting discrete data and refer the reader to detailed reduction computation to \cite{GarciaEtxebarria:2024fuk,Yu:2023nyn,Franco:2024mxa}.

The bulk type IIB fields in the CS coupling give rise to the following discrete
SymTFT fields:
\begin{equation}
C_4 \;\to\; e_3, \qquad
B_2 \;\to\; b_1, \qquad
C_2 \;\to\; c_1 ,
\label{eq:IIB-discrete-reduction}
\end{equation}
where $e_3$, $b_1$, and $c_1$ are $\mathbb Z_N$-valued cochains on the 5D
SymTFT spacetime.\footnote{Here we assume the 5D manifold supporting the SymTFT has no non-trivial torsional structure, thus ordinary cochains are good enough.} The resulting 5D interaction takes the form \cite{Heckman:2022xgu}
\begin{equation}
S_{\text{cubic}}
=
\frac{2\pi i}{N}
\int_{M_5}
\kappa\;
e_3 \cup b_1 \cup c_1 ,
\label{eq:IIB-cubic}
\end{equation}
where $\kappa$ is determined by the torsion
pairing of the link geometry and will not be important for our purposes.

The kinetic terms of the ten-dimensional bulk fields lead, after reduction, to
BF-type couplings pairing each discrete background field with a dual field of
complementary degree. Alternatively, this can be understood as the flux-noncommutativity in the phase space under Hamiltonian formalism (see, e.g., \cite{Freed:2006yc,Apruzzi:2021nmk}). This leads to the introduction of the SymTFT field pairs
\begin{equation}
(e_3,e_1), \qquad (b_1,b_3), \qquad (c_1,c_3),
\end{equation}
where $e_1$, $b_3$, and $c_3$ represent background gauge fields canonically dual to $e_3$, $b_1$, and $c_1$ in the
5D TFT. Combining the BF-type couplings arising from the kinetic terms with the cubic
interaction \eqref{eq:IIB-cubic}, we arrive at the SymTFT action for finite symmetries
\begin{equation}
S_{\text{SymTFT}}
=
\frac{2\pi i}{N}
\int_{M_5}
\Big(
e_1 \cup \delta e_3
+
b_1 \cup \delta b_3
+
c_1 \cup \delta c_3
+
\kappa\;
e_3 \cup b_1 \cup c_1
\Big).
\label{eq:SymTFT-IIB}
\end{equation}

The non-Abelian finite global symmetry follows directly from the equations of motion of
the above SymTFT: Varying the action \eqref{eq:SymTFT-IIB} with respect to the field $e_3$ yields the
constraint
\begin{equation}
\delta e_1 \;=\; \kappa\, b_1 \cup c_1 \qquad.
\label{eq:eom-extension}
\end{equation}
This equation implies that the background field $e_1$ is no longer closed in the
presence of non-trivial $b_1$ and $c_1$ backgrounds. Instead, its failure to be closed
is controlled by a bilinear expression in the latter. This is precisely the form in \eqref{eq:pure-beta-constraint}, which build a non-Abelian symmetry group symmetry $G$ via the following pure beta-extension, fitting into a short exact sequence
\begin{equation}
1 \rightarrow \mathbb Z_N^{(e)} \rightarrow G \rightarrow \mathbb Z_N^{(b)} \times \mathbb Z_N^{(c)} \rightarrow 1 ,
\label{eq:group-extension}
\end{equation}
where the extension class is determined by the coefficient $\kappa$ appearing in the
cubic SymTFT coupling.

We emphasize that the appearance of the non-Abelian symmetry is not restricted to our assumption with $\mathrm{Tor}H^2=\mathrm{Tor}H^4=\mathbb{Z}_N$. Rather, it follows inevitably from the presence of the cubic Chern--Simons coupling
in type IIB string theory once the torsion sector is taken into account.
\subsection*{Example: $\mathbb C^3/\mathbb Z_3$ and the Group $\Delta_{27}$}
\label{app:C3Z3}

To be more illustrative, we specialize the general discussion to the canonical
example of D3-branes probing the orbifold singularity $X=\mathbb C^3/\mathbb Z_3$.
This case provides a concrete realization of the non-Abelian symmetry generated by
the SymTFT mechanism described above.

The link of the singularity is $\partial X = S^5/\mathbb Z_3$, whose torsion
cohomology gives rise to discrete background fields valued in $\mathbb Z_3$. In the
notation of the previous discussion, this corresponds to setting $N=3$ in the SymTFT
action \eqref{eq:SymTFT-IIB}. The reduction of the type IIB Chern--Simons couplings on
$\partial X$ yields a non-vanishing cubic interaction with coefficient
\begin{equation}
\kappa = 1 \qquad (\mathrm{mod}\; 3).
\end{equation}

The equation of motion \eqref{eq:eom-extension} then takes the explicit form
\begin{equation}
\delta e_1 \;=\; b_1 \cup c_1 \qquad (\mathrm{mod}\; 3),
\label{eq:C3Z3-eom}
\end{equation}
implying that the $\mathbb Z_3$ symmetry of $e_1$ forms a non-trivial
extension of the $\mathbb Z_3^{(b)} \times \mathbb Z_3^{(c)}$ symmetry of $b_1$ and $c_1$. The resulting finite global symmetry group $G$ fits into the exact sequence
\begin{equation}
1 \rightarrow \mathbb Z_3 \rightarrow \Delta_{27} \rightarrow \mathbb Z_3 \times \mathbb Z_3 \rightarrow 1 ,
\end{equation}
with non-trivial extension class. Equivalently, the generators corresponding to
$b_1$ and $c_1$ fail to commute, with their commutator given by the central element
associated with $e_1$.

This reproduces the discrete non-Abelian symmetry originally identified in the
worldvolume theory of D3-branes probing $\mathbb C^3/\mathbb Z_3$, as discussed in
early string-theoretic analyses such as \cite{Gukov:1998kn,Berasaluce-Gonzalez:2012abm,Garcia-Valdecasas:2019cqn}.

Finally, we note that non-Abelian finite global symmetries arising from pure
beta-extensions appear quite broadly in string-theoretic realizations of
QFTs. Beyond the 4D examples discussed here,
related structures can be found in 3D ABJ theories \cite{Bergman:2024its}
and in 2D quiver gauge theories \cite{Yu:2023nyn,Franco:2024mxa}, suggesting that the
SymTFT mechanism described in this Appendix has wide applicability.

\newpage

\bibliographystyle{utphys}
\bibliography{NonAbSymTFT}

\end{document}